\documentclass[11pt,a4paper]{article}
\pdfoutput=1

\usepackage{jcappub}
\usepackage{caption}
\usepackage{subcaption}
\usepackage{rotating}
\usepackage{color}
\usepackage{bm}
\usepackage{hyperref} 
\usepackage{url}
\usepackage{float}

\newcommand{\be}{\begin{equation}}
\newcommand{\ee}{\end{equation}}

\numberwithin{equation}{section}

\definecolor{rossos}{rgb}{0.7,0,0.3}
\definecolor{violachiaro}{rgb}{1,0.6,1}
\definecolor{rossochiaro}{rgb}{1,0.6,0.6}
\definecolor{verdechiaro}{rgb}{0.6,1,0.6}
\definecolor{giallochiaro}{rgb}{1,1,0.6}
\definecolor{bluscuro}{rgb}{0.15, 0.2, 0.9}
\definecolor{verdes}{rgb}{0.1, 0.5, 0.1}
\definecolor{gold}{rgb}{1,0.84,0}
\definecolor{forestgreen}{rgb}{0.13,0.55,0.13}

\definecolor{oucrimsonred}{rgb}{0.6, 0.0, 0.0}
\definecolor{persianblue}{rgb}{0.11, 0.22, 0.73}
\definecolor{forestgreen}{rgb}{0.13,0.35,0.13}
 \hypersetup{colorlinks, citecolor=oucrimsonred, linkcolor=persianblue, urlcolor=oucrimsonred}

\begin{document}

\title{A critical reassessment of particle Dark Matter limits from dwarf satellites}

\author[a,b]{Piero Ullio}
\author[a,b]{and Mauro Valli}

\affiliation[a]{SISSA, Via Bonomea, 265, 34136 Trieste, Italy}
\affiliation[b]{INFN, Sezione di Trieste, Via Bonomea 265, 34136 Trieste, Italy}

\emailAdd{piero.ullio@sissa.it, mauro.valli@sissa.it}

Ô\date{\today}
 
\abstract{
Dwarf satellite galaxies are ideal laboratories for identifying particle Dark Matter signals. When setting limits on particle Dark Matter properties from null searches, it becomes however crucial the level at which the Dark Matter density profile within these systems is constrained by observations. In the limit in which the spherical Jeans equation is assumed to be valid for a given tracer stellar population, we study the solution of this equation having the Dark Matter mass profile as an output rather than as a trial parametric input. Within our new formulation, we address to what level dwarf spheroidal galaxies feature a reliable mass estimator. We assess then possible extrapolation of the density profiles in the inner regions and -- keeping explicit the dependence on the orbital anisotropy profile of the tracer population -- we derive general trends on the line-of-sight integral of the density profile squared, a quantity commonly dubbed $J$-factor and crucial to estimate fluxes from prompt Dark Matter pair annihilations.

Taking Ursa Minor as a study case among Milky Way satellites, we perform Bayesian inference using the available kinematical data for this galaxy. Contrary to all previous studies, we avoid marginalization over quantities poorly constrained by observations or by theoretical arguments. We find minimal $J$-factors to be about 2 to 4 times smaller than commonly quoted estimates, approximately relaxing by the same amount the limit on Dark Matter pair annihilation cross section from gamma-ray surveys of Ursa Minor. At the same time, if one goes back to a fixed trial parametric form for the density, e.g. using a NFW or Burkert profile, we show that the minimal $J$ can hardly be reduced by more than a factor of 1.5.
} 
\keywords{Dark Matter, dwarf spheroidal galaxies, $J$-factors, WIMPs}
\maketitle


\section{Introduction}\label{intro}

The dwarf spheroidal satellites (dSphs) of the Milky Way are a prime target for Dark Matter (DM) indirect detection searches \cite{Strigari:2006rd,Strigari:2007at}. First of all, they are relatively close to us and have fairly large DM densities~\cite{2002MNRAS.333..697L,Battaglia:2013wqa,WalkerReviewMWdSphs}, and hence are expected  to be among the brightest DM-induced emission. From the point  of view of background contaminants, they seem to be ideal objects as well: Intrinsic emission from standard astrophysical sources can generally be neglected (they host old low-luminosity stellar populations and tiny -- most often below  detection sensitivities -- amounts of gas~\cite{2012AJ_McConnachie}); at the  same time, most dwarfs are located at intermediate or high galactic  latitudes where Galactic foregrounds are suppressed. Multi-wavelength campaigns have therefore been promoted to search for DM signals, with  some of the most impressive results obtained with $\gamma$-ray  telescopes: e.g., the Fermi collaboration has recently published  updated limits on weakly interacting massive particles (WIMPs), excluding pair annihilating cross sections at the level of WIMP thermal relic cross sections for DM masses lighter than about  100 GeV \cite{Ackermann:2015zua,Ahnen:2016qkx}.
Limits from these searches can be meaningfully translated into Particle Physics constraints (see e.g. \cite{Martinez:2009jh,Cirelli:2015bda,Liem:2016xpm}). Moreover, the compelling DM content of dwarf spheroidals has also triggered general interest from the community in order to assess the nature of the DM particle (see for example the pioneer work in \cite{1983ApJ...266L..21L} and the more recent \cite{Diez-Tejedor:2014naa,Domcke:2014kla} and \cite{Zavala:2012us,Kaplinghat:2015aga}).

A few of the proposed signals (including the $\gamma$-ray flux from WIMP pair annihilations just mentioned or, e.g., the X-ray signal  from sterile neutrino decays \cite{Adhikari:2016bei})  are connected to prompt  emission from DM particles \cite{Cirelli:2010xx}. In these cases the flux predictions can be conveniently factorized into a term depending on the DM particle physics embedding (specifying, e.g., for a WIMP: the mass, the annihilation cross section and the emission yields), and a term depending on the distribution of DM in the  dwarf. For DM pair annihilation signals, the latter -- usually dubbed $J$-factor -- is defined as an angular and line-of-sight (l.o.s.) integral of the  square of the DM density profile:
\begin{equation}
  \label{eq:j_psi}
  J \equiv \int_{\Delta\Omega} d\Omega \int_{\rm l.o.s.} d\ell \, \rho^2(\vec{x})\,.
\end{equation}
The tight constraints on particle DM properties claimed from dwarf surveys reflect the assumption that fairly small observational and theoretical uncertainties affect these astrophysical factors:  e.g. in the analysis of~\cite{Ackermann:2015zua} mentioned above, limits are derived exploiting the full ensemble of known dwarfs and introducing a likelihood in which the $J$-factor dependence for each dwarf $i$ follows a log-normal distribution of given central value $\overline{\log_{10}(J_{i})}$ and width $\sigma_i$.  For most of the so-called classical dwarfs -- namely the only 8 dwarfs known before the first discoveries of ultra-faint ones as a byproduct in large scale structure surveys \cite{2015ApJ...807...50B} --  the assumed values of  $\sigma_i$ are of the  order of 0.2 \cite{Ackermann2011,Ackermann:2013yva,Ackermann:2015zua}. This translates into  an uncertainty on $J$ of about a factor of 1.5. At a superficial level, looking at Eq.~(\ref{eq:j_psi}) and assuming as known the distance of  the object as well as -- most crucially -- the shape of the DM density profile, one would deduce that the normalization of the density  profile can be inferred from observations with an uncertainty at the 20-25\% level.

Indeed, once a specific approach has been adopted in determining such normalization, it is in general true that, in case of the classical dwarfs, the quality of kinematical data is adequate to provide fairly small statistical errors \cite{Martinez:2013els}. On the other hand, it is a much more delicate issue to address intrinsic systematic errors of the theoretical models and their impact on parameter determinations, including the normalization and more critically the $J$-factor itself. Analyses in the literature give contradictory results: e.g. \cite{Bonnivard:2014kza} presents a comprehensive discussion of the impact of different theoretical assumptions on interpreting kinematical data within the framework of the Jeans equation (a moment projection of the collision-less Boltzmann equation) in the spherical symmetric limit; they conclude that systematic biases and uncertainties on the $J$-factor for classical dwarfs are up to a factor of 3 to 4, including a rather mild impact of a factor of 2.5 from the effect of the dwarf being a triaxial system rather than a spherical one. On the other hand, the authors in \cite{Hayashi:2012si} (see also Ref.~\cite{Hayashi:2015yfa}) show that the impact of axisymmetric models  for non-spherical DM structures can be much more dramatic on the mass at a reference radius, and hence the normalization of the profile, yielding uncertainties of factors as large as 10 even for the classical dwarfs (Ref.~\cite{Hayashi:2012si} does not discuss the impact on the $J$-factors, but, roughly speaking, the scaling should go as the square of the normalization of the profile). 

In this work, while still assuming as theoretical playground the Jeans equation for a spherically symmetric system, we aim to discuss the impact of the method that has been adopted in its solution by the vast majority of recent analyses. This goes into two steps: The first is to introduce parametric forms for the quantities appearing in the equation, namely the DM mass profile and the number density and velocity anisotropy profiles of the stellar populations used as dynamical tracers; The second is to sample the relative parameter space via Monte Carlo techniques in order to perform Bayesian inference, despite some loose theoretical and observational guidance. In particular, it is well known that the stellar anisotropy profile introduces patterns of degeneracies in the result and is unfortunately scarcely constrained by observations. Several recent studies seem to indicate a minor impact on the $J$-factor estimates \cite{Walker:2011fs,Charbonnier:2011ft,GeringerSameth:2011iw,Mazziotta:2012ux,Martinez:2013els,Geringer-Sameth:2014yza,Geringer-Sameth:2014qqa,Bonnivard:2015xpq}, however they mostly refer to blind analyses involving a marginalization over a parameter space and integration measure which, not being driven by observations or by theory, are essentially an arbitrary choice. 

We propose here instead to examine the problem under a different perspective, exploiting an approach in which the Jeans equation is so-to-speak ``inverted", rewriting the DM mass profile in a form in which its dependence on the stellar anisotropy profile becomes explicit. This method was originally outlined in two parallel analyses, see~\cite{Mamon:2010MNRAS,Wolf:2009tu}. In this work we re-derive the inversion formula in a new compact form, suitable for numerical analyses, and use it for the first time to discuss $J$-factor estimates. After showing some general trends, we examine, as study case, Ursa Minor, the closest among the classical dwarfs and hence one of those entering critically in the limits concerning the DM particle properties. Results for the $J$-factor are given without the need to marginalize over unknown parameters.  


\section{Mass models from the spherical Jeans equation}

Mass models for dwarf satellites of the Milky Way are most commonly derived exploiting a stellar population as a dynamical tracer of the underlying gravitational potential well (and hence of the dominant mass component, namely the DM mass profile).  Supposing that the tracers belong to a non-rotating pressure-supported population in dynamical equilibrium, one can assume that the stellar density function obeys a time-independent collision-less Boltzmann equation, to be solved projecting out velocity moments. Going to the limit in which the stellar and DM components are spherically symmetric, the second moment projection reduces to a single Jeans equation~\cite{1992ApJDejongheMerritt,bt08}, usually recast in the form:
\begin{equation}
  \label{eq:jeansdiff}
  \frac{dp}{dr}+ \frac{2\,\beta(r)}{r} p(r) = - \nu(r)\,\frac{G_N \mathcal{M}(r)}{r^2}\,,
  \quad
  {\rm with}
  \quad
  p(r) \equiv \nu(r) \sigma_r^{2}(r)\,.
\end{equation}
This equation shows that the radial dynamical pressure $p(r)$, the product of the tracer number density profile $\nu(r)$ and the radial component of the velocity dispersion tensor $\sigma_r^{2}(r)$, can be expressed in function of $\nu(r)$ itself, as well as of the total mass profile $\mathcal{M}(r)$ and the orbital velocity dispersion anisotropy $\beta(r)$. The latter involves also the other two diagonal components of the velocity dispersion tensor $\sigma_{\theta}^{2}$ and $\sigma_{\phi}^{2}$, being defined as:
\begin{equation}
  \label{eq:beta}
  \beta(r) \equiv 1-\frac{\sigma^{2}_{\theta}(r)+\sigma^{2}_{\varphi}(r)}{2\,\sigma^{2}_{r}(r)} \, .
\end{equation}
$\beta(r)$ parametrizes the deviation of the velocity ellipsoid from a sphere of radius squared $\sigma^{2}_{r}=\sigma^{2}_{\theta}=\sigma^{2}_{\varphi}$ \cite{bt08,Strigari:2013iaa}. By definition, $\beta(r)$ can cover the range $(-\infty,1]$, where the lower (upper) extreme corresponds to tracers moving on purely circular (radial) orbits. The formal solution of Eq.~(\ref{eq:jeansdiff}) is: 
\begin{equation}
  \label{eq:radpress}
  p(r) = G_{N} \int_r^\infty dr^\prime \frac{ \nu(r^\prime)  \mathcal{M}(r^\prime)}{{r^\prime}^2} \,\exp\left[2 \int_{r}^{r'} dr'' \frac{\beta(r'')}{r''}  \right] \,.
\end{equation}
The difficulty in fully exploiting this approach is that, despite the assumption of dynamical equilibrium and spherical symmetry at the bases of Eq.~(\ref{eq:jeansdiff}), the problem still involves three unknown functions: $\mathcal{M}(r)$, $\nu(r)$ and $\beta(r)$, to be inferred from only two quantities connected to observations: The first is the stellar surface density:
\begin{equation}
  \label{eq:starSB} 
  I(R) = \int_R^\infty dr \frac{2 r}{\sqrt{r^2-R^2}} \, \nu(r)\,,
\end{equation}
(here and everywhere in the following  ``$R$'' refers to the l.o.s. projected radius, while ``$r$'' is the radius in the spherical coordinate system centered on the dwarf) which, assuming constant stellar luminosity over the whole system, is proportional to the surface brightness as mapped in photometric surveys (see, e.g., \cite{Irwin1995}). The second quantity one can derive from observations is the l.o.s. velocity dispersion $\sigma_{los}(R)$, which traces the only velocity component accessible to spectroscopic measurements \cite{Battaglia:2013wqa,WalkerReviewMWdSphs}. The l.o.s. velocity dispersion profile can be expressed in terms of the radial dynamical pressure~\cite{Binney&Mamon82}:
\begin{equation}
  \label{eq:Pproj}
  \sigma_{los}^2(R) =  \frac{1}{I(R)} \int_R^\infty dr \frac{2 r}{\sqrt{r^2-R^2}} \,
  \left[1-\beta(r)\frac{R^2}{r^2}\right] \, p(r)\,.
\end{equation}

The mapping of the three unknowns into two observables is usually done by introducing parametric forms for the three unknowns: The template for $\nu(r)$ is typically related to a $I(R)$ supported in stellar photometric studies, 
such as the Plummer~\cite{plummer}, the King~\cite{king} and the Sersic~\cite{sersic} profiles.
$\mathcal{M}(r)$ usually stems from DM density profile $\rho(r)$ motivated by: 
\begin{itemize}
\item[-] numerical N-body simulations of hierarchical clustering in cold DM cosmologies \cite{Navarro:1995iw}, such as, e.g., the Navarro-Frenk-White (NFW) profile~\cite{Navarro:1996gj} (with a $1/r$ singularity towards the center of the system and a scale radius $r_n$ to set the transition into the $1/r^3$ scaling at large radii);
\item[-] phenomenological studies \cite{Salucci:2000ps,Salucci:2011ee} on the distribution of DM in galaxies, such as, e.g., the Burkert profile~\cite{burkert} (in this case the characteristic scale $r_b$ sets the size of the inner constant density core, before the transition again into the $1/r^3$ regime at large radii).
\end{itemize}
Finally for what regards the stellar anisotropy profile $\beta(r)$, templates assumed in the literature (see e.g. \cite{1979PAZh577O,1985AJ90.1027M,Baes:2007tx}) reflect more simplicity arguments rather than profound physical motivations, ranging from some constant value to functions connecting two asymptotic values at large and small radii, eventually with some parameter setting the sharpness of the transition. 

Attempts to break the degeneracy in Eq.~(\ref{eq:jeansdiff}) between $\mathcal{M}(r)$ and $\beta(r)$ using higher velocity moments \cite{1990AJ.....99.1548M,2002MNRAS.333..697L,Lokas:2004sw,2009MNRAS.394L.102L,Richardson:2012ig,Richardson:2013lja} or the determination of multiple tracer populations \cite{Walker:2011zu,Amorisco:2011hb,Strigari:2014yea}, together with the progress of N-body simulations \cite{2016MNRAS.457.844F,2016MNRAS.457.1931S,2016arXiv160205957W}, may represent a promising future opportunity to fully overcome current study limitations due to the l.o.s. measurements available for these systems \cite{Richardson:2013hga}.
Given the large parameter space at hand, one needs an efficient scanning technique and careful addressing of error propagation: For $\nu(r)$ (or directly for $I(R$)) a frequentist fit of data is usually implemented, in case of the classical dwarfs  most often referring to the data compilation in~\cite{Irwin1995}. On the other hand, all recent analyses explore the parameter space connected to  $\mathcal{M}(r)$ and $\beta(r)$ introducing a likelihood addressing the matching of the theoretical model for $\sigma_{los}(R)$ with data, and employ a  Markov Chain Monte Carlo (MCMC) sampling in the context of Bayesian inference. After a choice of priors and integration measures -- which in case of the anisotropy function, as for the choice of the functional form, seem essentially arbitrary -- one derives posteriors for  the parameters defining $\mathcal{M}(r)$ or $\rho(r)$, as well as for derived quantities such as the $J$-factor introduced in Eq.~(\ref{eq:j_psi}) above. 
While this procedure gives for the classical dwarfs posteriors on $J$ with small error bars, it is not transparent what is the impact of having selected given parametric forms, priors and integration measures, especially in the case of $\beta(r)$ for which a robust physical guidance is still missing. This is one of the issues we wish to investigate in this paper, exploring an alternative approach in which $\mathcal{M}(r)$ is a derived quantity and the dependance of the result on the assumed form for $\beta(r)$ is kept explicit. 


\section{General trends from an inversion formula}\label{Sec:OurInversion}                  

The method we build on here has been already outlined in two parallel analyses, see~\cite{Mamon:2010MNRAS,Wolf:2009tu}. The starting point relies on the observation that the two available observables, namely Eqs.~(\ref{eq:starSB}) and~(\ref{eq:Pproj}) above, correspond to the Abel transform $f$ of a function $\widehat{f}$:
\begin{equation}
  f(x) = \mathbf{A}[\widehat{f}(y)] = \int_x^\infty \frac{dy}{\sqrt{y-x}} \, \widehat{f}(y)
 \quad \quad \Leftrightarrow \quad \quad 
  \widehat{f}(y) = \mathbf{A}^{-1}[f(x)] = -\frac{1}{\pi} \int_y^\infty \frac{dx}{\sqrt{x-y}} \, \frac{df}{dx}\,.  
\end{equation}
In fact, looking back at Eq.~(\ref{eq:starSB}) the surface density $I(R^2)$ is the Abel transform of the number density profile $\nu(r)$ (everywhere in the following we will use $\widehat{I}(r^2)$ to indicate the number density profile instead of $\nu(r)$). In a similar fashion, introducing the projected dynamical pressure, $P(R) \equiv \sigma_{los}^2(R) \, I(R)$,
this expression can be manipulated (see Appendix~\ref{App:JeansInversion} for a detailed derivation) inverting it into a formula for the radial dynamical pressure:
\begin{equation}
  p(r) = [a_\beta(r)-1] \int_{r}^{\infty} dr^\prime \mathcal{H}_{\beta}(r,r^\prime) \, \frac{d\widehat{P}}{dr^\prime}    
  \label{eq:pofr}   
\end{equation}
where $\widehat{P}(r^2)$ is the inverse Abel transform of $P(R^2)$ and we defined:
\begin{equation}
 a_\beta(r) \equiv - \frac{\beta(r)}{1-\beta(r)}\,,
 \quad 
 {\rm and}
 \quad
 \mathcal{H}_{\beta}(r,r^\prime) \equiv \exp\left(\int_{r}^{r^\prime} dr^{\prime\prime} \frac{a_\beta(r^{\prime\prime})}{r^{\prime\prime}}\right)\,.
   \label{eq:ourinvdef}
\end{equation}
Inserting this result into the Jeans equation~\cite{bt08}, Eq.~(\ref{eq:jeansdiff}), one can find the mass profile:
\begin{equation}
  \mathcal{M}(r) = \frac{r^2}{G_{N}\,\widehat{I}(r)} \left\{- \frac{d\widehat{P}}{dr} [1-a_\beta(r)]
  - \frac{a_\beta(r) \, b_\beta(r)}{r}  \int_{r}^{\infty} dr^\prime \mathcal{H}_{\beta}(r,r^\prime) \,
     \frac{d\widehat{P}}{dr^\prime} \right\}\,,
  \label{eq:ourinv}   
\end{equation}
with
\begin{equation}
 b_\beta(r) \equiv 3 - a_\beta(r) + \frac{d \log a_\beta}{ d \log r}\,.
   \label{eq:ourinvdef2}
\end{equation}
An expression equivalent to Eq.~(\ref{eq:ourinv}) can be extracted from Ref.~ \cite{Wolf:2009tu}, while Ref.~ \cite{Mamon:2010MNRAS} gives explicit formulas for several simple anisotropy models; it is however the first time such a compact form in terms of observables is given, showing that the mass profile depends on the anisotropy profile only through the function $a_\beta(r)$ as defined above. 
The expression in Eq.f(\ref{eq:ourinvdef2}) shows that, at given $a_\beta(r)$, the mass profile can be properly reconstructed if the projected dynamical pressure can be efficiently constrained from data.
From the expression just derived one can read out the behaviour of the mass function in some special limits that will be useful in the discussion below. First of all, in case of isotropic stellar orbits,
namely  $\beta(r) \rightarrow 0$ for any $r$ (and hence $a_\beta(r) \rightarrow 0$):
\begin{equation}
  \mathcal{M}_{\beta=0}(r)  =   - \frac{r^2}{G_{N}\,\widehat{I}(r)} \frac{d\widehat{P}}{dr}\,.
  \label{eq:isomass}   
\end{equation} 
For circular orbits instead, i.e. $\beta(r) \rightarrow -\infty$ for any $r$ (and hence $a_\beta(r) \rightarrow 1$, $b_\beta(r) \rightarrow 2$):
\begin{equation}
  \mathcal{M}_{\beta \to -\infty}(r)  =  - \frac{2}{G_{N}\, \widehat{I}(r)} \int_{r}^{\infty} d r' r'  \frac{d\widehat{P}}{dr^\prime} \ .
  \label{eq:circmass}   
\end{equation} 
Eq.~(\ref{eq:ourinv}) is derived under the hypothesis $\beta \neq 1$. To take the exact radial orbit limit it is simpler 
to notice that the radial pressure for $\beta = 1 $ (i.e. $a_\beta \rightarrow -\infty$) takes the form:
\begin{equation}
p_{\beta=1}(r)= - r \frac{d \widehat{P}}{dr} \ ,
\end{equation}
and replacing this into the Jeans equation, Eq.~(\ref{eq:jeansdiff}), one finds:
\begin{equation}
  \mathcal{M}_{\beta=1}(r) = \frac{1}{G_{N}\, \widehat{I}(r)} \frac{d}{dr} 
  \left( r^3 \frac{d\widehat{P}}{dr} \right)\,.
  \label{eq:radmass}   
\end{equation} 

\subsection{A mass estimator for dwarf galaxies?}\label{Sec:MassEstimator}

As first noticed in MCMC analyses, regardless of what is assumed for the stellar velocity anisotropy $\beta(r)$, all models fitting the l.o.s. velocity dispersion profile tend to have approximately the same mass at a scale corresponding to about the surface brightness half-light radius~\cite{strigari2007m,Strigari:2008ib,2008ApJ672904P,2008ApJ6871460P,Walker2009,Walker2009err,Amorisco:2010ns,Campbell:2016vkb}. In Ref.~\cite{Wolf:2009tu} a rationale for the existence of such a mass estimator is provided through an analytic manipulation of the solution of the Jeans equation. Briefly recapping their argument, it is useful to consider the difference between the mass profile $\mathcal{M}(r)$  for a generic anisotropy profile and $\mathcal{M}_{\beta=0}(r)$;  after some algebra one finds:
\begin{equation}
  \mathcal{M}(r) - \mathcal{M}_{\beta=0}(r)  = 
  - \frac{\beta(r) \, r\, \sigma_r^2}{G_N} 
  \left( \frac{d \log  \widehat{I}}{ d \log r} + \frac{d \log \sigma_r^2}{ d \log r} + \frac{d \log \beta}{ d \log r} + 3  \right) \ .
  \label{eq:wolf21}
\end{equation} 
Among the terms within brackets on the r.h.s., towards the outskirts of the dwarf, the logarithmic derivative of the stellar number density $\widehat{I}(r)$ rapidly varies from close to zero to a  negative number. In the same region, the l.o.s. velocity dispersion is generally close to being flat and also $\sigma_r^2$ is arguably not too rapidly varying. If one now assumes that also $\beta(r)$ does not have a sharp change in that region, the difference in Eq.~(\ref{eq:wolf21}) is approximately zero at the radius $r_{*}$ defined as: 
\begin{equation}
  \left. - \frac{d \log \widehat{I}}{ d \log r} \right|_{r=r_{*}} = 3 \, .
  \label{eq:wolfrstar}   
\end{equation} 
Since this condition does not depend on $\beta$, $\mathcal{M}$ almost matches $\mathcal{M}_{\beta=0}$  at $r_{*}$ regardless of  the stellar anisotropy:
\begin{equation}
   \mathcal{M}(r_{*}) \simeq  \mathcal{M}_{\beta=0}(r_{*}) \simeq
   \frac{3}{G_{N}} \langle \sigma_{los}^{2} \rangle\, r_{*} \equiv \mathcal{M}_{*}  \,,
  \label{eq:wolfmstar}   
\end{equation}
where in the second step the symbol $\langle \ \rangle$ stands for a weighted average on the stellar number density, and $\langle \sigma_{los}^2 \rangle$ has been factorized out in computing $\widehat{P}(r)$ and implementing it in Eq.~(\ref{eq:isomass}).  

\begin{figure}[!t!]
  \begin{subfigure}[b]{0.4\textwidth}
    \centering
    \includegraphics[scale = 0.4]{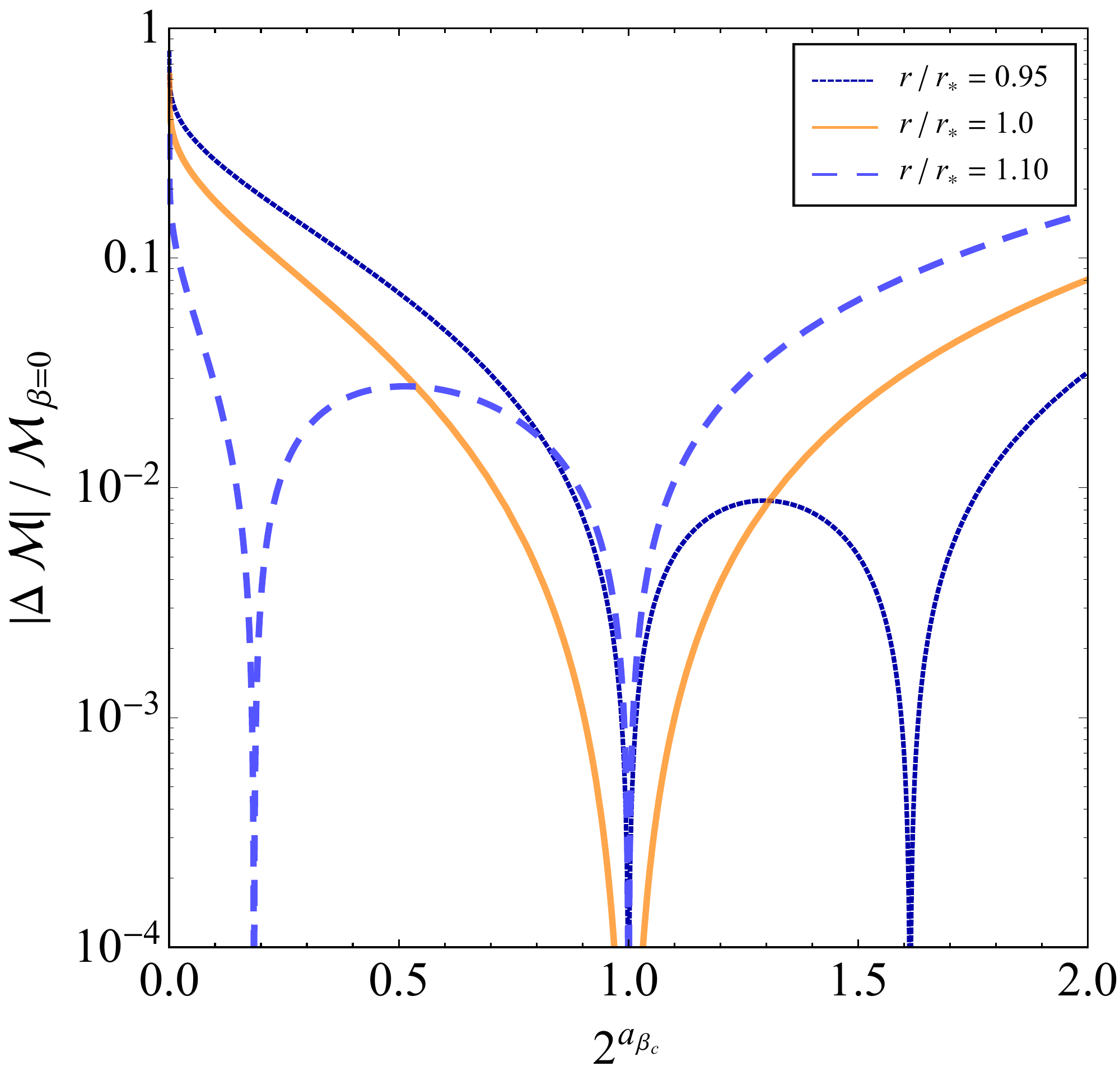}
  \end{subfigure}
\hfill
  \begin{subfigure}[b]{0.6\textwidth}
    \centering
    \includegraphics[scale = 0.38]{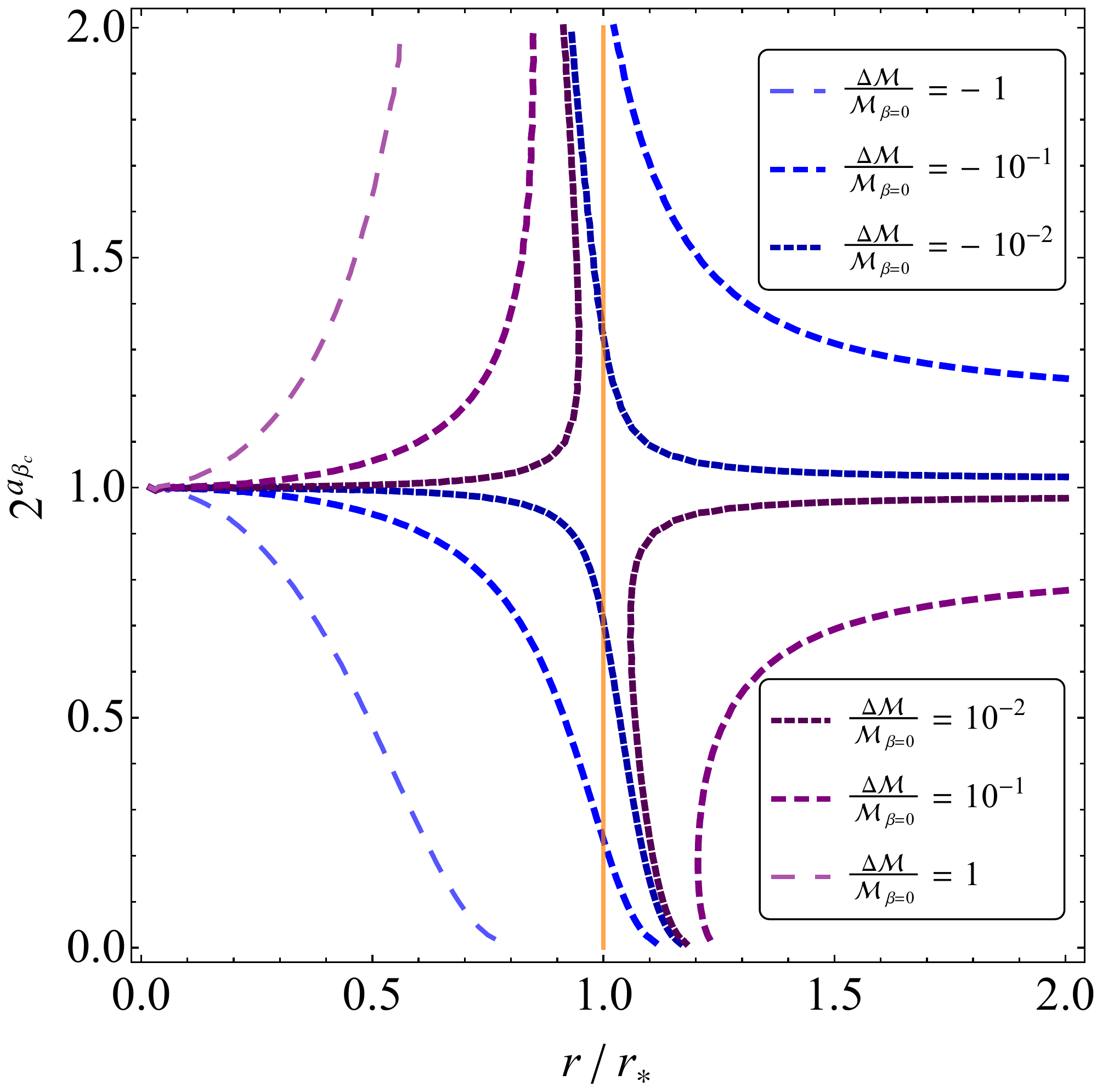}
  \end{subfigure}
  \caption{\underline{Left panel:} \textit{Difference in mass profiles,  $\Delta \mathcal{M}(r) \equiv \mathcal{M}_{\beta=\beta_c}(r) - \mathcal{M}_{\beta=0}(r)$, relative to the isotropic case,  $\mathcal{M}_{\beta=0}(r)$, for the  model with constant $\sigma_{los}$, Plummer stellar profile and constant orbital anisotropy, as a function of $2^{a_{\beta_{c}}}$. The orange, light blue and dark blue curves correspond, respectively, to a radius equal to  $r_{*} = \sqrt{3/2} \, R_{\textrm{\tiny{1/2}}}$, $5\%$ lower, $10\%$ larger.}  \underline{Right panel:} \textit{Isolevels for fixed relative mass difference in the plane $2^{a_{\beta_c}}$ versus $r/r_{*}$ within the same set of assumptions. The vertical orange line indicates $r_{*}$ as expected mass estimator.}}
    \label{fig:mass_estimator}
\end{figure}

We can check here this result with the formulas derived at the beginning of this Section. We start from a simple model where $\sigma_{los}^2(R)$ is assumed to be just a constant $\sigma_{los}^2$, and the stellar profile is described by a Plummer model~\cite{plummer}, a case in which the Abel transform can be performed analytically: 
\begin{equation}
  I(R^2)=\frac{I_{0}} {\pi R_{\textrm{\tiny{1/2}}}^2} \frac{1}{(1+R^2/R_{\textrm{\tiny{1/2}}}^2)^{2}} 
\quad \quad \Leftrightarrow \quad \quad
   \widehat{I}(r^2)=\frac{3  I_{0}} {4 \pi R_{\textrm{\tiny{1/2}}}^3} \frac{1}{(1+r^2/R_{\textrm{\tiny{1/2}}}^2)^{\frac{5}{2}}} \,.
  \label{eq:plummer}
\end{equation}
Under these two working hypotheses, the mass profile in the isotropic case has also a simple analytical form:
\begin{equation}
  \mathcal{M}_{\beta=0}(r)  = \frac{r\,\sigma_{los}^2}{G_{N}} \frac{5 \; r^2/R_{\textrm{\tiny{1/2}}}^2}{1+r^2/R_{\textrm{\tiny{1/2}}}^2} \,, 
  \label{eq:0sigmaconstmass}   
\end{equation} 
while Eq.~(\ref{eq:wolfrstar}) gives $r_{*} = \sqrt{3/2} \, R_{\textrm{\tiny{1/2}}}$. 
Assuming also a constant anisotropy profile, $\beta(r)=\beta_c$, in the left panel of Fig.~\ref{fig:mass_estimator} we show the relative difference in mass $|\Delta \mathcal{M}(r)|/\mathcal{M}_{\beta=0}(r) \equiv |\mathcal{M}_{\beta=\beta_c}(r) - \mathcal{M}_{\beta=0}(r)|/\mathcal{M}_{\beta=0}(r)$ versus the quantity $2^{a_{\beta_c}}$, useful to have circular and radial stellar anisotropies  equally spaced in the segment  $[0,2]$ (i.e. $0$ corresponds to radial, $1$ to isotropic and $2$ to circular orbits). 
The solid orange line corresponds to $r=r_*$ and shows that in the specific simple model under consideration the goodness of $r_{*}$ and $\mathcal{M}_{*}$ as mass estimator is within a level of about $8\%$ going to circular orbits, while it degrades to $10\%$ and larger towards radial orbits (in the purely radial limit derived in Eq.~(\ref{eq:radmass}) the discrepancy reaches the value of $75\%$). Also shown in the plot is the relative mass difference at $r=0.95\, r_{*}$ and $r=1.1\, r_{*}$ for which there is a better match, respectively, in the radial and circular regimes, as well as a larger discrepancy in the opposite regimes. In the right panel of Fig.~\ref{fig:mass_estimator} we show the isolevels for fixed relative mass difference in the plane $2^{a_{\beta_c}}$ versus $r/r_{*}$; one can see that -- still in the same model introduced above -- the mass difference is minimized along a curve that is slightly tilted with respect to the estimator proposed in Eq.~(\ref{eq:wolfmstar}), $r= r_{*}$. On the other hand moving away from $r_{*}$  the match rapidly diminishes in one of the two regimes; for example, taking $r= R_{\textrm{\tiny{1/2}}} \simeq 0.82 \, r_{*}$ as estimator radius is in our example a significantly worse choice, with relative mass differences at the level of $15\%$ for purely circular orbits and raising up to the level of 150\%  in the radial regime. The $3$D half-light radius proposed in \cite{Wolf:2009tu}, namely $r_{1/2} \simeq 1.3 \, R_{\textrm{\tiny{1/2}}} \simeq 1.06 \, r_*$ for the Plummer case, is a better choice.

\begin{figure}[!t!]
  \begin{subfigure}[b]{0.5\textwidth}
    \centering
    \includegraphics[scale=0.37]{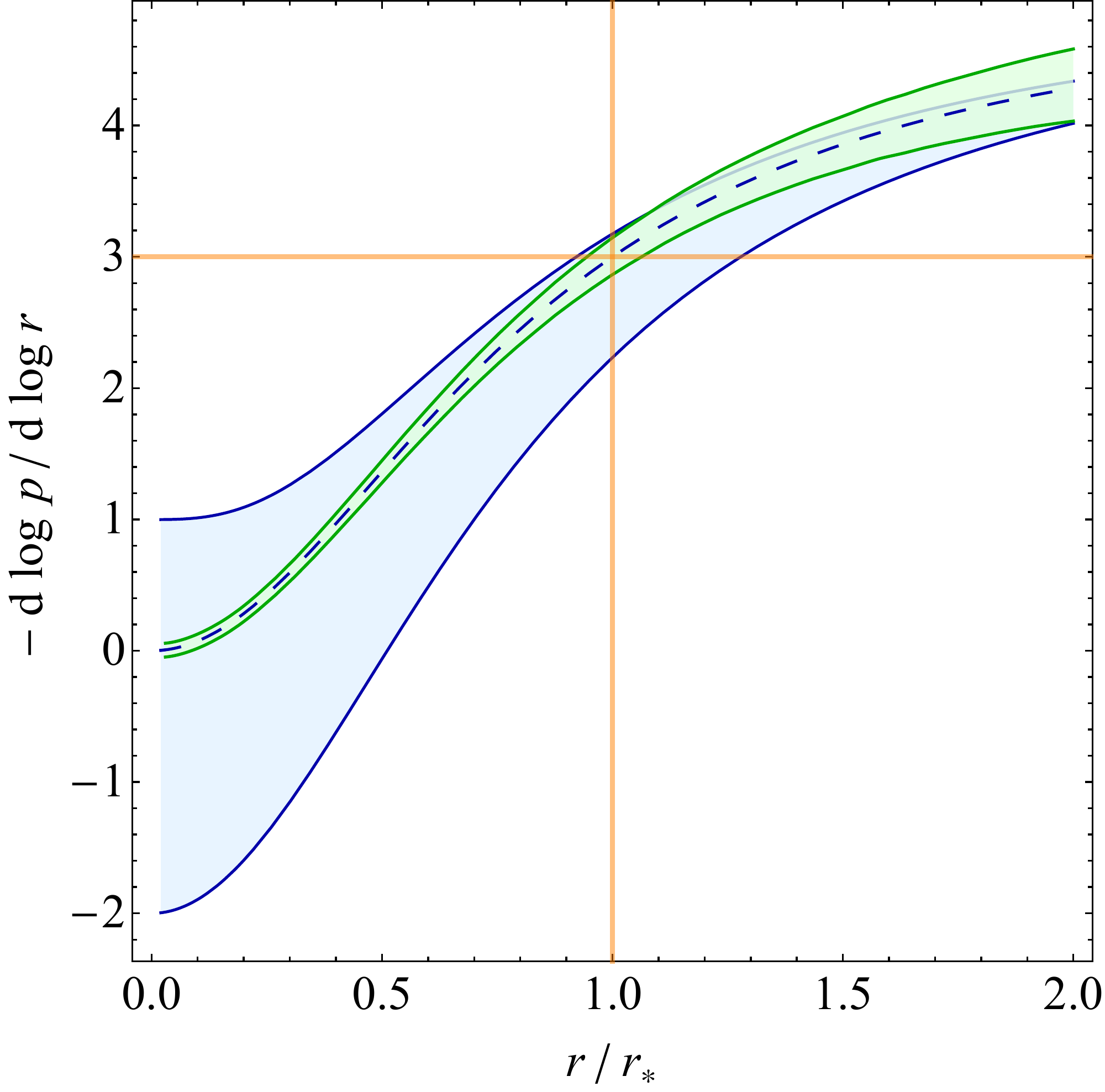}
  \end{subfigure}
\hfill
  \begin{subfigure}[b]{0.5\textwidth}
    \centering
    \includegraphics[scale=0.36]{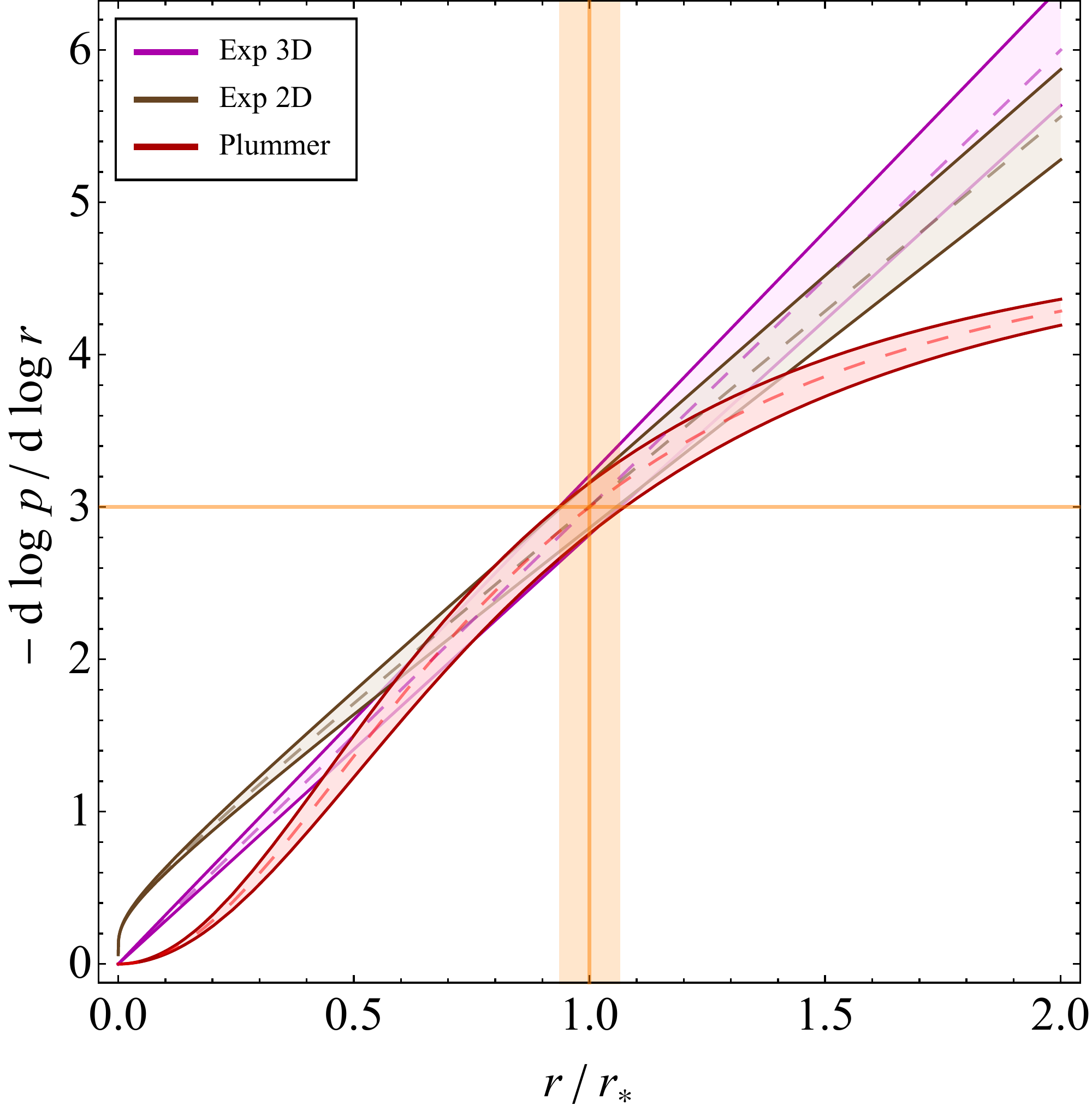}
  \end{subfigure}
  \caption{\underline{Left panel:} \textit{Logarithmic derivative of the dynamical pressure for a simple model with constant orbital anisotropy and Plummer stellar profile versus $r/r_{*}$. The blue band is relative to the case of constant l.o.s. velocity dispersion and $\beta_{c}$ varied in $(-\infty, 1]$, while the green one is obtained assuming a linear form in $R/R_{\textrm{\tiny{1/2}}}$ with slope $\pm \, 6\%$ and $\beta=0$.  }  \underline{Right panel:} \textit{Same quantity assuming constant $\sigma_{los}$ and $\beta=0$, but for the stellar profiles introduced in Eq.~(\ref{eq:plummer}), Eq.~(\ref{eq:exp2D})-(\ref{eq:exp3D}). The bands reflect typical uncertainty on the characteristic scale parameter of the stellar template.}}
    \label{fig:radpress}
\end{figure}

The failure of $\mathcal{M}_{*}$ as an exact estimator stems from the fact that even assuming that $\sigma_{los}$ is constant, for $\beta \neq 0$ there is still a non-negligible radial dependence in $\sigma_r$ (for $\beta = 0$ one trivially gets $\sigma_r(r) = \sigma_{los}$). This is shown in Fig.~\ref{fig:radpress} where we plot the logarithmic derivative of the radial dynamical pressure versus $r/r_{*}$. This quantity is related to $\sigma_r(r)$ via:  
\begin{equation}
  \frac{d \log p}{ d \log r} =  \frac{d \log \sigma_r^2}{ d \log r} + \frac{d \log \widehat{I}}{ d \log r}\,,
\end{equation} 
and hence, for $\beta = 0$ and constant $\sigma_{los}$, it coincides with the logarithmic derivative of the stellar number density $\widehat{I}$. In Fig.~\ref{fig:radpress} $-d \log  \widehat{I} / d \log r$ is plotted with a dashed line; by definition it crosses the value of $3$ at $r_{*}$. The blue band on the left panel shows the span in the logarithmic derivative for $p(r)$ when varying $\beta_{c}$ in the whole range of $(-\infty ,1]$ (respectively, upper and lower boundary of the band). 
The intersection of the band with the horizontal line at the value of $3$ gives the shift on $r$ needed to get $\Delta \mathcal{M} = 0$; the one with the vertical line at $r=r_{*}$ gives instead the magnitude of the departure of $\mathcal{M}_{*}$ from being an exact mass estimator.
Although the assumptions in the model we considered may appear rather drastic, the trends displayed are actually general. First, the hypothesis of constant $\sigma_{los}$ is not critical. 
Taking into account that available kinematical informations (see, for example, the binned data in \cite{Walker2009}) suggest l.o.s. velocity dispersions to be nearly flat in the region around $r_{*}$, we can parametrize  $\sigma_{los}(R)$ via the linear expression $c_{0} + c_{1} \, R /R_{\textrm{\tiny{1/2}}} $ and vary the slope $c_{1}/c_{0}$ in a generous range encompassing trends usually reported in literature, as e.g. those in \cite{Walker2009} (note however that the error associated to an overall normalization, while propagating on $\mathcal{M}_{\star}$, does not enter in relative mass differences).
As can be seen in the left panel of Fig.~\ref{fig:radpress} the impact of this uncertainty is marginal with respect to the one due to the orbital anisotropy.
The same conclusion holds when considering  also the second ingredient at hand, the modeling of the stellar distribution. As alternatives to the Plummer model, we introduce the two following exponential templates:
\begin{equation}
  I(R^2) = \frac{I_{0}}{2 \pi R_{e}^{2}}  \exp\left(-\frac{R}{R_{e}}\right) 
\quad \quad \Leftrightarrow \quad \quad
  \widehat{I}(r^2) = \frac{I_{0}}{2 \pi^{2} R_{e}^{2}} \, K_{0}\left( \frac{r}{R_{e}}\right) \,,
\label{eq:exp2D}
\end{equation}
for which $r_{*} \simeq 2.54\, R_{e} \simeq 1.51 \, R_{\textrm{\tiny{1/2}}}$, and:
\begin{equation}
  I(R^2) = \frac{I_{0}}{4 \pi r_{e}^{2}} \, \frac{R}{r_{e}} \, K_{1}\left( \frac{R}{r_{e}}\right)
\quad \quad \Leftrightarrow \quad \quad
  \widehat{I}(r^2) = \frac{I_{0}}{8 \pi r_{e}^{3}}\exp\left(-\frac{r}{r_{e}}\right) \,,
\label{eq:exp3D}
\end{equation}
for which $r_{*} = 3 \, r_{e} \simeq 1.48 \, R_{\textrm{\tiny{1/2}}}$ ($K_{n}(x)$ is the modified Bessel function of the second kind).
The functions above qualitatively reproduce typical realizations of a multi-parameter template like the Sersic profile, while differing substantially in the inner region from the Plummer model (the King model is instead qualitatively equivalent to the Plummer).  Back to the working hypothesis of constant $\sigma_{los}$ and $\beta=0$, we plot in the right panel of Fig.~\ref{fig:radpress} the logarithmic derivative of $p(r)$ as a function of $r/r_{*}$ for the three stellar profiles. 
The uncertainty bands displayed are obtained by the generation of a set of surface brightness mock data with binning and associated errors matching typical photometric maps of the 8 classical dSphs as in \cite{Irwin1995}.
Although within a given stellar profile the impact on $\mathcal{M}_{*}$ from our estimated uncertainty on $r_{*}$  is negligible as shown in Fig.~\ref{fig:radpress}, a higher impact derives from the mis-reconstruction of the stellar profile within a wrongly assumed parametric form, see the shift in $r_{*}$ with respect to the observable radius $R_{\textrm{\tiny{1/2}}}$ between the Plummer model and the two exponential profiles.

To summarize  this part of the discussion, we find that the uncertainty on the dwarf mass estimator is dominated by the kinematical determination of the normalization on $\sigma_{los}$ as long as models with radial-like tracer orbits are not included in the analysis. While general criteria hinting for unphysical phase-space densities in connection to radial-like tracer orbits are present in literature (see e.g. \cite{2010MNRAS.408.1070C,2011ApJ...726...80V}), a rigorous theorem for the exclusion of these scenarios holds only at the center of the system \cite{An:2005tm}.

\subsection{Extrapolating to inner radii: densities profiles} 
While in the standard approach to solve the Jeans equation, Eq.~(\ref{eq:jeansdiff}), physical mass profiles are automatically obtained imposing a ``physical'' parametric ansatz, there is no a priori guarantee that the procedure proposed here gives physical outputs. 
\begin{figure}[!t!]
  \centering
   \includegraphics[scale=0.37 ]{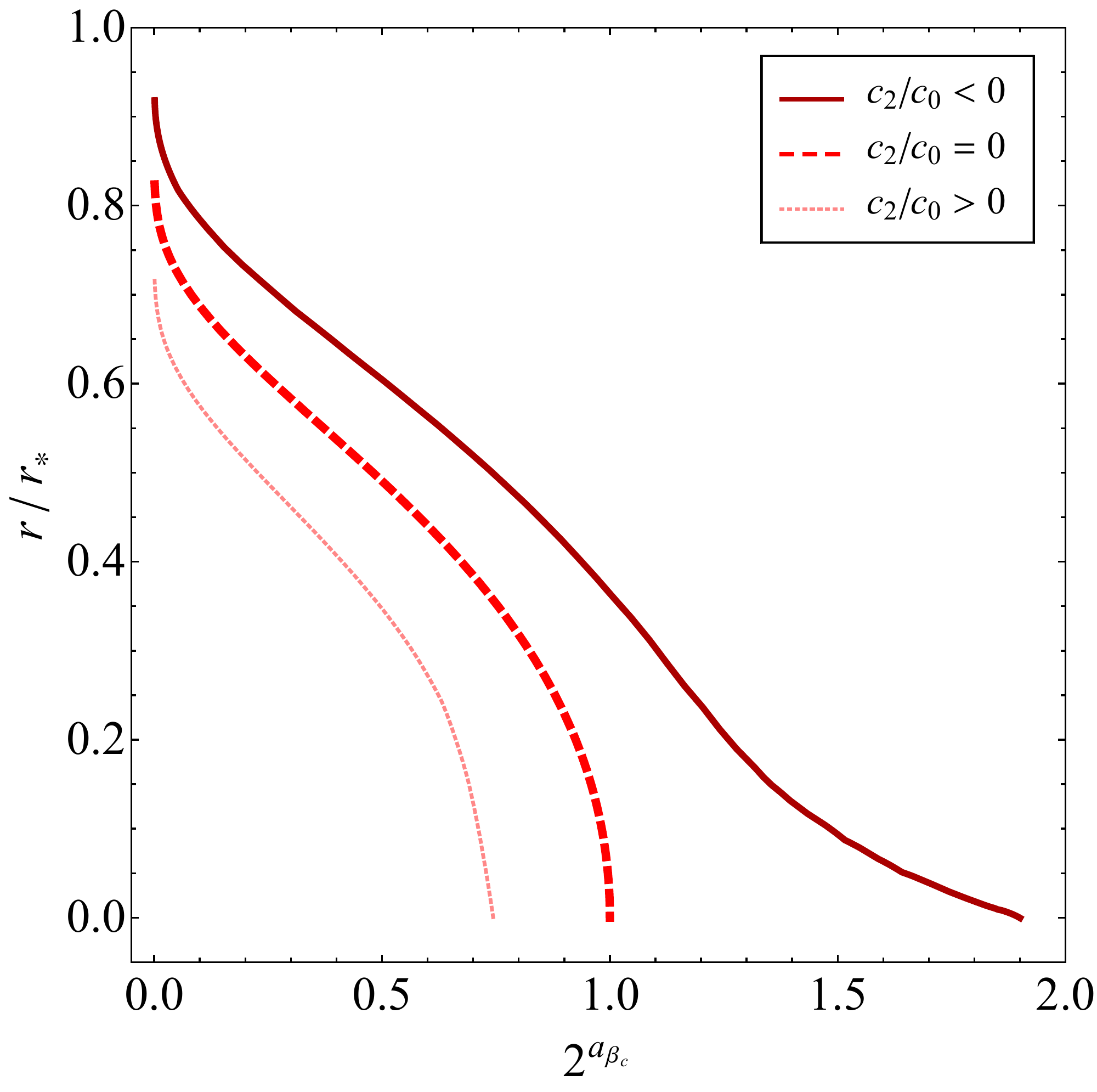}
  \caption{\textit{Isolevel for $\mathcal{M}=0$ in the plane $\beta_{c}\,$-$\,r/r_{*}$ assuming a Plummer stellar template: The red dashed line corresponds to the case of $constant$ $\sigma_{los}$ and delimits the region above the curve where $\mathcal{M} > 0$ from the one where the mass becomes negative. Exploiting  a quadratic expression in $R$ for $\sigma_{los}(R)$, we plot also a dark red (light red dashed) line that corresponds to a convex (concave) tilt in the extrapolation of  the constant $\sigma_{los}$ value towards inner radii, using as a reference $c_{2}/c_{0} = 0.5$ and $c_{1}/c_{0}=-0.8$, ($c_{2}/c_{0} = -1.4$ and $c_{1}/c_{0}=2.3$). }}
  \label{fig:MassZeros}
\end{figure} 
A first basic check is on the positivity of the solution of Eq.~(\ref{eq:ourinv}) at any radius. E.g. in the simplified model introduced above
one finds a non-trivial constraint on the allowed range of orbital anisotropies. This is shown in Fig.~\ref{fig:MassZeros} in the plane $2^{a_{\beta_{c}}}\,$-$\,r/r_{*}$, where isolevels for $\mathcal{M}=0$ are displaced in the limit of constant orbital anisotropy, marking the minimum radius at which a solution of Eq.~(\ref{eq:ourinv}) is positive. In particular, the red dashed line corresponds to the case of constant $\sigma_{los}$ and Plummer surface brightness.
One can see that, within this setup, positive mass solutions can be extrapolated down to $r = 0$ only when $\beta_{c}\le0$. On the other hand, the extrapolation to $r \rightarrow 0$ critically depends on what is assumed for the extrapolation of $\sigma_{los}(R)$ for $R \to 0$. To sketch this effect we introduce as sample parametrization for the l.o.s. velocity dispersion the form: $ \sigma_{los}(R) = c_{0} + c_{1} (R/R_{\textrm{\tiny{1/2}}}) + c_{2} (R/R_{\textrm{\tiny{1/2}}})^{2}$ iff $R/R_{\textrm{\tiny{1/2}}} \le 1/2$, constant iff $R/R_{\textrm{\tiny{1/2}}} > 1/2$. Fig.~\ref{fig:MassZeros} shows that an inner concave tilt of $\sigma_{los}(R)$ forces to restrain to progressively more negative values of $\beta_{c}$ (with  $a_{\beta}$ approaching 1), while a concave one allows for radially anisotropic stellar velocity profiles.

In general, the positivity of the mass is not the only condition we would like to supplement the Jeans inversion with: e.g. the mass profile should not decrease going to larger radii (i.e., up to the cutoff of the profile). In the following, we will be actually more restrictive and define as a physical outcome of our inversion the model satisfying the following requirements:
\begin{equation}
\mathcal{M}(r) > 0 \  , \ \rho(r) = \frac{1}{4 \pi r^{2}} \frac{d \mathcal{M}}{d r} > 0 \ , \ \frac{d \rho}{d r} \le 0 \ \ \ \ \forall \ r > 0 \ .
  \label{eq:phys_conds}
\end{equation}
Assuming that the density of the DM profile $\rho(r)$ provides the dominant component to the dSph potential well, the third condition ensures the potential well to be monotonic and hence provides a necessary condition of stability for the model. \\
Checking a posteriori these conditions, $\rho(r)$ is simply obtained taking the derivative of the mass function, Eq.~(\ref{eq:ourinv}):
\begin{eqnarray}
4 \pi G_{N} \, r^{2} \,\rho(r)  \; & =  & - \left[1-a_{\beta}\right]  \frac{d}{dr} \left(\frac{r^{2}}{\widehat{I}} \frac{d  \widehat{P}}{d  r} \right) + a_{\beta}  \left[ \frac{d\log a_{\beta}}{d \log r} + b_{\beta} \right] \left(\frac{r}{\widehat{I}} \frac{d  \widehat{P}}{d  r} \right)  \nonumber \\
 &&  - \, a_{\beta} \, b_{\beta} \, \left[ 1 - a_{\beta} - \frac{d \log \widehat{I}}{d \log r} + \frac{d \log a_{\beta} b_{\beta}}{d \log r} \right] \, \left(\frac{1}{\widehat{I}} \int_{r}^{\infty} dr^\prime \mathcal{H}_{\beta}(r,r^\prime) \,
     \frac{d\widehat{P}}{dr^\prime} \right)
    \label{eq:rhojeansinv}
\end{eqnarray}
Among the three terms on the r.h.s., only the first contributes in the limit of  isotropic motion of the tracers ($\beta\rightarrow 0$ or equivalently $a_{\beta}\rightarrow 0$); assuming also that the l.o.s. velocity dispersion is constant, one simply finds:
\begin{equation}
  \rho_{\beta=0}(r) =  \frac{\sigma_{los}^{2}}{4 \pi G_{N} \, r^{2}} \frac{d}{dr} \left(-r \frac{d \log{\widehat{I}}}{d \log{r}} \right) \ .
  \label{eq:rhobeta=0}
\end{equation}
For such profile to have a core (namely: ${d\log \rho_{\beta=0}}/{d \log r} \rightarrow 0$ for $r\rightarrow 0$), the logarithmic slope of the stellar density profile needs to scale as $r^{2}$ towards the center of the system. E.g., considering a multi-parameter stellar template like the Zhao profile \cite{Zhao:1995cp}:
\begin{equation} 
 \widehat{I}(r) = \frac{\widehat{I}_{0}}{\left( \frac{r}{r_{s}} \right)^{\gamma} \left[ 1 + \left(\frac{r}{r_{s}}\right)^{\alpha} \right]^{\frac{\delta-\gamma}{\alpha}}} \ ,
 \label{eq:stellarZhao}
\end{equation}
where $\gamma$ and $\delta$ represent the inner and outer slope of the profile, and $r_{s}$ and $\alpha$ the scale radius and the smoothness of the transition between the inner and outer scaling, the logarithmic slope is:
\begin{equation}
\frac{d \log \widehat{I}}{d \log r} = - \gamma - (\delta - \gamma) \frac{\left(\frac{r}{r_{s}}\right)^{\alpha}}{1+ \left(\frac{r}{r_{s}}\right)^{\alpha}} \ ,
\label{eq:logstellarZhao}
\end{equation}
and a cored profile is obtained only in case $\gamma = 0$ and $\alpha = 2$. For any $\gamma > 0$ the scaling of the DM profiles jumps to $1/r^{2}$, as in the isothermal sphere model.

\begin{figure}[!t!]
  \begin{subfigure}[b]{0.5\textwidth}
    \centering
    \includegraphics[scale=0.38]{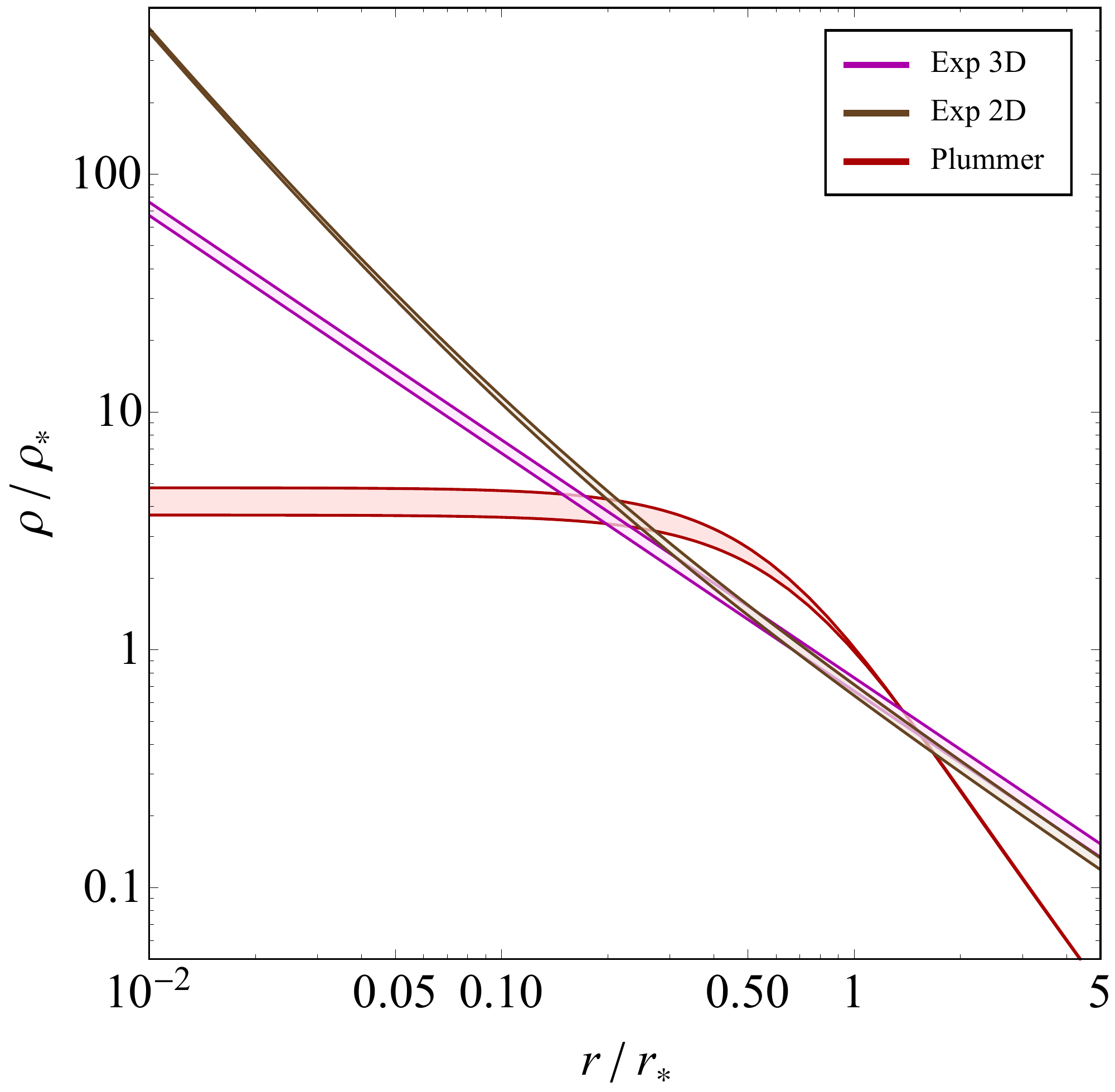}
  \end{subfigure}
\hfill
  \begin{subfigure}[b]{0.5\textwidth}
    \centering
    \includegraphics[scale=0.38]{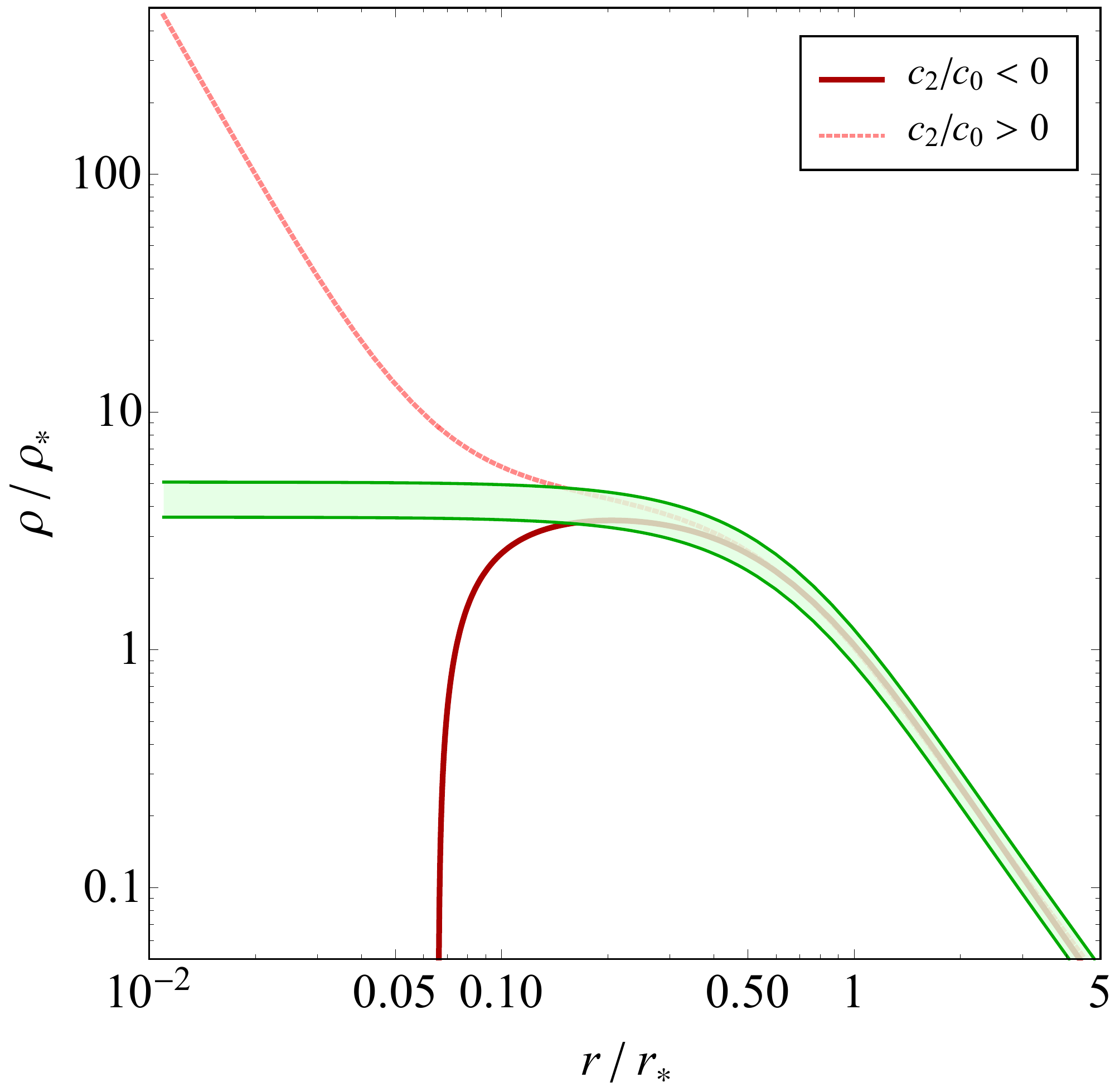}
  \end{subfigure}
  \caption{\underline{Left panel:} \textit{Density profile of the system in the model with constant $\sigma_{los}$ and isotropic tracer orbits, for the choices of stellar template: a Plummer model, an exponential profile for the stellar density and for the surface brightness. The error bands reflect the uncertainty on the stellar model assumed as in Fig.~\ref{fig:radpress}. Each profile is normalized to $\rho_{*} = \rho(r_{*})$ in the Plummer case.}  \underline{Right panel:} \textit{Density profile assuming the Plummer surface brightness and $\beta=0$, parametrizing $\sigma_{los}(R)$ with a linear expression in $R$ (green band) and angular coefficient varied as in Fig.~\ref{fig:radpress}. In the same plot, the effect on $\rho(r)$ due to a departure from a constant $\sigma_{los}$ for radii  $R \le R_{\textrm{\tiny{1/2}}}/2$, slightly tilting its constant profile according to a convex (concave) quadratic ansatz in $R$, i.e. using $c_{2}/c_{0} = 0.1$ and $c_{1}/ c_{0}=-0.3$ ($c_{2}/c_{0} = -0.1$ and $c_{1}/ c_{0}=0.3$).  }}
    \label{fig:RhoStellarAndSigma}
\end{figure}

Looking back at Eq.~(\ref{eq:plummer}), one sees that the Plummer model belongs exactly to the class of the Zhao profiles  providing a core in $\rho_{\beta=0}(r)$ if $\sigma_{los}$ is constant. This is also shown in the left panel of Fig.~\ref{fig:RhoStellarAndSigma}, where we plot the result for the inversion within the assumptions of constant $\sigma_{los}$ and $\beta=0$. In the same plot we show also the resulting case of the two exponential stellar templates introduced in Eq.(\ref{eq:exp2D})-(\ref{eq:exp3D}). Note that an exponential surface brightness implies a logarithmically divergent stellar number density and hence a scaling of the inner density of $\left(r \log r \right)^{-2}$, while the (cored) exponential stellar density gives a $1/r$ inner density scaling, standing in between the cored case and singular isothermal sphere. 
Indeed, the Plummer model is the simplest stellar template we can consider which provides a core in the density profile. 
Since cored profiles are of particular importance when estimating minimal values for the l.o.s. integral of squared densities (the $J$-factor defined in the Introduction), it is the case we will concentrate on in the following.

Before discussing the case of $\beta \neq 0 $,  it is interesting to assess how the density profile is affected by a mild departure from the approximation of constant $\sigma_{los}$. 
The green band in the right panel of  Fig.~\ref{fig:RhoStellarAndSigma} shows the very mild impact on $\rho(r)$ when we implement the linear scaling of the l.o.s. velocity dispersion already introduced in Fig.~\ref{fig:radpress}.
For the quadratic scalings already implemented in Fig.~\ref{fig:RhoStellarAndSigma} -- even using much milder concave and convex tilts for $\sigma_{los}(R)$ -- one sees instead a rather drastic change in $\rho(r)$, with a sharp enhancement of the inner density for a convex perturbation and an unphysical solution induced by the concave tilt. Associating the trends seen in the left and right panels of the figure, one can deduce that a cored profile, standing also at the border with unphysical solutions, is obtained in the inversion procedure only via a fine adjustment between the trend imposed by the choice of stellar number density profile and that from the $R \to 0$ scaling of $\sigma_{los}(R)$. 
 
We are now in the position to address the implications on $\rho(r)$ of a non-vanishing orbital anisotropy, considering first of all the case in which it does not depend on the radial coordinate, i.e. $\beta(r)= \beta_{c}$. Looking back at Eq.~(\ref{eq:rhojeansinv}), also the second and the third term on the r.h.s. give a contribution to $\rho(r)$; the second term however has the same $r \to 0$ scaling  as the first, hence does not alter the discussion just presented for $\rho_{\beta=0}(r)$. The third term instead introduces a non-trivial dependence on $a_{\beta_{c}}$ of the inner radial slope. Assuming  $a_{\beta_{c}} \ne 1$, the scaling can be read out from the corresponding logarithmic derivative diminished by 2 (taking into account the $r^2$ factorized on the l.h.s. of Eq.~(\ref{eq:rhojeansinv})):
\begin{equation}
\frac{d}{d \log r} \left[ \log \left(\frac{1}{\widehat{I}} \int_{r}^{\infty} dr^\prime \mathcal{H}_{\beta}(r,r^\prime) \,
     \frac{d\widehat{P}}{dr^\prime} \right) \right] - 2 = - 2 - a_{\beta_{c}} - \frac{d \log \widehat{I}}{d \log r} 
     - \frac{r^{a_{\beta_{c}}} \frac{d\widehat{P}}{dr} }{\int_{r}^{\infty} dr^\prime {r^\prime}^{a_{\beta_{c}}} \,
     \frac{d\widehat{P}}{dr^\prime}}\,.
      \label{eq:scalingbetac}
\end{equation} 
In general the term $-2 - a_{\beta_{c}}$ is the most relevant, driving $\rho(r)$ to a scaling that is even more singular than the singular isothermal sphere in case of circularly anisotropic profiles; the term in the logarithmic derivative of $\widehat{I}$ can at most mitigate the singularity in case $\widehat{I}$ itself is singular. The last term on the r.h.s. of Eq.~(\ref{eq:scalingbetac}) is in general less relevant; for constant $\sigma_{los}$ and a Plummer $\widehat{I}$, in the limit $r \to 0$ it is equal to 0 if $a_{\beta_{c}} \ge -2$, and  to $a_{\beta_{c}} + 2$ for $a_{\beta_{c}} < -2$. In this last case, one would apparently get a cored profile; note however that $\rho(r)$ is obtained by summing this contribution to the first two terms in Eq.~(\ref{eq:rhojeansinv}), and, in the same limit, these drive $\rho(r)$ to an unphysical result, with the negative mass solution already discussed above. Analogously to what is shown in Fig.~\ref{fig:RhoStellarAndSigma}, a proper readjustment of the inner radial scaling of $\sigma_{los}(R)$ would be needed, requiring however a even more severe tuning to get physical solutions with a core \cite{Evans:2008ik}, since  $\sigma_{los}(R)$ impacts also on the scaling in Eq.~(\ref{eq:scalingbetac}).

A further subtle point regards the limit of $a_{\beta_{c}} \to 1$. When $\beta_{c}$ approaches extreme negative values, the logarithmic derivative of $\widehat{I}$ appears as an extra multiplicative factor in the third term on the r.h.s. of Eq.~(\ref{eq:rhojeansinv}) and hence its radial scaling (the logarithmic derivative of the logarithmic derivative of $\widehat{I}$) should be added to Eq.~(\ref{eq:scalingbetac}). Referring again to the Zhao profile in Eq.~(\ref{eq:logstellarZhao}), this contribution is 0 if $\gamma \ne 0$, it is equal to $+\alpha$ if $\gamma = 0$. Back to the Plummer model and constant $\sigma_{los}$ one would then find a scaling of the density profile that goes like $r^{-2-a_{\beta_{c}}+\alpha} \to 1/r$, as opposed to $r^{-2 + a_{\beta_{c}}}$ valid for $0 < a_{\beta_{c}} < 1 $.
\begin{figure}[!t!]
  \begin{subfigure}[b]{0.5\textwidth}
    \centering
    \includegraphics[scale=0.55]{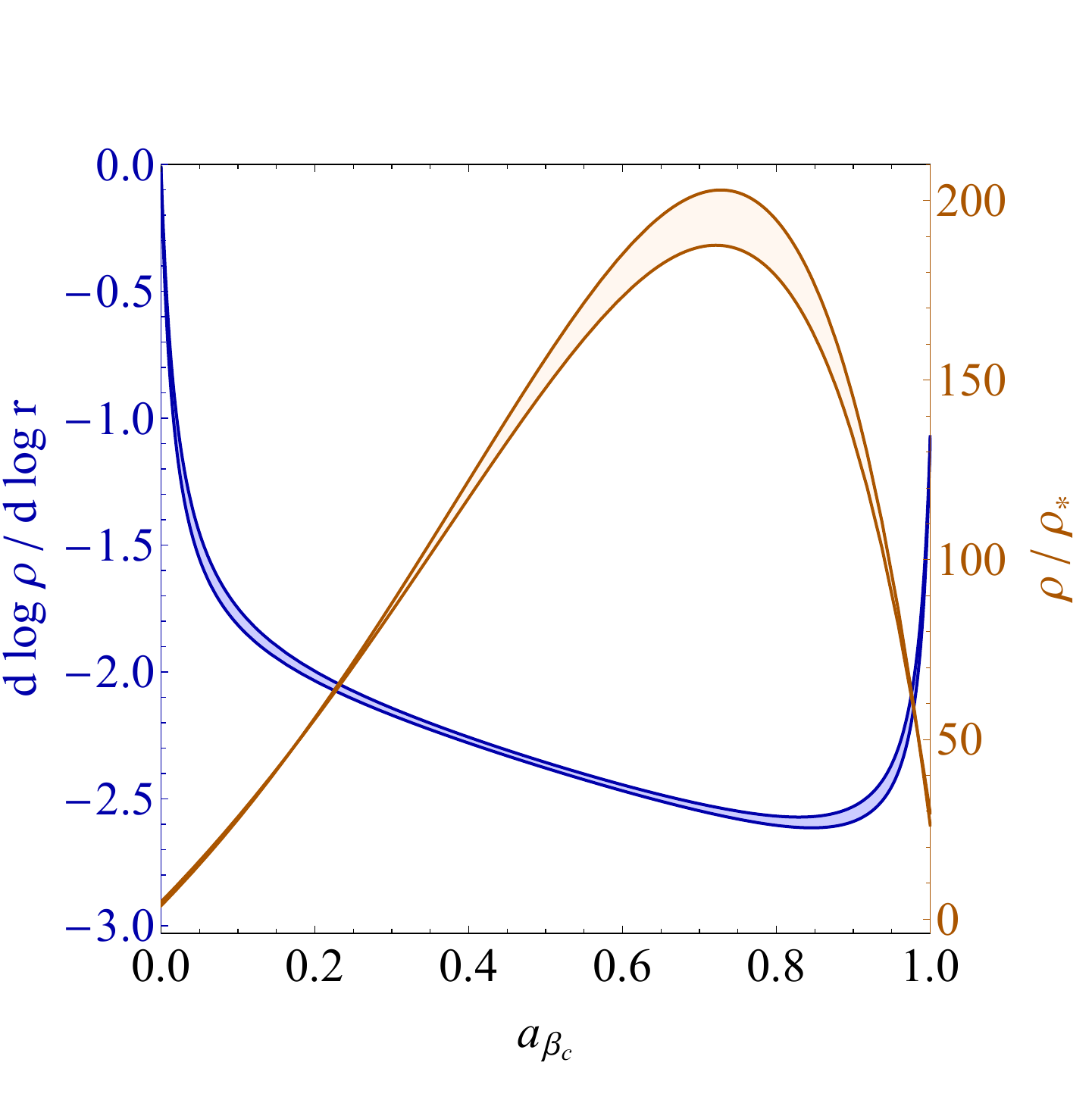}
  \end{subfigure}
\hfill
  \begin{subfigure}[b]{0.5\textwidth}
    \centering
    \includegraphics[scale=0.55]{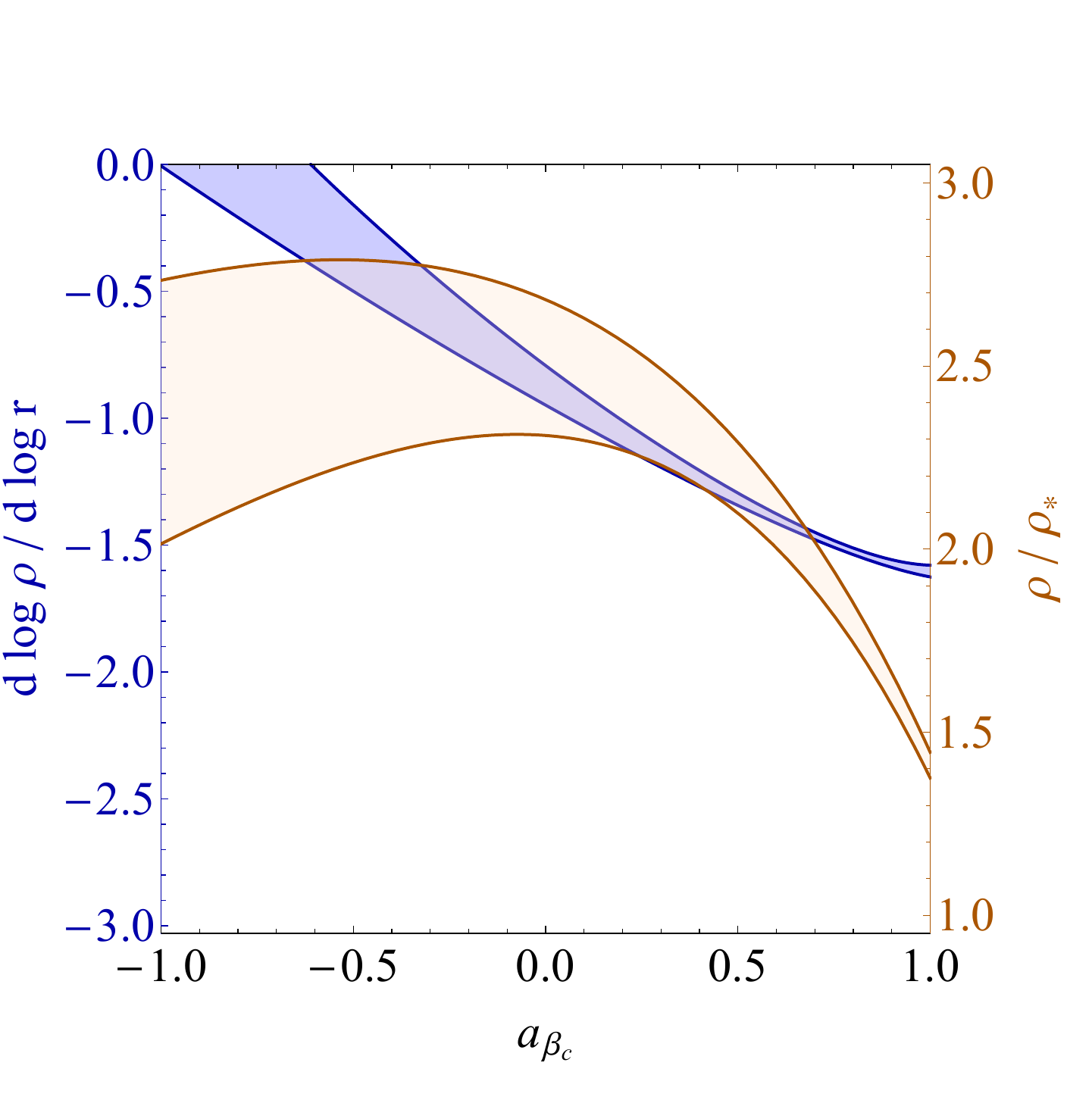}
  \end{subfigure}
  \caption{\underline{Left panel:} \textit{Logarithmic slope (blue colour) and density profile normalized at $\rho_{*} = \rho_{\beta=0}(r_{*}) $ (yellow colour) for a physical model implemented with Plummer stellar density and constant $\sigma_{los}$ as a function of the allowed values of $a_{\beta_{c}}$ and at the fixed ratio $r/r_{*}$ of $5\%$. The bands are related to the error on the characteristic scale radius of the Plummer profile.} \underline{Right panel:} \textit{Same kind of plot produced at the fixed ratio $r/r_{*}$ of $50\%$. Note that the values of $a_{\beta_{c}}$ for which the model is physical span in this case the larger range $\beta_{c} \in (-\infty, 1/2\,]$.}}
    \label{fig:RhoBetac}
\end{figure}
In the left panel of Fig.~\ref{fig:RhoBetac} we show the slope of the density at the fixed value $r=0.05 \, r_{*}$ as a function of all the set of $a_{\beta_{c}}$ that provide a physical solution (see Eq.~(\ref{eq:phys_conds})) in our Jeans inversion approach.  The resulting blue band in the plot highlights qualitatively the trend analyzed so far: Exploring even smaller  ratios of $r/r_{*}$, one would retrieve the linear $-2 - a_{\beta_{c}}$ scaling with sudden transition of $ d \log \rho /  d \log r $ to $0$ at $a_{\beta_{c}}=0$ and to $-1$ at $a_{\beta_{c}}=1$.
Once we leave the center of the dSph to move towards its outskirts, such a behaviour related to the allowed physical solutions gets relaxed: e.g. at $r= \, r_{*}/2$, within the uncertainty of the Plummer profile, physical densities are allowed up to $\beta_{c} = 1/2$, scaling with a power law index between $-1.5$ and $0$, as represented by the blue band in the right panel of Fig.~\ref{fig:RhoBetac}. 
Also shown in the same figure is the value of the profile at the two chosen radii normalized to  $\rho_{*} = \rho_{\beta=0}(r_{*})$. In the central region of the system there is a smooth, but sharp, variation of the density varying $a_{\beta_{c}}$, spanning roughly two orders of magnitude. On the other hand, at large radii the variation in $\rho(r)$ is within $50\%$. \\
The subtle limit  $a_{\beta_{c}} \to 1$ could have been inferred also looking at the behaviour of the mass profile in the case of purely circular stellar motion. In fact, for the model implementing a constant $\sigma_{los}$ and Plummer surface brightness Eq.~(\ref{eq:circmass}) goes as:
\begin{equation}
 \mathcal{M}_{\beta \to -\infty}(r) = \frac{2 \, R_{\textrm{\tiny{1/2}}} \, \sigma_{los}^{2}}{3 \, G_{N}} \, \left[ 2 \left(1 + \frac{r^2}{R_{\textrm{\tiny{1/2}}}^2} \right)^{5/2} - \frac{r^3}{R_{\textrm{\tiny{1/2}}}^3}  \left(5 + 2 \frac{r^2}{R_{\textrm{\tiny{1/2}}}^2} \right)
  \right]
    \label{eq:plummermasscirc}
\end{equation}
and in the limit of $r$ going to 0 becomes
\begin{equation}
\lim_{r \to 0} \mathcal{M}_{\beta \to -\infty}(r) =  \frac{4 R_{\textrm{\tiny{1/2}}} \, \sigma_{los}^{2}}{3  \, G_{N}} \ .
    \label{eq:bhcirc}
\end{equation}
We observe that the case of constant orbital anisotropy going to $-\infty$ is the only case when the condition of $\mathcal{M}(0)=0$ is not met in this model,
 mimicking the scenario of a black hole at the center of the dwarf supporting the velocity dispersion.
 With the logarithmic slope of $\rho(r)$ approaching $-3$, the mass profile gets larger and larger contribution close to $r=0$. The logarithmic divergence is avoided via the appearance of the black hole-like feature and hence a discontinuity in the density profile:
\begin{equation}
  \label{eq:rhocirc}
\rho(r)_{\beta \to -\infty}  =   \frac{5 \, \sigma_{los}^{2}}{6 \pi \,  R_{\textrm{\tiny{1/2}}}^{2} G_{N} } \left(\frac{R_{\textrm{\tiny{1/2}}}}{r}\right) \left[ 2 \left( 1+ \frac{r^{2}}{R^{2}_{1/2}}\right)^{3/2} 
-3  \frac{r}{R_{\textrm{\tiny{1/2}}}} - 2\frac{r^{3}}{R^{3}_{1/2}}  \right]
\end{equation}
holding a logarithmic slope equal to $-1$ at the center, as anticipated. 

We summarize the results of this section, briefly recapping what we have achieved so far with our method. 
First, we have highlighted the existence of possible unphysical solutions encoded in the general master formula,  Eq.~(\ref{eq:ourinv}), derived by inverting Eq.~(\ref{eq:jeansdiff}). 
Physical solution within this approach require the following two conditions:
\begin{itemize}
\centering
\item[\textit{ i)}] $\mathcal{M}(r) > 0 \ \ \ \ \ \ \ \ \,\forall \ \ r > 0 \, $ 
\item[\textit{ii)}] $\mathcal{M}(r') > \mathcal{M}(r) \ \ \forall \ \ r' \ge r$
\end{itemize}
Then, we have made a step forward deriving Eq.~(\ref{eq:rhojeansinv}) to study the density of non-rotating pressure-supported systems like dSphs. 
In order to deal with a physical density, we have supplemented the latter with the following third condition:
\begin{itemize}
\centering
\item[\textit{iii)}]  $\mathcal{\rho}(r') \leq \mathcal{\rho}(r) \ \ \ \ \ \ \forall \ \ r' \ge r$
\end{itemize}
We have primarily focussed our attention on the trends of the inner density profile of the system for several tracer density templates and l.o.s. velocity dispersion profiles, see Fig.~\ref{fig:RhoStellarAndSigma}, under the assumption of an isotropic tracer motion. Eventually, we have analyzed in details the benchmark scenario of $\sigma_{los}$ and stellar Plummer model, varying the orbital anisotropy $\beta_{c}$, as reported in Fig.~\ref{fig:RhoBetac}.

\subsection{$J$-factor scalings}
The simple form of $\rho(r)$ derived in the circular anisotropy limit may be taken as a good starting point to discuss $J$-factor trends in the inversion approach considered here.
Indeed, integrating the square of the profile in Eq.~(\ref{eq:rhocirc}) following the definition in Eq.~(\ref{eq:j_psi}), we can get  an analytic form for the $J$. 
In Appendix~\ref{App:JfactorComputation} we provide an expression for the $J$-factor, see Eq.~(\ref{eq:Joversimplified}), which is valid in the limit of distance $\mathcal{D}$ of the dwarf much larger with respect to the typical transverse size of these galaxies (note that the optimal angular aperture $\psi\sim 0.5^{\circ}$ is usually considered in literature, see e.g. \cite{Walker:2011fs,Ackermann2011,Bonnivard:2014kza}). 
In the case at hand it gives:
\begin{eqnarray}
J_{\beta \to -\infty} \;\; & =  &\;\; \frac{1}{\pi \mathcal{D}^{2}}\frac{ \sigma_{los}^{4}}{  R_{\textrm{\tiny{1/2}}} G^{2}_{N} } \frac{5 }{63} \, \Big[ 68 + 140 \, \mathcal{Z} + 245 \,\mathcal{Z}^3 + 168 \,  \mathcal{Z}^5 + 40 \mathcal{Z}^7 
- (68 + 40 \mathcal{Z}^2)\,(1 +  \mathcal{Z}^2)^{5/2}  \Big]  \nonumber \\
&   & \xrightarrow[\mathcal{Z} \to \infty]{ }  \   \frac{340}{63 \pi} \, \frac{R_{\textrm{\tiny{1/2}}}^{3}}{\mathcal{D}^{2}} \, \left( \frac{\sigma_{los}^{2}}{ \,  R_{\textrm{\tiny{1/2}}}^{2} G_{N}} \right)^{2} \ ,
\label{eq:Japproxcirc}
\end{eqnarray}
where the quantity in the last brackets is an energy density and  the dimensionless ratio $\mathcal{Z} = \mathcal{R}/R_{\textrm{\tiny{1/2}}}$ is introduced to take into account a possible finite size $\mathcal{R}$ of the spherical halo density $\rho(r)$. 

Plugging in Eq.~(\ref{eq:Japproxcirc}) typical values for dSphs, namely $\sigma_{los} \sim 10 $ km s$^{-1}$,  $R_{\textrm{\tiny{1/2}}} \sim 0.3$ kpc and $\mathcal{D} \sim  100 $ kpc, one gets a $J$-factor of about $10^{18}$ GeV$^2$ cm$^{-5}$. 
Notice that this is not a totally realistic case since the black hole mass one would infer from Eq.~(\ref{eq:bhcirc}) would be $\sim 10^{7}$ $M_{\odot}$, possibly consistent with kinematical observables, but likely too high to be found at the center of these galaxies \cite{2014AJ_dwarfs_BH_nearby,2013MmSAI_dwarfs_BH_nearby}\,. \\
Even in case of perfect isotropic motion of the stellar tracers an analytic expression for the density profile, Eq.~(\ref{eq:rhobeta=0}), and the $J$-factor can be provided. For the latter we have:
\begin{figure}[!t!]
  \centering
   \includegraphics[scale=0.37 ]{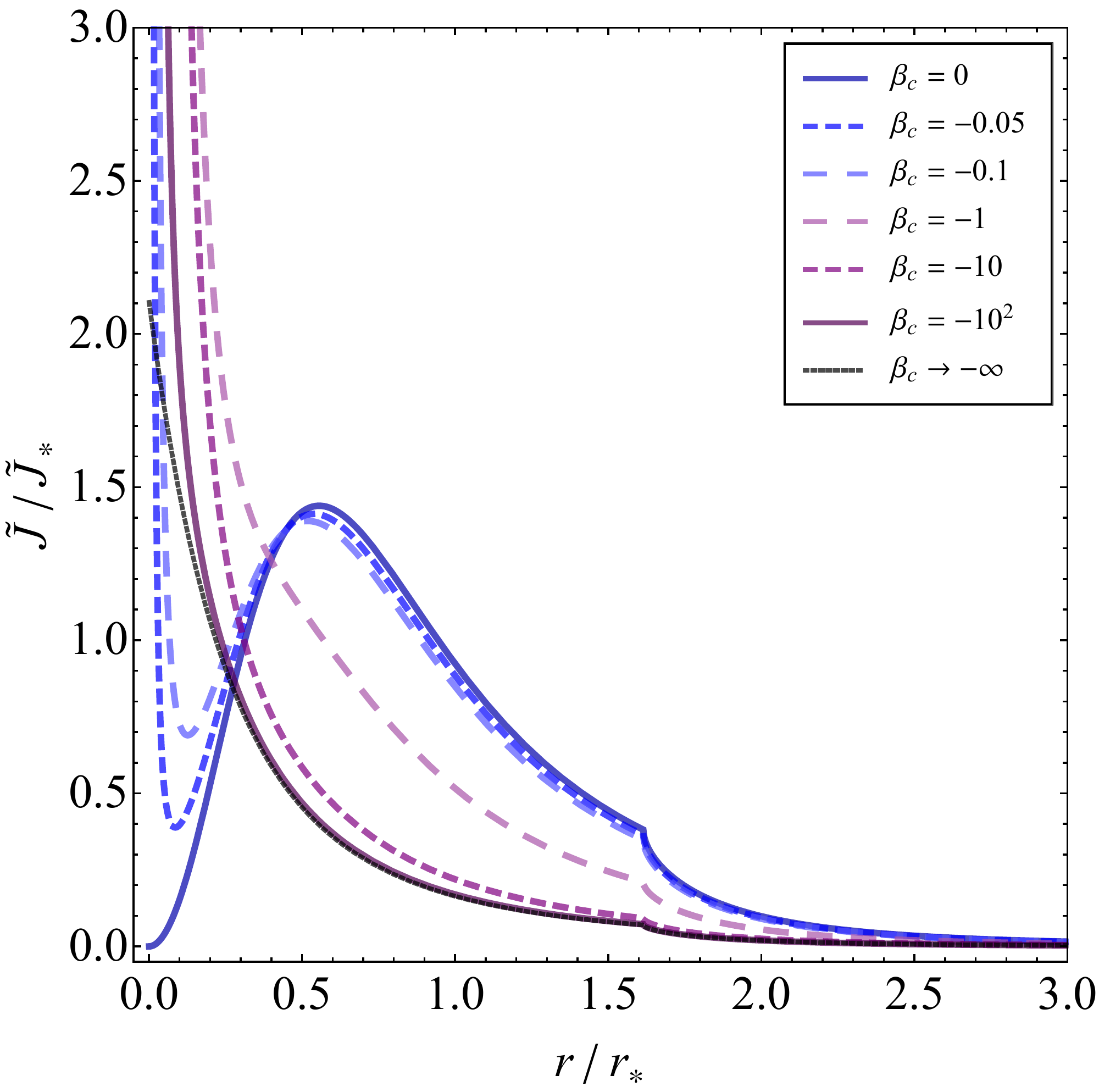}
  \caption{\textit{Integrand of the $J$-factor in radial coordinates according to Eq.~(\ref{eq:Jtilde}) for different realizations of constant orbital anisotropies in a model with constant $\sigma_{los}$ and with a Plummer stellar template. We assumed a distance $\mathcal{D} = 10^{2}$ kpc and an angular aperture $\psi_{max} \simeq 0.5^{\circ}$, normalizing $\widetilde{J}(r)$ at the reference value $\widetilde{J}_{*} = \widetilde{J}_{\beta=0}(r_{*})$.}}
  \label{fig:Jtilde}
\end{figure} 
\begin{equation}
J_{\beta = 0}=  \frac{25}{ 4 \pi \mathcal{D}^{2}}\frac{ \sigma_{los}^{4}}{  R_{\textrm{\tiny{1/2}}} G^{2}_{N} }\left[  \frac{ \mathcal{Z} \left(-3 + \mathcal{Z}^2 \right) \left(5 + 3 \mathcal{Z}^2\right) }{12 \left(1 +  \mathcal{Z}^2 \right)^3} + \frac{5}{4} \arctan  \mathcal{Z} \right] \xrightarrow[ \mathcal{Z} \to \infty]{} \frac{125}{32} \, \frac{R_{\textrm{\tiny{1/2}}}^{3}}{ \mathcal{D}^{2}} \, \left( \frac{\sigma_{los}^{2}}{ \,  R_{\textrm{\tiny{1/2}}}^{2} G_{N}} \right)^{2} \ ,
\label{eq:Japproxiso}
\end{equation}
that is a factor of $2$ larger than the $\beta \to -\infty$ case.
The picture for constant orbital anisotropies related to physical solutions (still taking $\sigma_{los}$ constant and Plummer profile) is shown in Fig.~\ref{fig:Jtilde}: 
we plot a function $\widetilde{J}(r)$ appearing as a linear measure in the $J$-factor computation (the area under each curve represents the J-factor, up to the normalization factor $\widetilde{J}_{\beta=0}(r_{\star})$), see Eq.~(\ref{eq:Jtilde}). \\
Then, the general trend for the $J$-factor at constant $\sigma_{los}$ can be drawn: Starting from the minimum $J$-value corresponding to the case of $\beta=0$, the $J$-factor increases going to smaller values of $\beta_{c}$, with an inner cusp appearing and at the same time a reduction of the contribution at larger radii, up to the exact circular limit, when the central cusp is suddenly reduced in correspondence to the discontinuity of $\rho(r)\,$. 

As anticipated, the picture above is expected to provide a rather conservative estimate of the $J$-factor: leaving a cored stellar template in favor of a more general Zhao profile, Eq.~(\ref{eq:stellarZhao}), with $\gamma>0$, or an exponential template, like the Sersic one, would enhance the inner density profile and hence the $J$-value; moreover, an inner convex tilt in the profile of the l.o.s. velocity dispersion would go in the same direction. A concave perturbation to a flat $\sigma_{los}$ would offer a way out to lower  the $J$-value, but would apply only to the cases of circular-like stellar orbits, where the density is relatively cuspy, to not generate unphysical outputs like the one previously encountered in Fig.~\ref{fig:RhoStellarAndSigma}. Therefore, discarding the ``extreme'' solution of the Jeans inversion at $\beta \to -\infty$, the lowest $J$-factor emerging from a physical model in Fig.~\ref{fig:Jtilde} corresponds to the case of perfectly isotropic stellar motion, Eq.~(\ref{eq:Japproxiso}).
Within the same framework, this conclusion can be modified once we follow a more conservative approach in extrapolating the inner density of $\rho(r)$.
Indeed, one may question whether the basic assumptions involved in the derivation of Eq.~(\ref{eq:jeansdiff}) itself, most importantly the spherical symmetry of the system, should be trusted down to exceedingly small radii, imposing by construction a choice of coordinates which are singular in the origin and also extrapolating $\sigma_{los}$ in a region which is simply not accessible with data. An alternative is to introduce a saturation scale $r_{c}$ so that $\rho(r\le r_{c}) = \rho(r_{c})$; in Fig.~\ref{fig:Jtilde} one sees that a relatively small inner cutoff, i.e. $r_{c}/{r_{*}} \simeq 0.1$, has an important impact on the inner density profiles with $-\infty < \beta_{c} < 0$ and hence on the corresponding $J$-factors. 
Focussing on the area below each curve of Fig.~\ref{fig:Jtilde} one can easily visualize that in the case of $\beta_{c} \simeq 0$, the $J$-factor is almost insensitive to an inner cutoff, 
while for a  negative orbital anisotropy, a very small $r_{c}$ can significantly enhance the $J$-value with respect to what can be obtained with a more conservative cut on the inner density.
In the exact circular orbit limit the $J$-factor is again quite insensitive to the choice of $r_{c}$.
Note that with the implementation of $r_{c} \neq 0$, the density $\rho(r)$ related to very negative orbital anisotropies is now smoothly tracking the -- previously discontinuous -- case of circular orbits. Therefore, the flattening of the density due to the a non-vanishing inner cut allows for lower $J$-factors than the one obtained at $\beta = 0$. 
At the same time, one should not forget that lower orbital anisotropies in this context would imply growing black-hole-like features up to the questionable point that an important contribution to the total mass of the system comes from the center. \\ Eventually, note that -- in contrast to the approach followed in this work -- a physical black hole at the center of the dSph  may be instead modeled so to strengthen the DM annihilation signal from the galaxy. This possibility has been explored in, e.g., \cite{Gonzalez-Morales:2014eaa,Wanders:2014xia}.

Focussing on the minimum $J$-value one can obtain, we can now address the impact  of a radial dependence in the orbital anisotropy function.
In what follows we will assume the  orbital anisotropy profile to be well described by the rather general form provided in \cite{Baes:2007tx}:
\begin{equation}
  \beta(r) = \frac{\beta_{0} + \beta_{\infty} \left( \frac{r}{r_{\beta}} \right)^{\eta_{\beta}}}{1+\left( \frac{r}{r_{\beta}} \right)^{\eta_{\beta}}} \ , 
  \label{eq:radial_general_beta}
\end{equation}
offering an interpolation between the tracer behaviour at the center of the system, set by $\beta_{0}$, and the one towards the outer 
part, set by $\beta_{\infty}$, with characteristic scale and  sharpness between the two regimes respectively determined by $r_{\beta}$ and $\eta_{\beta}$. \\
Note that, on general grounds, for solution to be physical satisfying the conditions in Eq.~(\ref{eq:phys_conds}), a sharp radial dependence in $\beta(r)$ (such that its derivatives would even impact on the scalings discussed looking at Eq.~(\ref{eq:scalingbetac})) can be implemented only together equally sharp variations of $\sigma_{los}(R)$ in $R$; we will provide an explicit example in the next Section, enlightening also the level of tuning involved. Here we will consider instead smoother behaviour for $\beta(r)$, assuming then, without loss of generality, a constant $\sigma_{los}$ and referring to the trends illustrated in Fig.~\ref{fig:RhoStellarAndSigma} for its extensions.

We have seen in Section~\ref{Sec:MassEstimator} that a notion of a mass estimator for the system is to some extent available thorough the mass enclosed in $r_{*}$. We can exploit this information to find the density profile that minimizes the $J$-factor. This can be done introducing a simple broken power-law ansatz for $\rho(r)$:
\begin{equation}
\rho(r) = \rho_0 \left\{ \left(\frac{r_{1}}{r} \right)^{\alpha_{1}}  \theta_{H} \left(r_{1}-r\right) +
\sum_{i=2}^n \left[ \prod_{j=2}^{i-1} \left( \frac{r_{j-1}}{r_{j}} \right)^{\alpha_{j}} \left(\frac{r_{i-1}}{r} \right)^{\alpha_{i}} 
 \theta_{H} \left(r_{i}-r\right)  \, \theta_{H}(r-r_{i-1}) \right] \right\}
\end{equation}
with $\alpha_{i}$ being the logarithmic derivative of the profile within the radial interval $[r_{i-1},r_i]$,  $\theta_{H}(r)$ the Heaviside step function, and the normalization $\rho_0$ is derived from the condition $\mathcal{M}(r_*)=\mathcal{M}_{*}$. 
For a given $\mathcal{M}_{*}$ (as following from the approximate relation in Eq.~(\ref{eq:wolfmstar})), $J$-factor can be minimized as a function of the $\alpha_{i}$ and $r_i$.
\begin{figure}[!t!]
  \begin{subfigure}[b]{0.5\textwidth}
    \centering
    \includegraphics[scale=0.35 ]{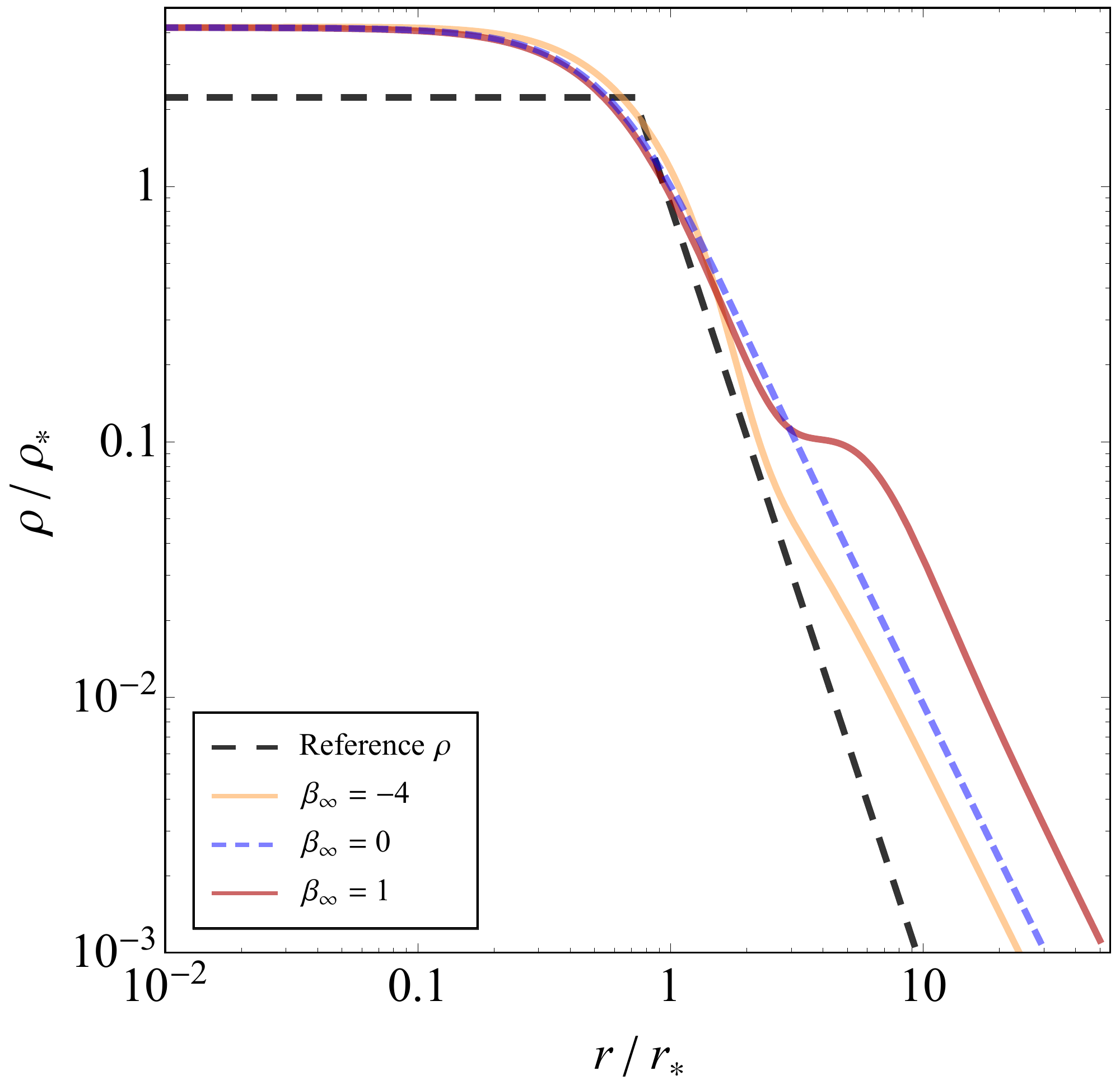}
  \end{subfigure}
\hfill
  \begin{subfigure}[b]{0.5\textwidth}
    \centering
    \includegraphics[scale=0.35 ]{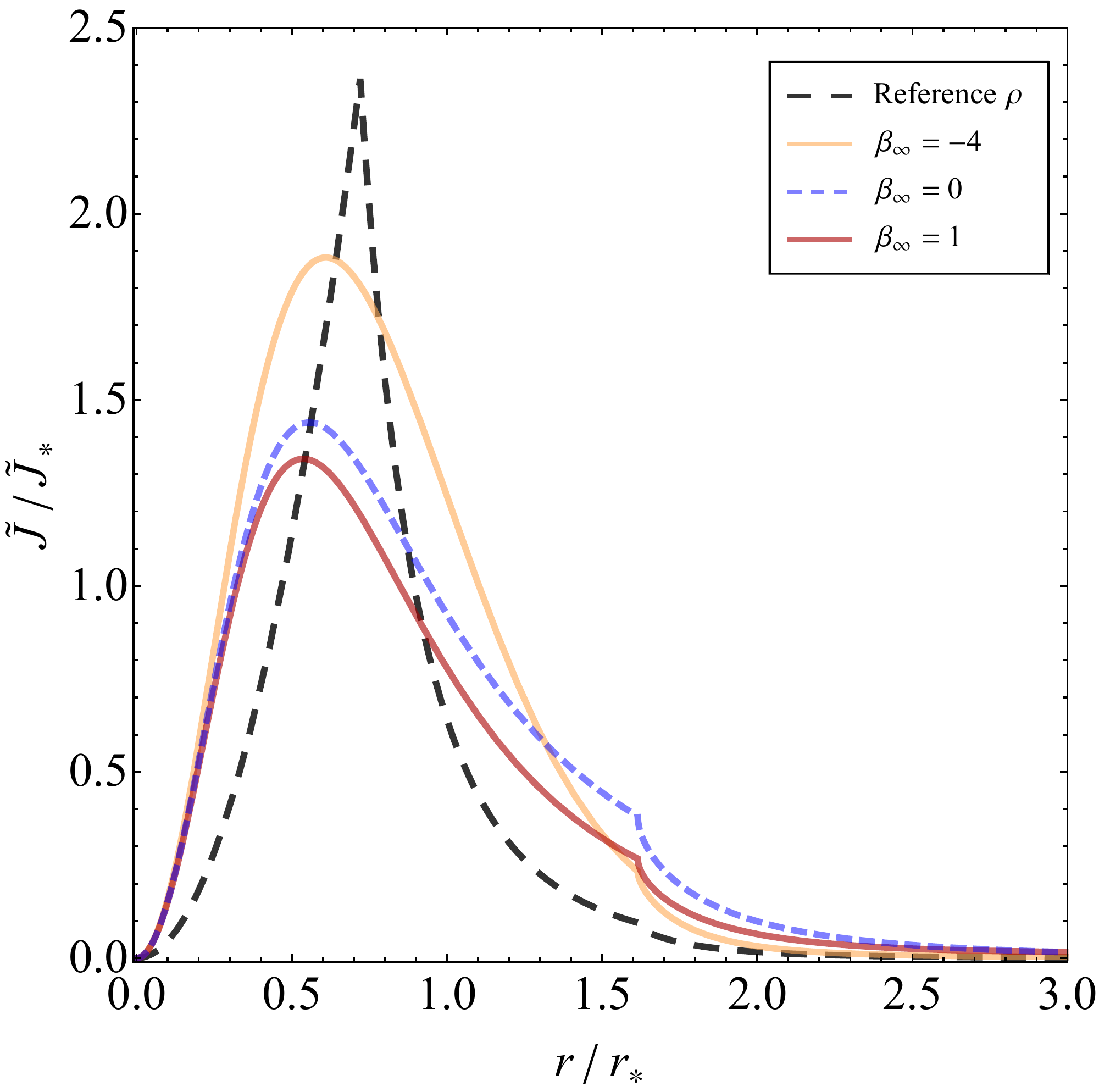}
  \end{subfigure}
  \caption{\underline{Left Panel:} \textit{Reference density profile obtained by minimizing the $J$-factor at given $\mathcal{M}_{*}$. In the same panel, three solutions of Eq.~(\ref{eq:rhojeansinv}) matching the reference $\rho(r)$ in the simple model of constant $\sigma_{los}$ and with Plummer stellar profile and with orbital anisotropy profile $\beta(r)$ given in Eq.~(\ref{eq:radial_general_beta}), defined by the set $r_{\beta}/r_{*} = 3, \eta_{\beta}=3, \beta_{0}=0$ and $\beta_{\infty} = -4,0,1$,  light orange, dashed blue and dark red line respectively. All the profiles are normalized to $\rho_{*}=\rho_{\beta=0}(r_{*})$.} \underline{Right Panel:} \textit{Corresponding $\widetilde{J}(r)$ function, Eq.~(\ref{eq:Jtilde}), whose integral yields the $J$-factor, for the same set of density profiles. We assumed a distance $\mathcal{D} = 10^{2}$ kpc and an angular aperture $\psi_{max} \simeq 0.5^{\circ}$, normalizing $\widetilde{J}(r)$ at the reference value $\widetilde{J}_{*} = \widetilde{J}_{\beta=0}(r_{*})$.}}
  \label{fig:RhoBetaRadJmin}
\end{figure}
In Fig.~\ref{fig:RhoBetaRadJmin} we show the outcome of this procedure in the sample case of 4 power laws indices; we imposed as  constraints $-3\footnote{Here we are also assuming the mass profile to grow at least logarithmically with the radius, as e.g. for a NFW profile.} \le \alpha_i \le 0$ and $r_{2} \le r_{\star} < r_{3}$: Not too surprisingly the result is that the profile minimizing the $J$-factor at given $\mathcal{M}_{*}$,  drawn with a black dashed line,  is cored for $r \lesssim r_{*} $ and it drops to $0$ as fast as possible for larger $r$; we checked explicitly that such result is independent of number of power laws assumed to model $\rho(r)$. \\ 
This result provides an independent check about our findings at constant orbital anisotropy: as long as the volume integral of the density profile at $r_{*}$ encloses the whole mass in $r_{*}$, i.e. no black-hole-like feature is present, the physical configuration that has the minimum $J$-factor happens to be at $\beta=0$, namely when $\rho(r)$ features an inner core. \\
In Fig.~\ref{fig:RhoBetaRadJmin} we report with a dashed blue curve the density profile obtained from the Jeans inversion procedure with $\beta=0$, constant $\sigma_{los}$, Plummer stellar profile and same $\mathcal{M}_{*}$ of the reference density. The good agreement with the latter in the innermost part of the profile does not  leave so much room for improvements. Indeed, exploiting the orbital anisotropy form of  Eq.~(\ref{eq:radial_general_beta}),  a good match to the black dashed line requires $\beta_{0}=0$ to generate an inner core. Moreover, one needs $r_{\beta} \gtrsim /r_{*}$ to have such a core as extended as in the reference case.  Consequently, one ends up to require $\eta_{\beta} \gtrsim 1 $ (but not too large in order to not invalidate conditions in Eq.~(\ref{eq:phys_conds})) to get an appreciable departure from isotropy with $\beta_{\infty} \neq 0$. 
We display this set of results in the left and right panel of Fig.~\ref{fig:RhoBetaRadJmin}: 
\begin{itemize}
\item In the left panel we show two different configurations of the density profile with $\beta(r) \neq 0$ in the outskirts of the dSph, through a mildly circular-like or a purely radial value assigned to $\beta_{\infty}$. Both profiles do not provide a dramatic improvement in matching the reference density. 
\item In the right panel we see that the $J$-value corresponding to the case of $\beta_{\infty} <0 $ has slightly increased, while the radial case gives slightly lower values of  $J$ (an effect anyhow at the per mille level). Note however that  in the limit of $\beta_{\infty}  = 1$  the density is almost turning to an unphysical profile, since an unphysical ripple is starting to appear.
\end{itemize}
We can then conclude this Section stating that in the search for density  profiles that minimize the $J$-factor -- within a given mass $\mathcal{M}_{*}$ at the radius $r_*$, as constrained by kinematical data -- a radial dependence in the unknown orbital anisotropy profile does not significantly alter the picture previously outlined assuming constant $\beta_{c}$.

 
\section{The study case of Ursa Minor}\label{sec:UMi}

\subsection{A few generalities on the dwarf}
Ursa Minor is most often referred to as the target for which both signal and background are most reliably estimated, providing a very competitive limit on annihilating DM models. As discussed in the first analysis on Milky Way satellites by the Fermi collaboration~\cite{Ackermann2011}  (assuming NFW DM profiles and a fixed angular acceptance of solid angle $\Delta\Omega = 2.4 \times 10^{-4}$ sr), and in agreement  with previous analyses \cite{Strigari:2006rd}, Draco and Ursa Minor are the prime targets among the classical dSphs. Ref.~\cite{Cholis:2012am}, assuming instead a Burkert profile and performing an optimization of the angular acceptance, discussed uncertainties in the gamma-ray background determination, concluding that Ursa Minor is the favorite target from this point of view, followed by Sextans which however provides less stringent constraints, while large uncertainties lie in Draco and Sculptor. 
All this motivated us to consider Ursa Minor as a suitable sample object for the purpose of our study. 
We briefly summarize here the main characteristics of this satellite, specifying some of the choices for the data set considered in the phenomenological analysis that follows.  \\
The wide field photometry study in \cite{carrera2002} finds that Ursa Minor hosts a predominantly old stellar population, with virtually all the stars formed before 
10 Gyr ago, and 90\% of them formed before 13 Gyr ago, making it the only dSph Milky Way satellite hosting a pure old stellar population. Using the magnitude 
of the horizontal branch stars and comparing with Hipparcos data on globular clusters the authors determined the distance of Ursa Minor from the Sun to be  
$\mathcal{D} = 76~\pm~4$~kpc, in agreement with the determination from \cite{Bellazzini2002}, but  larger than the mid 1980's value of $66 \pm 3$ kpc quoted in 
\cite{Olszewski1985,Cudworth1986}. As pointed out in \cite{Piatek2005}, the difference is mainly due to the absolute magnitude calibration of the horizontal branch; 
standing in between are the values of  $70 \pm 9$ kpc \cite{Nemec1988} and $69 \pm 4$ kpc \cite{Mighell1999}. Here we choose to adopt  the mean value $\mathcal{D} = 66$~kpc from the old determination of \cite{Olszewski1985,Cudworth1986}, since most often the same has been done in the most recent literature discussing J-factor uncertainties, see, e.g.,~\cite{Bonnivard:2015xpq}. 
Note that, while we will be mostly concerned about relative shifts on J-factor determinations connected to the solution of the Jeans equation, switching from 66~kpc to 76~kpc would imply an overall decrease in values quoted below of about 25\%.
\\
Regarding the stellar surface brightness, the one of Ursa Minor shows the largest ellipticity among all classical dwarfs (excluding Sagittarius that is suffering heavy 
tidal disruption), with mean value of $\epsilon \equiv 1-b/a$ (where $b/a$ is the minor over major axis ratio) estimated in \cite{Irwin1995} to be $0.56\pm0.05$. 
Nevertheless, most analyses treat Ursa Minor as a spherically symmetric system, with stellar surface brightness to be fitted with a template, 
most often via the Plummer, the King or the Sersic model. We will follow \cite{Walker2009,Walker2009err} which suggest to adopt the Plummer model, also in view of the discussion on stellar number density profiles in the previous Section. While the value of the normalization parameter $I_{0}$ in Eq.~(\ref{eq:plummer}) does not need to be specified in the Jeans analysis as well as in its inversion, 
as projected half-light radius  we assume $R_{\textrm{\tiny{1/2}}} = 0.30 \pm 0.02$ kpc, estimate originally obtained in \cite{Irwin1995} from a geometric average of the corresponding half-brightness radii along the semi-major and  semi-minor axis of the projected stellar profile. \\
In what follows, the computation of the l.o.s. integral of $\rho^{2}(r)$ of  Ursa Minor will always refers to pointing to the center of the system with the optimal angular aperture  $\psi_{max} = \arctan(2 \, R_{\textrm{\tiny{1/2}}} / \mathcal{D}) \simeq 0.5^{\circ}$, as most often adopted in literature (see e.g. \cite{ Ackermann:2015zua, Bonnivard:2015xpq}). J-factors will be computed according to  Eq.~(\ref{eq:Jradial}), integrating up to an estimated outer radius $\mathcal{R}=1$ kpc (changing this to an arbitrarily larger value, as 
a negligible numerical impact, generally at the per mille level).

\subsection{Jeans inversion with Ursa Minor data} \label{sec:UMi_fit}

The starting point of our phenomenological analysis on Ursa Minor is a $\chi^{2}$ fit  of the binned l.o.s. velocity dispersion data from \cite{Walker2009}. We consider two possibilities for the fit: {\sl i)} The standard approach in which $\sigma_{los}(R)$ is computed solving the Jeans equation, see Eq.~(\ref{eq:radpress})-(\ref{eq:Pproj}), for given parametric forms of the DM density $\rho(r)$ and of the orbital anisotropy $\beta(r)$ (in the following we will refer to this procedure as \textit{parametric fit}); {\sl ii)}  A direct fit of the data within a given functional form for $\sigma_{los}(R)$ (in the following:  \textit{$\sigma_{los}\,$-driven fit}). 

\begin{figure}[!t!]
  \centering
   \includegraphics[scale=0.37 ]{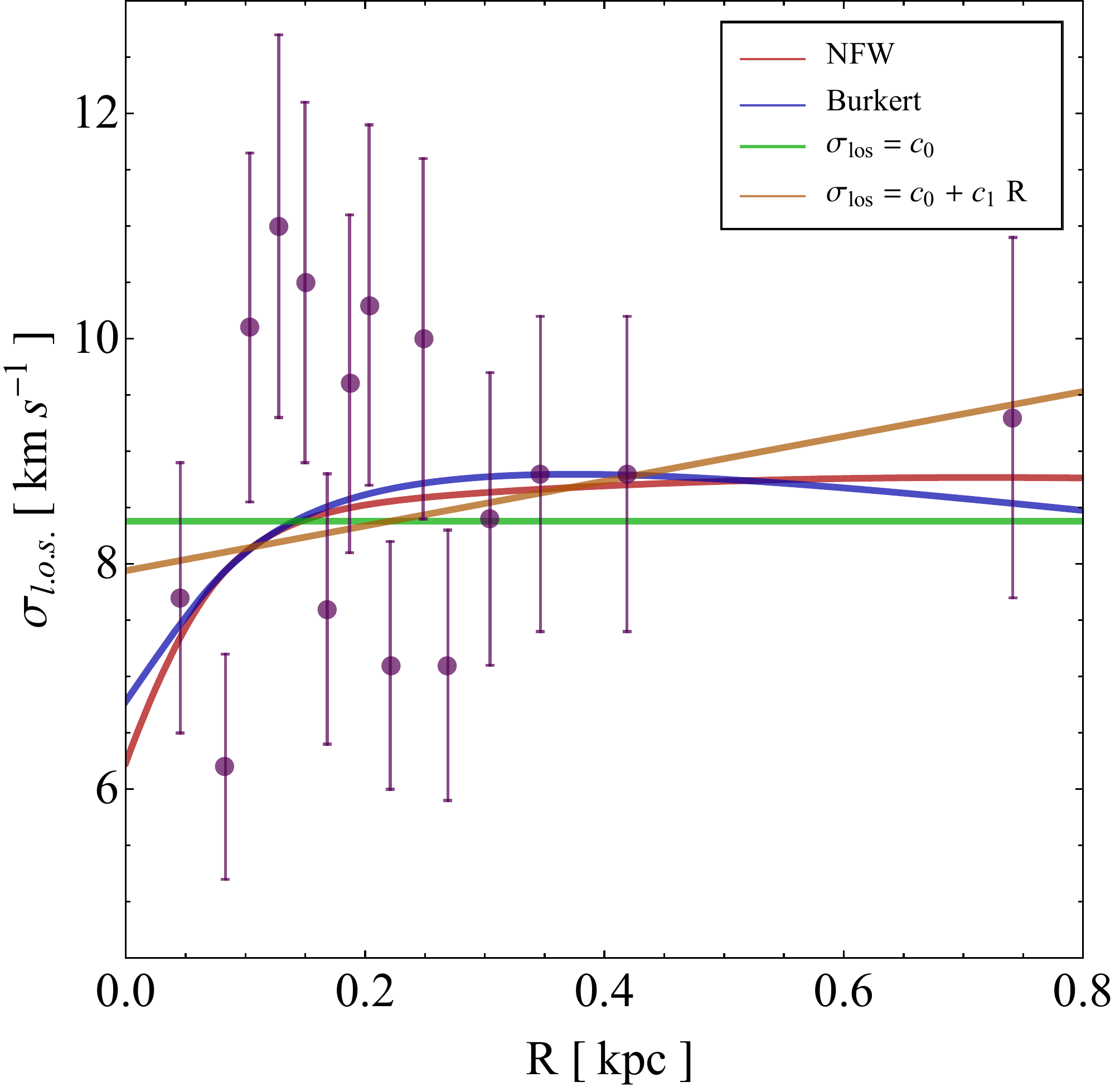}
  \caption{\textit{Binned l.o.s. velocity dispersion data for Ursa Minor dwarf galaxy from \cite{Walker2009}. Best-fit curves are also shown according to four benchmark cases: 
  a parametric fit with a NFW or Burkert density profile together with the assumption of constant orbital anisotropy,  
  and a $\sigma_{los}\,$-driven fit assuming $\sigma_{los}(R)$ to be a constant or a linear function in $R$.}}
  \label{fig:fit_UMi}
\end{figure}

The result of the fit according to four different benchmark cases is shown in Fig.~\ref{fig:fit_UMi}.
The two parametric fits correspond to a  cuspy and a cored $\rho(r)$, namely a NFW and Burkert halo density together with the assumption of constant orbital anisotropy. The other two cases considered are of the two simplest of the $\sigma_{los}\,$-driven kind, namely constant $\sigma_{los}(R)$ and a linear regression in $R$. The best-fit parameters and the corresponding $\chi^{2}_{\textrm{red}} \equiv \chi^{2}/\textrm{n.d.f.}$ are given in Table~\ref{tab:fit_UMi}; as it can be seen all the four benchmarks provide fairly good fits and comparable $\chi^{2}_{\textrm{red}}$.

\begin{table}[ht] 
\caption{\textit{Best-fit values for the parameters involved in the fit of Ursa Minor data from \cite{Walker2009} for the four reference cases under scrutiny. We minimize a $\chi^{2}$ estimator under the assumption of Gaussian distributed data, using  MINUIT package \cite{MINUIT} and estimating the confidence level (C.L.) intervals for the fitted parameters with the MINOS algorithm.}} 
\centering 
\begin{tabular}{l c c c c} 
\hline \hline
Benchmark & Parameters & Mean value & 68\% C.L. & $\chi^{2}_{\textrm{red}}$ \\ [0.5ex]	 
\hline \hline
 NFW & 
\begin{tabular}{ l }
  $r_{n}$ [kpc]  \\
  $\rho_{n}$ [GeV] \\
  $\beta_{c}$  \\ 
\end{tabular}
 & 
 \begin{tabular}{ c }
  $0.61$  \\ 
  $2.59$  \\  
  -0.83  \\  
\end{tabular}
& 
\begin{tabular}{ c }
  $[0.14 \, , \, 2.94]$  \\  
  $[0.30 \, , \, 35.68]$  \\  
  $[-3.02 \, , \, -0.19]$  \\  
\end{tabular} 
& 
\begin{tabular}{ c }
  $1.41$  \\ 
\end{tabular} \\ 
\hline
 Burkert & 
\begin{tabular}{ l }
  $r_{b}$ [kpc]  \\ 
  $\rho_{b}$ [GeV]  \\  
   $\beta_{c}$  \\
\end{tabular}
 & 
 \begin{tabular}{ c }
  0.28  \\  
  12.77  \\  
  -0.36  \\ 
\end{tabular}
& 
\begin{tabular}{ c }
  $[0.12 \, , \, 0.54]$  \\  
  $[5.59 \, , \, 55.13]$  \\  
  $[-1.63 \, , \, 0.10]$  \\  
\end{tabular} 
& 
\begin{tabular}{ c }
  $1.44$  \\ 
\end{tabular} \\  
\hline 
 $\sigma_{los} = c_{0} $ & 
\begin{tabular}{ l }
\ \ \  $c_{0}$  [km s$^{-1}$] \\ 
\end{tabular}
 & 
 \begin{tabular}{ c }
  $8.38$  \\  
\end{tabular} 
& 
\begin{tabular}{ c }
  $[8.03 \, , \, 8.73]$  \\ 
\end{tabular} 
& 
\begin{tabular}{ c }
  $1.32$  \\ 
\end{tabular} \\ 
\hline
 $\sigma_{los} = c_{0} + c_{1} \, \frac{R}{\mathcal{R}}$ & 
\begin{tabular}{ l }
 \ \ \    $c_{0}$  [km s$^{-1}$] \\ 
  \ \ \   $c_{1}$  [km s$^{-1}$] \\ 
\end{tabular}
 & 
 \begin{tabular}{ c }
   $7.94$   \\  
   $1.99$  \\  
\end{tabular} 
& 
\begin{tabular}{ c }
  $[7.32 \, , \, 8.56]$  \\  
  $[-0.29 \, , \, 4.27]$  \\  
\end{tabular} 
& 
\begin{tabular}{ c }
  $1.35$  \\ 
\end{tabular} \\ 
\hline  
\end{tabular} \\
\label{tab:fit_UMi} 
\end{table}

\begin{figure}[!t!]
  \begin{subfigure}[b]{0.5\textwidth}
    \centering
    \includegraphics[scale=0.4 ]{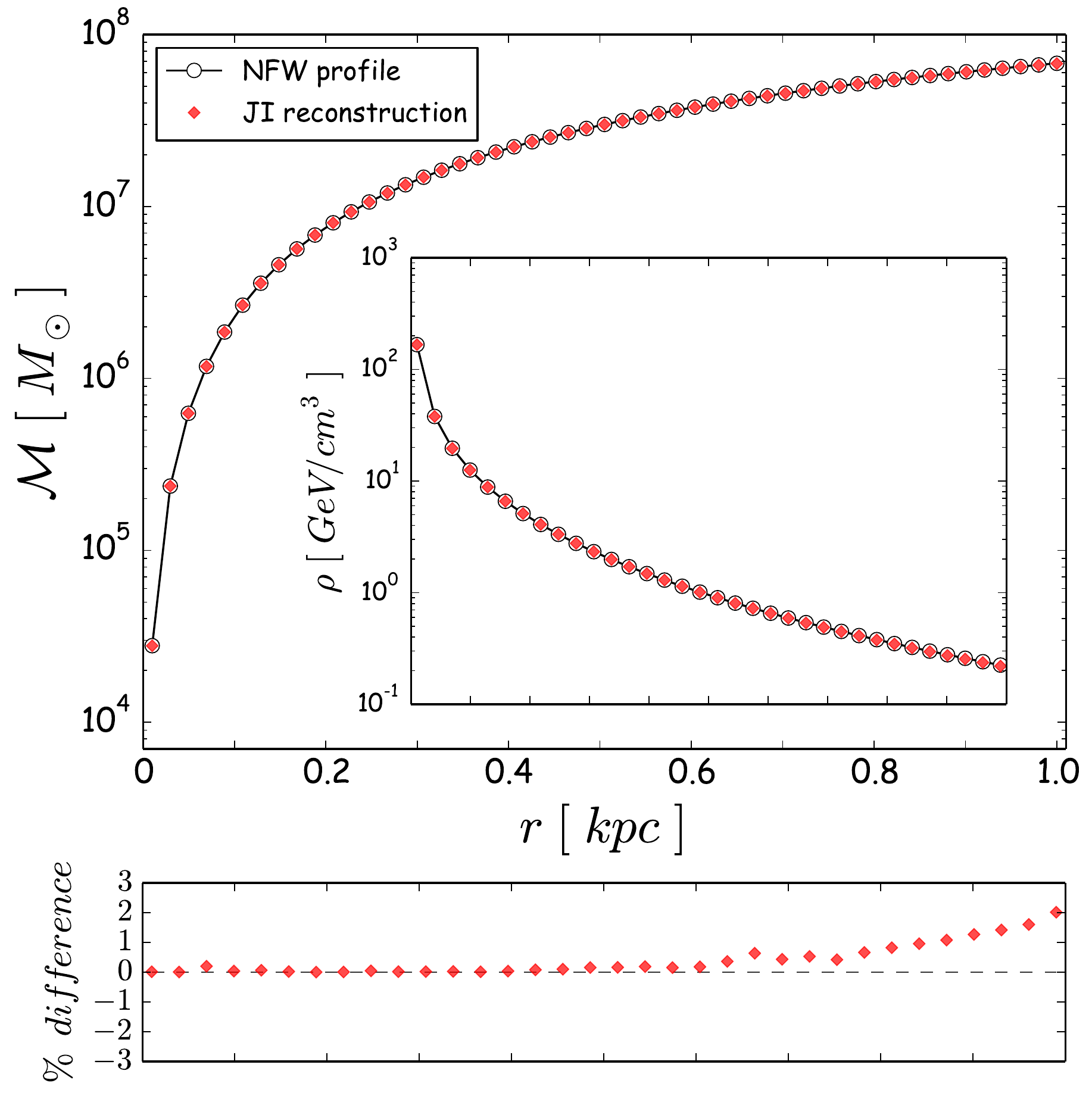}
  \end{subfigure}
\hfill
  \begin{subfigure}[b]{0.5\textwidth}
    \centering
    \includegraphics[scale=0.4 ]{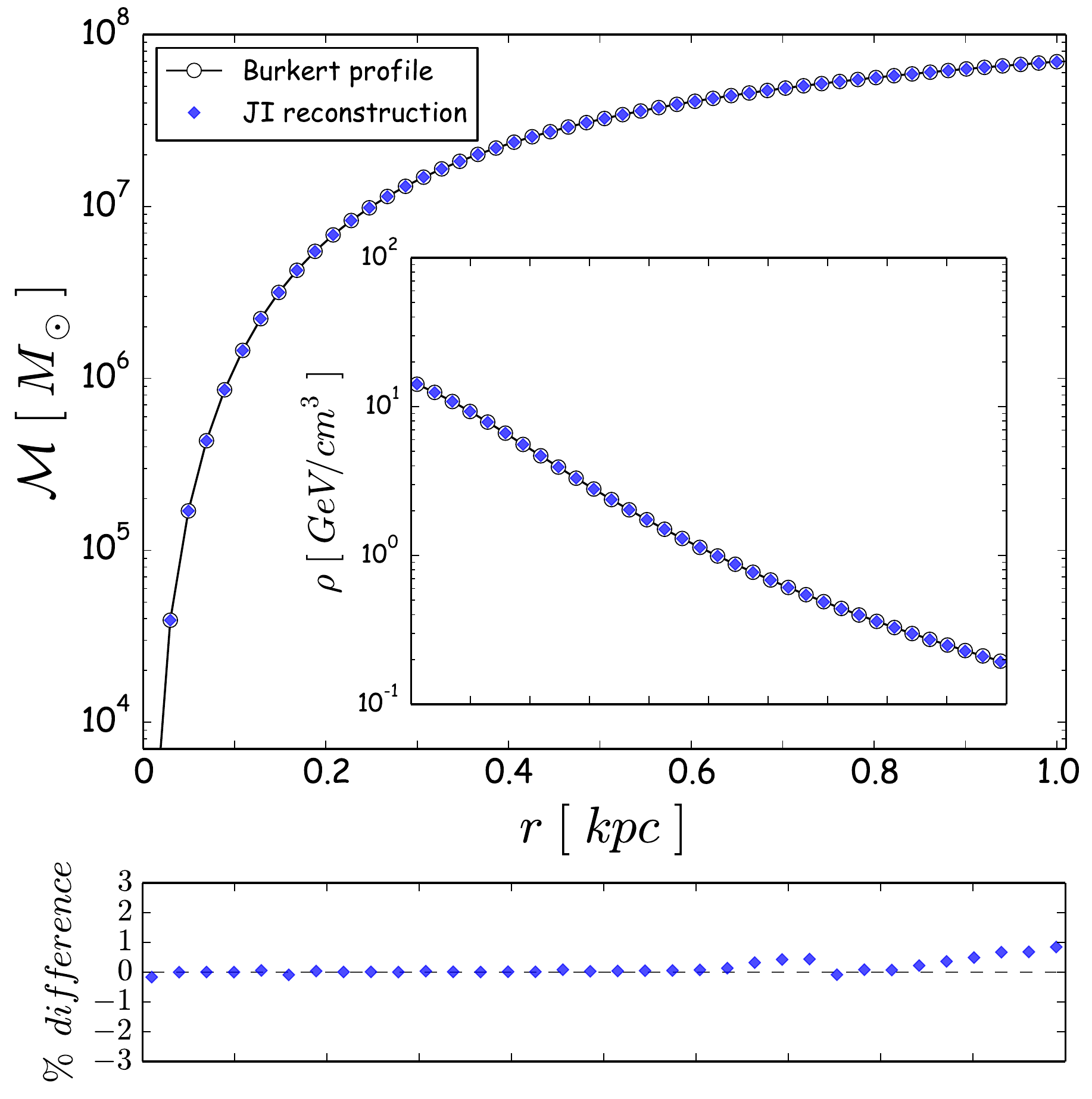}
  \end{subfigure}
  \caption{\textit{Mass and the density profiles reconstructed via the Jeans inversion (JI) algorithm taking as input the parametric fit of $\sigma_{los}(R)$, in case of the NFW (left panel) and Burkert (right panel) best-fits (see Table~\ref{tab:fit_UMi}). Percentage differences for the reconstructed density profiles are shown in the insets.}}
  \label{fig:CorrectMassRec}
\end{figure}

Fits of the l.o.s. velocity dispersion data are taken as an input for the procedure of  inversion of the Jeans equation, to reconstruct -- at given $\beta(r)$ -- the DM mass and density profiles, and hence study how the minimal $J$-factor of Ursa Minor, i.e. the 2$\, \sigma$ lower-limit of the posterior distribution of $J$,  depends on the assumed $\beta(r)$. The different choices of $\sigma_{los}(R)$ have been considered to check whether the parametric fit from an a priori physical model and/or the $\sigma_{los}\,$-driven fit (agnostic, but non-necessarily corresponding to a physical model) may be introducing a bias in the analysis. Incidentally, the parametric fits also allow for a cross-check on the accuracy of our numerical implementation of the inversion procedure. In Fig.~\ref{fig:CorrectMassRec} we show profiles reconstructed via Eq.~(\ref{eq:ourinv}) compared against the initial NFW and Burkert parametric forms. The latter are given as an input to derive the parametric fits for $\sigma_{los}(R)$, which are displayed in Fig.~\ref{fig:fit_UMi}. To perform this exercise we used the best-fit values specified in Table~\ref{tab:fit_UMi}. The displayed reconstruction of the mass profile for both cuspy and cored cases comes with a level of accuracy better than the per mille in the whole range of the binned dispersion data of Ursa Minor, i.e. $40-750$ pc; relative differences  above few percent arise only at  inner radii much smaller than $10$ pc. A similar level of accuracy is found for the reconstruction of the inner density profile, for which we also show the relative percentage difference in the insets below the two plots. The density profile of parametric fits can be conveniently evaluated via an iterative difference quotient algorithm \cite{NumRecipes1992} applied to the mass profile.

\subsection{Inversion and MCMC with constant anisotropies}

When investigating the impact of orbital anisotropy on density profiles and $J$-factors the emphasis will be on discussing minimum values consistent with kinematical observables. In fact, since the DM velocity-averaged pair annihilation cross section $\langle \sigma \, v \rangle$ accessible to gamma-ray observations scales with the measured flux $\mathcal{F}$ as
\begin{equation}
 \langle \sigma \, v \rangle \sim \mathcal{F} / J \ ,
 \label{eq:sigmav}
\end{equation}
the lowest $J$-value allowed by Ursa Minor data can be directly linked to how much the  upper-bound on $ \langle \sigma \, v \rangle$, reported e.g. in \cite{Ackermann:2013yva} for this galaxy, can be relaxed.  As shown in Section~\ref{Sec:OurInversion}, the minimum $J$-value turns out to be weakly affected by a radial dependence of the anisotropy profile as long as $\sigma_{los}(R)$ is mildly varying in $R$ as well. Thus we can restrict our phenomenological analysis to the simple case of $\beta(r)=\beta_{c}$ without loss of generality in the conclusions.

\begin{figure}[!t!]
  \begin{subfigure}[b]{0.5\textwidth}
    \centering
    \includegraphics[scale=0.37]{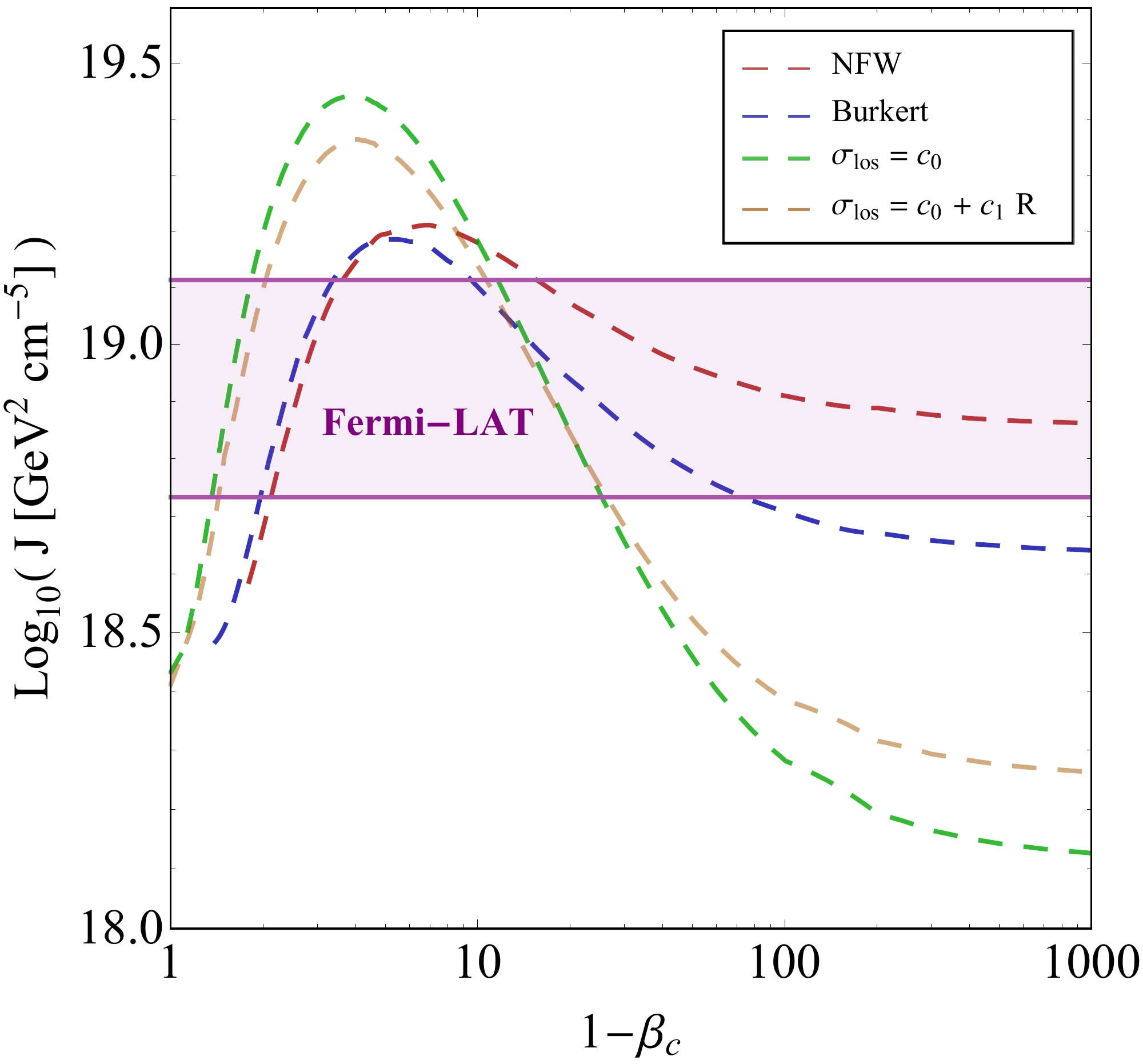}
  \end{subfigure}
\hfill
  \begin{subfigure}[b]{0.5\textwidth}
    \centering
    \includegraphics[scale=0.36]{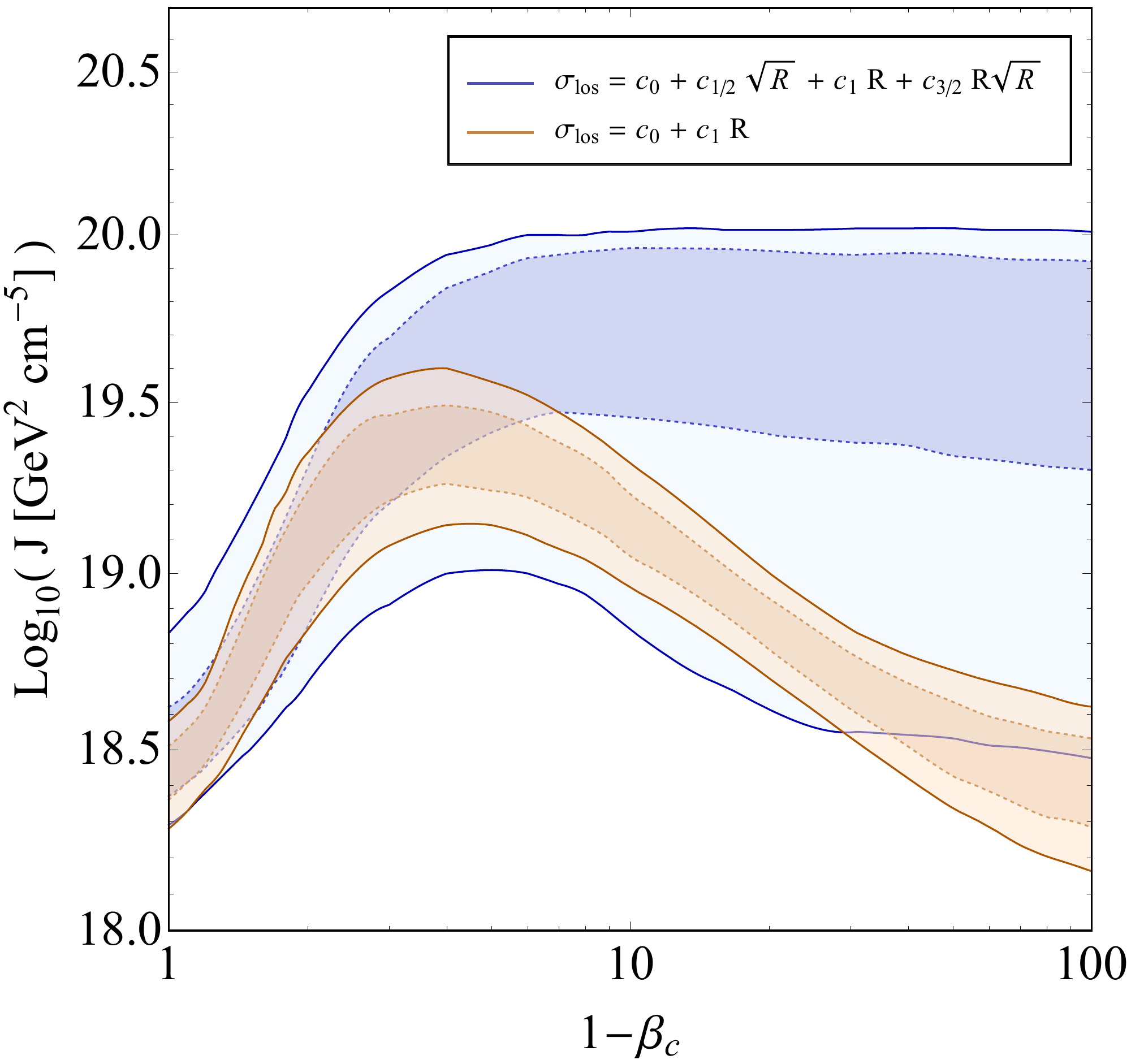}
  \end{subfigure}
  \caption{\underline{Left panel:} \textit{$J$-factor as a function of constant orbital anisotropy $\beta_{c}$ for the four benchmark cases at hand. An inner cut of $10$ pc is applied to all the physical densities as described in text.  In the plot also the representative $1\sigma$ band for the $J$-factor of Ursa Minor assumed by Fermi-LAT in the latest analysis regarding limits on pair annihilations of DM particles.} \underline{Right panel:} \textit{$68\%$ and $95\%$ probability region associated to the $J$-factor as a function of $\beta_{c}$ from the MCMC we performed -- in the context of the Jeans inversion approach proposed -- with the BAT library \cite{Caldwell:2008fw}. We considered two different parameterizations of $\sigma_{los}(R)$, as reported in the legend.}}
  \label{fig:J_VS_beta_const}
\end{figure}

Starting with the four benchmark cases for $\sigma_{los}(R)$ shown in Fig.~\ref{fig:fit_UMi}, in the left panel of Fig.~\ref{fig:J_VS_beta_const} we show results for $\log_{10}J$ as a function of the value assumed for $\beta_{c}$ and in the range corresponding to models satisfying the set of conditions for a physical model,  see Eq.~(\ref{eq:phys_conds}): for the parametric fits
only values of $\beta_{c}$ lower or equal than the one assumed for computing $\sigma_{los}(R)$ are allowed; for the
$\sigma_{los}\,$-driven fits, the Jeans inversion procedure gives physical models up to $\beta_{c} =0$. Note that the exclusion for all the four benchmarks of constant radial-like anisotropy profiles, matches with the requirement of $\beta(r \to 0) \le 0$ to obtain a positive stellar phase-space density at the center of the system  \cite{An:2005tm}. The behavior of $\log_{10} J$ as a function of $\beta_{c}$ is qualitatively the same for all the four scenarios: For a given stellar surface density and $\sigma_{los}(R)$ when starting from  $\beta_{c}$ close to 0 and going to progressively larger circular anisotropy the density profiles becomes progressively more concentrated and hence the $J$-factor grows, up to the level one starts to see the turnaround in logarithmic slope already seen the left panel of Fig.~\ref{fig:RhoBetac} when getting close to the pure circular orbit limit. The decrease in $J$-factor at this turning point becomes even more pronounced since we are taking here the conservative view of not extrapolating the profile obtained from the inversion procedure all the way to $r\to 0$. We rather introduce a inner density cutoff $\rho(r<r_c)= \rho(r_c)$, with $r_{c}=10$~pc as sample value avoiding  an extrapolation to radii smaller than the order of magnitude of the radius in the innermost bin of $\sigma_{los}$ data.
The two benchmark $\sigma_{los}(R)$ obtained from parametric fits are both characterized by a concave inner tilt; this makes the density profile shallower than those for the $\sigma_{los}\,$-driven cases at small negative values of $\beta_{c}$ and hence we find lower $J$-factors. On the other hand the same concave tilt partially washes out the cancellation we discussed for purely circular orbits, as well as makes the effect of the internal cutoff less severe, and hence drives a less pronounced decrease of the $J$-factor at very large negative $\beta_{c}$.
In the left panel of Fig.~\ref{fig:J_VS_beta_const} we show also the $1 \, \sigma$ band for the Ursa Minor $J$-factor adopted in the Fermi-LAT analysis of Ref.~\cite{Ackermann:2015zua} (scaled to the same dwarf distance $\mathcal{D}$ adopted here): the minimum $J$-values for the parametric fit cases are obtained for the same $\beta_{c}$  implemented to generate the $\sigma_{los}(R)$ profiles and are  within $2 \, \sigma$ with respect to Fermi quoted values (we take the Fermi band as visual guide only and do not intend to make any statistical statement at this point).
For the case of $\sigma_{los}\,$-driven fits, i.e. when assuming a constant $\sigma_{los}(R)$ or a linear regression, the minimum $J$ turns out  to correspond to the circular orbit limit; in particular to the sample benchmark with constant $\sigma_{los}(R)$ the minimum $J$ is roughly $ 4 \, \sigma $ away from the nominal value in \cite{Ackermann:2015zua} for Ursa Minor, driving -- as naive estimate -- a relaxation of the extrapolated limit  on $ \langle \sigma \, v \rangle$ of a factor of few. On the other hand these correspond to rather extreme configurations, with, as explained in the previous Section, extreme cusps which would be developing in the very inner region of the system (even below the cutoff radius we are considering), finally shrinking to a $1/r$ profile and a central black hole. The black hole mass for $\beta_{c} \lesssim - 100$ would be of the order of $10^{6}~M_{\odot}$ or larger, and while it could be physically motivated that to have a flat density around them, see, e.g., scenarios in Ref.~\cite{Ullio:2001fb}, such masses are at the edge of observational limit for this system~\cite{2009ApJL_UMi_BH,2013_UMi_BH}.   

In order to provide a more robust statistical assessment of these findings, we also present here the results of  a Bayesian fit of Ursa Minor $\sigma_{los}$ binned data, computing the $J$-factor through 
the inversion formula for a finite grid of constant orbital anisotropies. 
We have exploited for the purpose two different parameterizations of $\sigma_{los}(R)$, namely the same linear expression in $R$ already introduced, and the following polynomial form (recall that $\mathcal{R}$ is the outer radius and we picked as reference value 1~kpc):
\begin{equation}
\label{eq:sigma_los_param}
\sigma_{los}(R) = c_{0} + c_{1/2}\, \sqrt{\frac{R}{\mathcal{R}}} + c_{1}\, \frac{R}{\mathcal{R}}+ c_{3/2}\, \frac{R}{\mathcal{R}} \sqrt{\frac{R}{\mathcal{R}}}  \ .
\end{equation}
This form has been chosen since it can nicely interpolate among the four benchmarks in Fig.~\ref{fig:fit_UMi} within a broader set of behaviours, including, 
eventually convex tilts at small $R$.\\
To perform our MCMC analysis we use the Bayesian Analysis Toolkit library \cite{Caldwell:2008fw}, assigning generous flat priors to the coefficients defining $\sigma_{los}(R)$, 
namely $c_{0} \in [-50,50]$, $c_{1} \in [-500,500]$, and $c_{1/2,\,3/2} \in [-250,250]$, and performing a total of  $10^{8}$ iterations distributed in $20$ chains. 
In each iteration of the MCMC we invert the Jeans equation for all the set of $\beta_{c}$ considered, computing the density profiles via Eq.~(\ref{eq:rhojeansinv}) and the $J$-factor when the physical 
conditions in Eq.~(\ref{eq:phys_conds}) are met.
As a result of this involved procedure, we obtain in the end the posterior probability density function of $\log_{10} J$ for each of the selected constant anisotropies, i.e. 
without any marginalization procedure over unknown parameters unrelated to observable quantities.

\begin{figure}[!t!]
  \centering
   \includegraphics[scale=0.37 ]{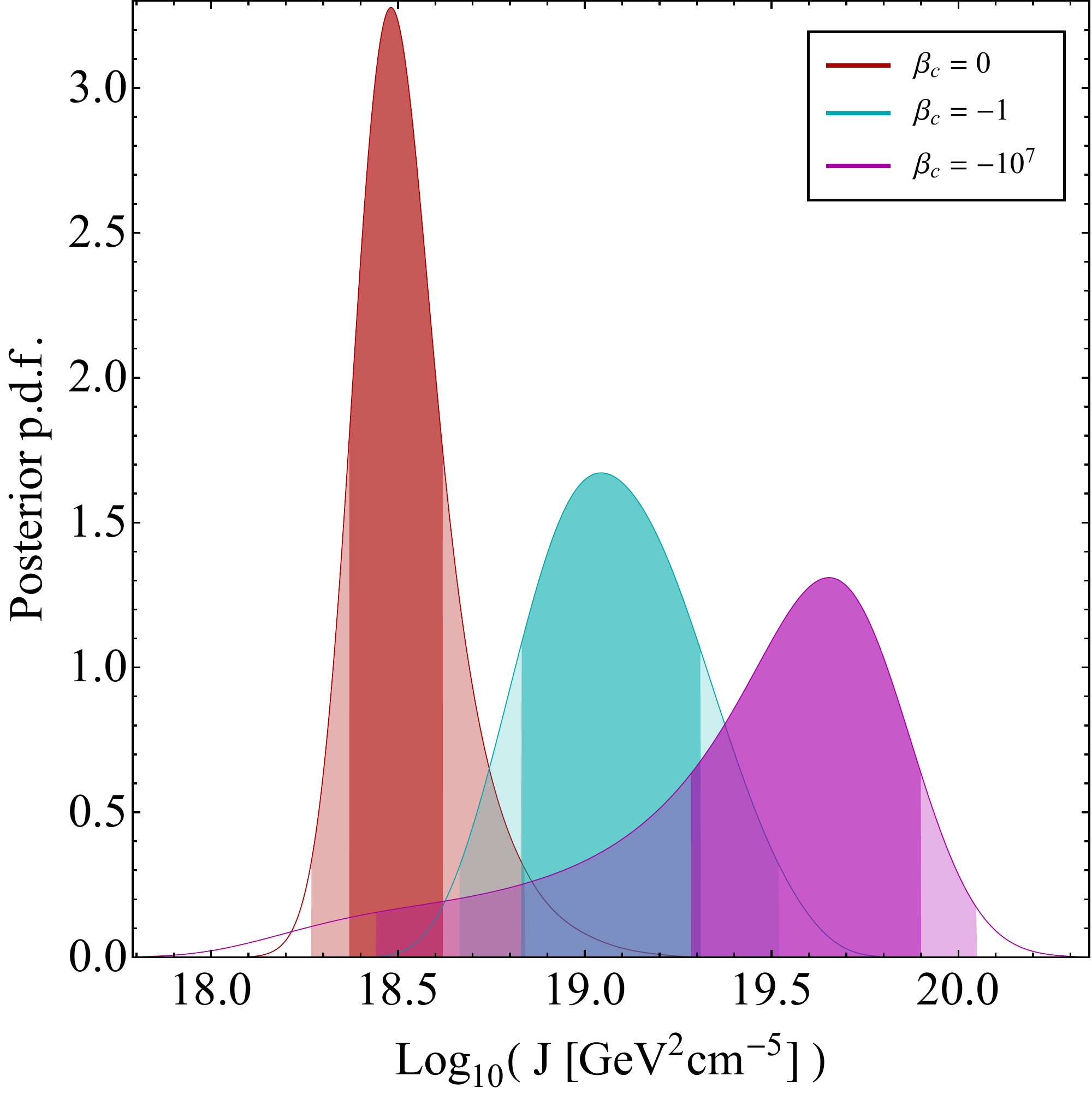}
  \caption{\textit{Posterior probability density function of $\log_{10}(J)$ for a constant orbital anisotropy $\beta_{c} = 0, -1, -10^{7}$ obtained through our inversion of the spherical Jeans equation. 
  The darker colored area in each distribution corresponds to the $68\%$ probability region, the shader one accounts for the $95\%$.} }
  \label{fig:log10Jhisto}
\end{figure} 

\begin{table}[b!] 
\caption{\textit{$68\%$ and $95\%$ minimum $J$ for the two different parameterizations of $\sigma_{los}(R)$ used in our MCMC and related to our Jeans inversion approach.
In the last column we report the relaxing factor one can naively derive for the constraints of DM particle properties comparing our  $J$-value at $95\%$ probability with the $2 \, \sigma$ Fermi-LAT minimum
value for Ursa minor in \cite{Ackermann:2015zua}, namely $\min J^{\textrm{Fermi}}_{@\,2\,\sigma}= 3.5 \times 10^{18}$ GeV$^{\,2}$ cm$^{-5}$ after the proper rescaling to the distance of the dSph used in this work.}}
\centering 
\begin{tabular}{c c c c c} 
\hline \hline
$\sigma_{los}(R)$  & $\min J_{@68\%}$  [GeV$^{2}$ cm$^{-5}$]    &  $\min J_{@95\%}$  [GeV$^{2}$ cm$^{-5}$]  & $\min J_{@95\%}/ \min J^{\textrm{Fermi}}_{@\,2\,\sigma}$ \\ [0.5ex]	 
\hline \hline
 $c_{0} + c_{1} R/\mathcal{R}$ & 
\begin{tabular}{ c } 
$1.29 \times 10^{18}$ 
\end{tabular}
 & 
 \begin{tabular}{ c }
$9.12 \times 10^{17}$  \\ 
\end{tabular}
& 
\begin{tabular}{ c }
3.85 \\
\end{tabular} \\
\hline
Eq.~(\ref{eq:sigma_los_param})  & 
\begin{tabular}{ c }
$2.34 \times 10^{18}$  
\end{tabular}
 & 
 \begin{tabular}{ c }
$1.86 \times 10^{18}$     \\ 
\end{tabular}
& 
\begin{tabular}{ c }
 1.88   \\
\end{tabular}\\
\hline  
\end{tabular} \\
\label{tab:J_MCMC_min} 
\end{table}

In the right panel of Fig.~\ref{fig:J_VS_beta_const}, we plot the $68\%$ and $95\%$ probability region of $\log_{10}J$ for orbital anisotropies in the range $-100 \lesssim \beta_{c} \le 0$, finding again for the linear parametrization of $\sigma_{los}(R)$ that the minimum value of $J$ happens in the circular orbit limit, with precise value sensitive to the choice of inner cutoff radius at $10$ pc.
While the more general parametrization in Eq.~(\ref{eq:sigma_los_param}) 
encodes the linear behaviour as well, the latter becomes now only a special realization of  it, and consequently populate the tail of the distribution in $\log_{10} J$ when one probes lower ad lower stellar anisotropies. This trend of the posteriors of $\log_{10} J$ is summerized in Fig.~\ref{fig:log10Jhisto}, where we show three illustrative cases, highlighting with the color code their $68\%$ and $95\%$ probability area (defined from the local mode of the distribution). Eventually, we can conclude that the lowest $J$-value is again found in correspondence to the isotropic stellar motion when considering the case of Eq.~(\ref{eq:sigma_los_param}), while it is provided by the limit of  circular-like orbits in case a linear form for $\sigma_{los}(R)$ is implemented (together with the caveats of the related black hole feature discussed above). For what concerns the bounds on the cross section of a DM annihilating pair, in Table~\ref{tab:J_MCMC_min} we report the naive maximum relaxation one can apply to these limits for the study case of Ursa Minor: We remark that at $2 \, \sigma$ one can naively relax the Ursa Minor upper-bound on $\langle \sigma \, v \rangle$ by a factor roughly ranging from 2 to 4.
 
\subsection{$J$-factors and $\beta(r)$ from NFW and Burkert profiles}

While the general impact on $J$ for spatially dependent orbital anisotropies has been qualitatively discussed in Section~\ref{Sec:OurInversion}, we try to address here a slightly different, though related issue: within a physically motivated ansatz for the DM density profile of the system, what is the orbital anisotropy profile compatible with the velocity dispersion data that, at the same time, provides the smallest $J$-factor possible for the galaxy? \\ 
Assuming a rather general form for the profile of $\beta(r)$, we can answer to this question quantitatively taking again Ursa Minor as our study case. 
Taking as reference parametric forms the NFW or Burkert  profiles, these are completely determined by only two parameters, namely a characteristic scale radius $r_{s}$ and a normalization $\rho_{0}$. As discussed in Section \ref{Sec:MassEstimator}, in case of nearly flat projected l.o.s. velocity dispersion profile in the outskirts of galaxy -- a condition fulfilled by Ursa Minor, see Fig.~\ref{fig:fit_UMi} -- the mass $\mathcal{M}_{*}$ enclosed within the radius $r_*$ (close to the half-light radius of the stellar profile and defined as the radius at which its logarithmic slope  is equal to -3, see Eq.~(\ref{eq:wolfrstar})) is nearly independent of the assumed orbital anisotropy profile. To a good approximation, we can then trade the normalization of the DM profile $\rho_{0}$ by $\mathcal{M}_{*}$:
 \begin{equation}
 \label{eq:normalization_of_rho}
 \mathcal{M_{*}} = 4 \pi \int_{0}^{r_{*}} d \tilde{r} \, \tilde{r}^{2} \, \rho(\tilde{r} ; \rho_{0}, r_{s}) \ \Rightarrow \ \rho_{0} = \rho_{0}( \mathcal{M}_{*}, r_{s}) \ .
 \end{equation}
This expression then sets the normalization of $\rho$ to be a function of $r_{s}$ and $\mathcal{M}_{*}$, with in turn the latter being set in terms of the normalization of $\sigma_{los}$. At fixed $\mathcal{M}_{*}$, the $J$-factor for the profile becomes only a function of $r_{s}$ and selecting the minimum $J$ fixes this parameter as well,  fully determining the density profile. Eventually, assuming a definite form for the stellar anisotropy, one is able to read the profile from a fit of the dispersion data.
\begin{figure}[!t!]
  \begin{subfigure}[b]{0.5\textwidth}
    \centering
    \includegraphics[scale=0.37 ]{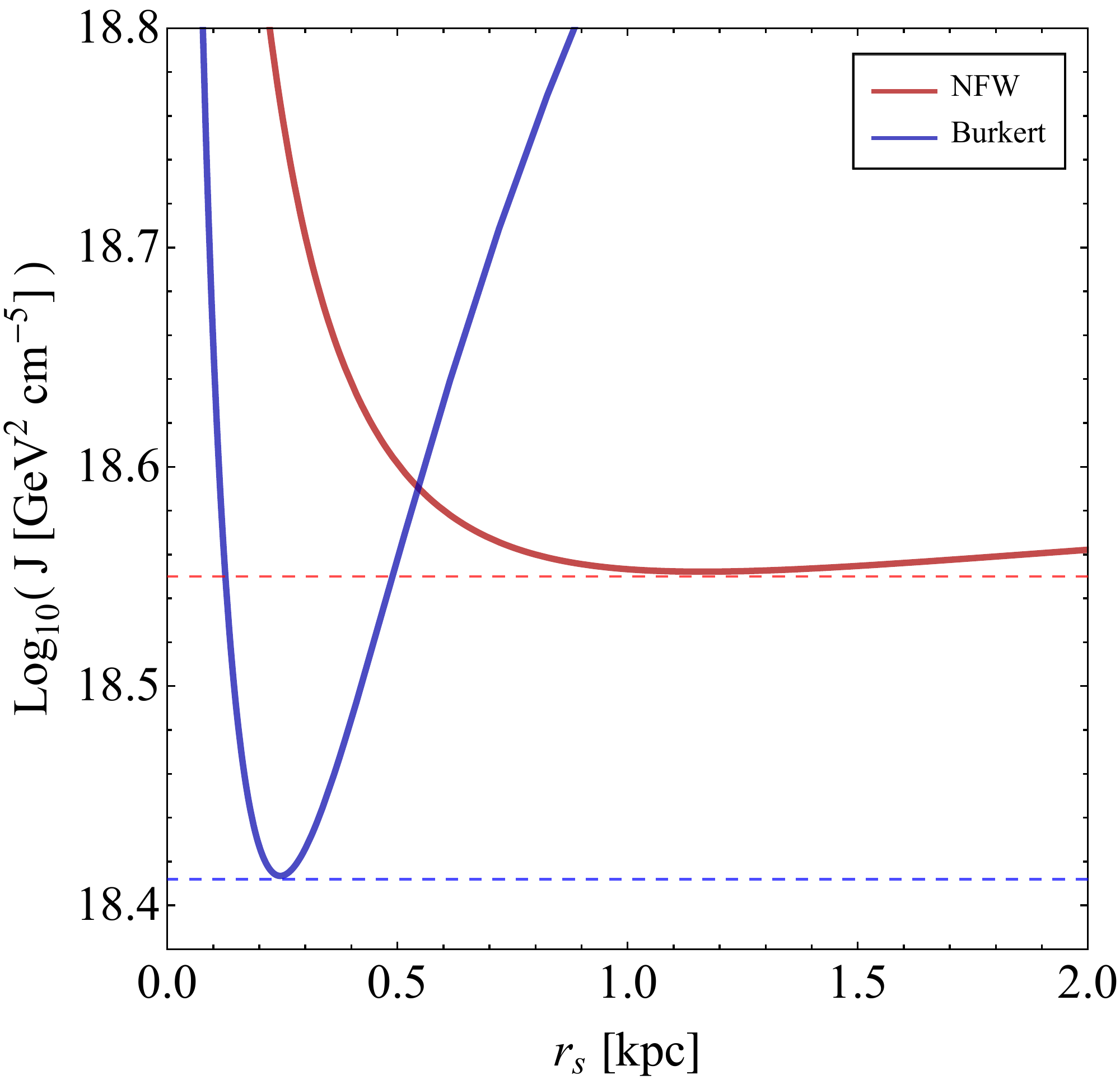}
  \end{subfigure}
\hfill
  \begin{subfigure}[b]{0.5\textwidth}
    \centering
    \includegraphics[scale=0.36]{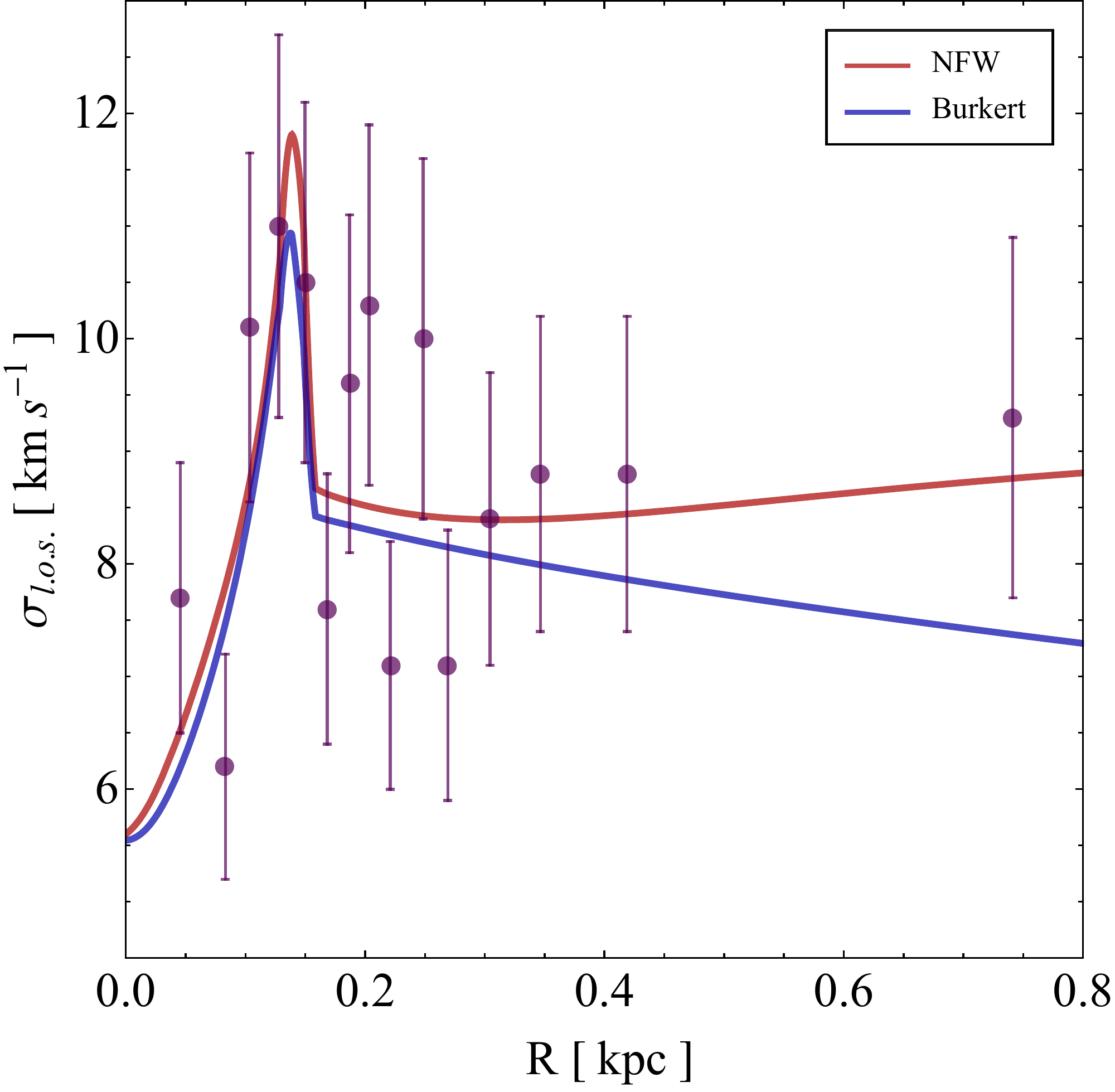}
  \end{subfigure}
  \caption{\underline{Left panel:} \textit{$J$-factor for the NFW and Burkert profiles as a function of the scale radius $r_{s}$, after fixing the density normalization assuming $\mathcal{M}_{*} = 1.82 \cdot 10^{7}\, M_{\odot}$.  The dashed horizontal lines highlight the minimum J-value compatible with such constraint.} \underline{Right panel:} \textit{The fit of the l.o.s. velocity dispersion data of Ursa Minor considered in this work using NFW and Burkert profiles that minimize the $J$-factor according to the constraint $\mathcal{M}_{*} = 1.82 \cdot 10^{7}\, M_{\odot}$. }}
  \label{fig:J_min_with_BURNFW}
\end{figure}

In Fig.~\ref{fig:J_min_with_BURNFW} we show the logarithm of $J$ as a function of the scale radius $r_{s}$ for both the NFW and Burkert cases fixing $\mathcal{M}_{*} = 1.82 \cdot 10^{7}\, M_{\odot}$, as follows from applying Eq.~(\ref{eq:wolfmstar}) to the Ursa Minor $\sigma_{los}$ binned data. Analogously to the results in Section~\ref{Sec:OurInversion}, for both the NFW and Burkert profiles, the minimum of $J$ corresponds to an intermediate value for the scale radius $r_{s}$ such that the density profile flattens as much as possible in the inner part, and falls off rapidly in the outskirts. Comparing the two cases, 
the lowest  $J$-value corresponds once again to the most cored of the two density profiles. The radius $r_{s}$ that minimizes $J$ at fixed $\mathcal{M_{*}}$ is reported in Table~\ref{tab:J_rs_min}. Having selected within this approach one NFW and one Burkert profile, we search for a compatible orbital anisotropy profile through Eq.~(\ref{eq:radpress})-(\ref{eq:Pproj}), using Ursa Minor velocity dispersions data and the $4$ parameter function in Eq.~(\ref{eq:radial_general_beta}) for $\beta(r)$. The corresponding  best-fit  $\sigma_{los}(R)$ are plotted in the right panel of Fig.~\ref{fig:J_min_with_BURNFW} and are generated in both cases by a $\beta(r)$ making, in correspondence to the deep in $\sigma_{los}(R)$,   a violent transition between a mild circular orbit  regime to a purely isotropic tracer motion. The nominal best-fit values of the anisotropy parameters are collected in Table~\ref{tab:J_rs_min}, together with the $\chi^{2}_{red}$ of the fit related only to the stellar anisotropy degrees of freedom. 
\begin{table}[t] 
\caption{\textit{Nominal best-fit values of the orbital anisotropy parameters fitted using NFW and Burkert profiles that minimize the $J$-factor according to the constraint $\mathcal{M}_{*} = 1.82 \, \times 10^{7}\, M_{\odot}$.}} 
\centering 
\begin{tabular}{c c c c c c c c} 
\hline \hline
$\rho(r)$  & $r_{s}$ [kpc]   &  $\log_{10} (J \, [GeV^{2}   cm^{-5}])$  & $\beta_{0}$ & $\beta_{\infty}$  & $r_{\beta}$ [kpc] & $\eta_{\beta}$ & $\chi^{2}_{red}$ \\ [0.5ex]	 
\hline \hline
 NFW & 
\begin{tabular}{ c } 
  $1.16$   
\end{tabular}
 & 
 \begin{tabular}{ c }
  $18.55$  \\ 
\end{tabular}
& 
\begin{tabular}{ c }
  -8.0  \\
\end{tabular} 
& 
\begin{tabular}{ c }
  $0$  \\
\end{tabular}
& 
\begin{tabular}{ c }
  $0.15$  \\
\end{tabular}
&
\begin{tabular}{ c }
   $\gtrsim 10^{3}$  \\
\end{tabular}
&
\begin{tabular}{ c }
  $0.98$  \\
\end{tabular}\\
\hline
 Burkert & 
\begin{tabular}{ c } 
  $0.25$   
\end{tabular}
 & 
 \begin{tabular}{ c }
  $18.41$  \\ 
\end{tabular}
& 
\begin{tabular}{ c }
  -5.0  \\
\end{tabular} 
& 
\begin{tabular}{ c }
  $0$  \\
\end{tabular}
& 
\begin{tabular}{ c }
  $0.15$  \\
\end{tabular}
&
\begin{tabular}{ c }
  $\gtrsim 10^{3}$  \\
\end{tabular}
&
\begin{tabular}{ c }
  $1.17$  \\
\end{tabular}\\
\hline  
\end{tabular} \\
\label{tab:J_rs_min} 
\end{table}

Since the value of $\log_{10}(J$ [GeV$^{2}$ cm$^{-5}$]$)$ for Ursa Minor reported by the Fermi collaboration in \cite{Ackermann:2013yva}  is respectively $18.92 \pm 0.19$ and $18.82 \pm 0.20$  for an assumed NFW and Burkert density (considering a distance $\mathcal{D}$ of $66$ kpc), according to our findings in Table~\ref{tab:J_rs_min}, the minimum possible $J$ compatible with a mass estimator of $1.82 \, \times 10^{7}\, M_{\odot}$ and a NFW or Burkert profile is essentially within the $2 \, \sigma$ range of the corresponding Fermi-LAT quoted value. 
An uncertainty to $\mathcal{M_{*}}$ can be naively associated  from a simple constant fit of the dispersion data, namely $\sigma_{ \mathcal{M_{*}}} / \mathcal{M_{*}} = 2 \, \sigma_{ \sigma_{los}}\, / \sigma_{los}$. 
Then, at $2 \, \sigma$ we find that the minimum mass estimator is $\mathcal{M_{*}} = 1.52 \, \times 10^{7}\, M_{\odot}$ and the corresponding $J$-factor in Table~\ref{tab:J_rs_min}  shits to the value of $2.49 \times 10^{18}$ GeV$^{2}$ cm$^{-5}$ for the NFW profile, $1.8 \times 10^{18}$ GeV$^{2}$ cm$^{-5}$ for the Burkert case. It follows that relaxation of the DM particle physics limits results only in a factor of $1.39$ or $1.46$, respectively for the cuspy or cored cases. Hence we conclude that, when assuming a definite functional form for the DM density profile, the Fermi-LAT bounds result to be quite robust even against the extreme case of a dramatic radial dependence in the tracer orbital anisotropy.


\section{Conclusions }

We have considered the spherical Jeans equation for a non-rotating pressure-supported tracer population, and reexamined a method to solve it in which the Dark Matter mass profile  -- accounting for the underlying potential well -- appears as an output rather than a trial parametric input as required in the standard approach. We obtained a very compact form for such explicit expression for the mass profile, showing that it depends just on the derivative of the inverse Abel transform of the radial dynamical pressure, and on the tracer orbital anisotropy profile $\beta(r)$, to be rewritten in terms of the function $a_\beta(r) \equiv -\beta(r)/(1-\beta(r))$.  We have exploited this new formulation to discuss, on general grounds, to what level dwarf spheroidal galaxies feature a reliable mass estimator -- the total mass enclosed within the radius at which the logarithmic derivative of tracer number density profile is equal to -3 (in the outskirts of the dwarf, at about 1.3 to 1.5 times the half-light radius of the projected surface brightness) -- regardless of the assumptions on the (unknown) orbital anisotropy profile. We have then turned to discuss the Dark Matter mass and density profiles in the inner region of the system, enlightening connections with assumptions on the inner scaling of the tracer number density profile and the shape of the tracer line-of-sight projected velocity dispersion. Having kept explicit the dependence on $\beta(r)$, we have illustrated the induced scalings on the density profile and hence on the angular and line-of-sight integral of the density squared, the quantity usually dubbed $J$-factor and entering the predictions of prompt Dark Matter pair annihilation signals, including the prompt gamma-ray signals the Fermi-LAT instrument has searched for and put constraints on. 

We applied the method to one sample case, the Ursa Minor Milky Way satellite. 
After discussing general trends on a few benchmarks, we have performed a Bayesian fit via an appropriate parametrization of the tracer line-of-sight projected velocity dispersion without involving a marginalization over unknown quantities. We compared $J$-factor at given assumed orbital anisotropy with values quoted in the literature; in particular, the emphasis has been to extract information about minimal $J$-factor compatible with Ursa Minor kinematical data.  In this respect we conclude: In a blind analysis, assuming no prior knowledge of the Dark Matter density profile, minimal  $J$-factor can be a factor of 2 to 4 smaller than commonly quoted estimates, relaxing by the same amount the limit on Dark Matter pair annihilation cross section deduced from gamma-ray surveys of Ursa Minor. At the same time if one goes back to a fixed trial parametric form for the density profile, such as the NFW or Burkert profiles, the shift in minimal $J$ is just a factor of about 1.4 to 1.5. 

The method illustrated here is applicable to any dwarf with adequate kinematical data, in particular to any of the so-called classical dwarfs. We expect it to be particularly valuable for those cases in which two distinct populations of dynamical tracers have been singled out, as well as plan to extend it to include higher velocity moments.       


\section*{Acknowledgements}

P.U. and M.V. acknowledge partial support from the European Union FP7 ITN INVISIBLES (Marie Curie Actions, PITN-GA-2011-289442), and partial support by the research grant ``Theoretical Astroparticle
Physics'' number 2012CPPYP7 under the program PRIN 2012 funded by the Ministero dell'Istruzione, Universit\`a e della Ricerca (MIUR).
The work of M.V. has also received funding from the European Research Council under the European UnionÕs Seventh Framework Programme (FP/2007-2013) / ERC Grant Agreements n. 279972 ÒNPFlavourÓ.

We acknowledge Eric Carlson, Daniele Gaggero, Manoj Kaplinghat, Stefano Profumo, Paolo Salucci, Louis Strigari, Tommaso Treu and Alfredo Urbano for valuable discussions.
We wish to thank the anonymous referee for the improvements in the revised version.
M.V. is also grateful to Federico Bianchini, Alejandro Castedo Echeverri and Serena Perrotta for comments and suggestions during the completion of the work.


\begin{appendix}
\section{Jeans inversion in spherical systems: the details}\label{App:JeansInversion}

In this appendix we review the ``inversion" of the Jeans equation for spherical systems.  
We start following the procedure outlined in \cite{Wolf:2009tu}, and we end with a general expression for the inverted mass profile 
$\mathcal{M}$  that holds for any generic l.o.s. velocity dispersion and surface brightness profile. Our derivation is valid for a generic
orbital velocity anisotropy, requiring only $\beta(r) \neq 1\ \forall \, r$. Inversion formulas for specific anisotropy models can be also found in \cite{Mamon:2010MNRAS}. 

A good starting point to derive the inversion formula is the definition of the projected dynamical pressure, $P(R) \equiv \sigma_{los}^2(R) \, I(R)$, see Eq.~(\ref{eq:Pproj}), split into two integrals with integrand modified by adding and subtracting the term $p(r)\, \beta(r) / \sqrt{r^{2}-R^{2}}$:
\begin{equation}
   P(R)= \int_{R^{2}}^{\infty} \frac{dr^{2}}{\sqrt{r^{2}-R^{2}}}  p(r) \left[ 1-\beta(r) \right]    
   + \int_{R^{2}}^{\infty} \frac{dr^{2}}{\sqrt{r^{2}-R^{2}}}  \left(r^{2}-R^{2}\right) \frac{p(r)  \beta(r)}{r^{2}} \ ;
\end{equation}
the second contribution on the r.h.s. can be rewritten as: 
\begin{equation}
  \int_{R^{2}}^{\infty} dr^{2} \sqrt{r^{2}-R^{2}}   
  \frac{d}{dr^{2}} \left[ - \int_{r^{2}}^{\infty} d\tilde{r}^{2}\, \frac{p(\tilde{r}) \beta(\tilde{r})}{\tilde{r}^{2}}  \right] \ ,
\end{equation}
and integrated by parts obtaining:
\begin{equation}
  \frac{1}{2} \int_{R^{2}}^{\infty} \frac{dr^{2}}{\sqrt{r^{2}-R^{2}}} \int_{r^{2}}^{\infty} d\tilde{r}^{2} \, \frac{p(\tilde{r}) \beta(\tilde{r})}{\tilde{r}^{2}} \ ,
\end{equation}
with the boundary term at $r^2 \to \infty$ in the integration by parts vanishing under the assumption that $p(r) \beta(r)$ drops to $0$ faster than $1/r$, i.e. the same assumption which had already to be valid for Eq.~(\ref{eq:Pproj}). Then, the projected dynamical pressure reads:
\begin{equation}
  \label{eq:P(R)Abeltransf}
  P(R) = \int_{R^{2}}^{\infty} \frac{dr^{2}}{\sqrt{r^{2}-R^{2}}} \left\{ p(r) \left[1-\beta(r)\right] + 
  \frac{1}{2}  \int_{r^{2}}^{\infty} d\tilde{r}^{2} \, \frac{p(\tilde{r}) \beta(\tilde{r})}{\tilde{r}^{2}}\right\} \ ;
\end{equation}
making it explicit that the quantity in the curly brackets is the inverse Abel transform of $P(R)$. Indeed, assuming that $P(R)$ vanishes at large $R$ faster than $1/R$, one is formally allowed to invert this expression to find:
\begin{equation}
  \label{eq:p(R)invAbel}
 p(r) \left[1-\beta(r)\right] + \frac{1}{2} \int_{r^{2}}^{\infty} d\tilde{r}^{2}\, \frac{p(\tilde{r}) \beta(\tilde{r})}{\tilde{r}^{2}} = 
 - \frac{1}{\pi} \int_{r^{2}}^{\infty} \frac{dR^{2}}{\sqrt{R^{2}-r^{2}}}\frac{dP}{dR^{2}} \ ,
\end{equation}
and performing another integration by parts,
\begin{equation}
 p(r) \left[1-\beta(r)\right]  + \frac{1}{2} \int_{r^{2}}^{\infty} d\tilde{r}^{2}\, \frac{p(\tilde{r}) \beta(\tilde{r})}{\tilde{r}^{2}} =
 \frac{2}{\pi} \int_{r^{2}}^{\infty} dR^{2} \sqrt{R^{2}-r^{2}} \frac{d^{2}P}{(dR^{2})^{2}} \ .
\end{equation}
We can now differentiate the equation above in the log measure $dr/r$ to get:
\begin{equation}
  \label{eq:p(r)diff}
  \left[ 1 - \beta(r) \right] \, r  \frac{d p}{dr} - \left[ \beta(r) + r \frac{d\beta}{dr} \right] p(r) =  
  - \frac{ 2 r^{2}}{\pi} \int_{r^{2}}^{\infty} \frac{dR^{2}}{\sqrt{R^{2}-r^{2}}} \frac{d^{2}P}{(dR^{2})^{2}} \ ,
\end{equation}
i.e. a first order differential equation for $p(r)$ analogous to Eq.~(\ref{eq:jeansdiff}), where the substantial difference lies
 on the presence of the second derivative of $P(R)$ in place of the first derivative of the gravitational potential. 
The formal solution for a physical radial pressure vanishing at infinity is:
\begin{equation}
p(r) = \int_r^{\infty} d\tilde{r}  \frac{ 2 \tilde{r}}{\pi \left[1-\beta(\tilde{r})\right]}  \,
 \exp \left\{-\int_r^{\tilde{r}}  dr^\prime \frac{\beta(r^\prime) + r^\prime \frac{d\beta}{dr^\prime}}{r^\prime\,\left[1-\beta(r^\prime)\right]} \right\} 
 \, \int_{\tilde{r}^{2}}^{\infty} \frac{dR^{2}}{\sqrt{R^{2}-\tilde{r}^{2}}} \frac{d^{2}P}{(dR^{2})^{2}}
\end{equation}
and, exchanging the order of integration, it can be rewritten as:
\begin{equation}
  \label{eq:pr(r)_fin}
  p(r) = \frac{1}{\pi \left[1-\beta(r)\right]} \int_{r^{2}}^{\infty} dR^{2} \frac{d^{2}P}{(dR^{2})^{2}}  \int_{r^{2}}^{R^{2}} \frac{d\tilde{r}^{2}}{{\sqrt{R^{2}-
  \tilde{r}^{2}}}} \mathcal{H}_{\beta}(r,\tilde{r}) \ ,
\end{equation}
where $\mathcal{H}_{\beta}(r,\tilde{r})$ was defined in Eq.~(\ref{eq:ourinvdef}). Finally, plugging this result in Eq.~(\ref{eq:jeansdiff}) we get:
\begin{equation}
   \label{eq:masterformula}
  \mathcal{M}(r) =  \frac{r^{2}}{G_{N} \nu(r)} \left\{ \frac{ 2 \, r}{\pi \left[ 1 -\beta(r) \right]} \int_{r^{2}}^{\infty} \frac{dR^{2}}{\sqrt{R^{2}-r^{2}}} \frac{d^{2}P}{(dR^{2})^{2}} 
  -  \left[ \frac{\frac{\beta(r)}{r} + \frac{ d\beta}{dr}}{1-\beta(r)} +2 \frac{\beta(r)}{r} \right] p(r) \right\} \  .
\end{equation}
Taking into account that: {\sl i)} $\nu(r)$ is the inverse Abel transform $\widehat{I}(r^2)$ of the surface brightness $I(R^2)$;
{\sl ii)}  the Abel integral transform satisfies the property that, if $f(x) = \mathbf{A}[\widehat{f}(y)]$, then  $df/dx = \mathbf{A}[d\widehat{f}/dy]$; {\sl iii)} it is convenient to introduce  $a_{\beta}(r)$ using the definition given in Eq.~(\ref{eq:ourinvdef}); 
it is then easy to rewrite the above inversion formula for the mass in the compact form given in Eq.~(\ref{eq:ourinv}). 
\\
In case the computation of the inverse Abel transform and its derivative becomes numerically challenging, alternatively the mass profile can be calculated as a single integral of the 
second derivative in $R^{2}$ of the projected  dynamical pressure over a kernel depending on $r$. In this form Eq.~(\ref{eq:masterformula}) just reads:
\begin{equation}
  \label{eq:mass_beta}
  \mathcal{M}(r)=   \frac{r}{ G_{N} \, \pi \, \nu(r) \left[ 1-\beta(r) \right]} \int_{r^{2}}^{\infty} dR^{2} \frac{d^{2}\,P}{(dR^{2})^{2}}\widetilde{W}_{\beta}(r,R) \,,
\end{equation}
with the kernel being:
\begin{equation}
  \label{eq:wtilde_beta}
  \widetilde{W}_{\beta}(r,R)=\frac{2  r^{2}}{\sqrt{R^{2}-r^{2}}}- \frac{\beta(r)}{1-\beta(r)} \left( 3+\frac{d \log \beta}{ d \log r} - 2 \, \beta(r) \right) \int_{r^{2}}^{R^{2}} \frac{d\tilde{r}^{2} }{\sqrt{R^{2}-\tilde{r}^{2}}}  \mathcal{H}_{\beta}(r,\tilde{r}) \,.
\end{equation}
\\
Eq.~(\ref{eq:mass_beta}) is the form which has been used to compute mass profiles corresponding to projected line-of-sight velocity dispersions derived from a trial parametric form of the DM density profiles, the ``parametric fit" cases we introduced at the beginning of Section~\ref{sec:UMi_fit}. To reach the exquisite precision level displayed in Fig.~\ref{fig:CorrectMassRec}, rather than computing $\sigma_{los}(R)$ alone,  it was actually useful to implement the analytic expression for $d^{2}P/ (d R^{2})^{2}$ one finds taking the definition  Eq.~(\ref{eq:Pproj}), supplemented by Eq.~(\ref{eq:radpress}), and performing a few manipulations:
\begin{eqnarray}
  \label{eq:ddP_R2_an}
  \frac{d^{2}\,P}{(dR^{2})^{2}} \;\; & = &\;\;  \frac{G_{N}}{2} \int_{R}^{\infty} \frac{dr}{\sqrt{r^{2}-R^{2}}}\Bigg\{\left(1-\frac{R^{2}}{r^{2}}\right)\frac{1}{R}
  \left( \frac{\partial K_{1}}{\partial R}+ \frac{\partial K_{2}}{\partial R}\right) \nonumber \\
  \;\; & \ &\;\; - \, \frac{K_{1}(r,R)+K_{2}(r,R)-2\,K_{3}(r,R)}{r^{2}} \nonumber \\
  \;\; & \  &\;\; -\,2 \sqrt{r^{2}-R^{2}}  \int_{R}^{r} \frac{d\tilde{r}}{\sqrt{\tilde{r}^{2}-R^{2}}} \left[ \left(1-\frac{R^{2}}{\tilde{r}^{2}}\right)\frac{1}{\tilde{r}\,R}
  \left( \frac{\partial K_{4}}{\partial R}- \frac{\partial K_{5}}{\partial R}\right) \right.\nonumber \\
  \;\; & \ &\;\; \left. -\, \frac{K_{4}(\tilde{r},R)-K_{5}(\tilde{r},R)}{\tilde{r}^{3}} \right]  \, \exp \left[{\,2 \int^{r}_{\tilde{r}} ds \frac{\beta_i(s)}{s}}\right] \Bigg\}\left[ \frac{\mathcal{M}_i(r) \nu(r)}{r^{2}}\right] \ ; 
\end{eqnarray}
where $\mathcal{M}_i(r)$ and $\beta_i(r)$ label, respectively, the mass and orbital anisotropy profiles taken as trial initial step, while the five integral kernels just introduced are the following dimensionless functions:
\begin{eqnarray}
  \label{eq:ddP_kernels}
  K_{1}(r,R) \; & = &\;   \left[ 1+ \beta_i(r) \left(2-3\frac{R^{2}}{r^{2}}\right)\right] \left(\frac{d\log \mathcal{M}_i}{d\log r}+\frac{d\log \nu }{d\log r }-3\right) \ ,  \nonumber\\ 
  K_{2}(r,R) \; & = &\;   \beta_i(r) \left[ 6 \frac{R^{2}}{r^{2}} + \frac{d\log \beta_i}{d\log r}\left(2-3\frac{R^{2}}{r^{2}}\right) \right] \ , \nonumber \\ 
  K_{3}(r,R) \; & = &\;   \beta_i(r) +\bar{\beta_i}(r) \left(8-9\frac{R^{2}}{r^{2}} \right) \ , \nonumber \\
  K_{4}(r,R) \; & = &\;  2 \left[ 1+\beta_i(r) \right] \left[\beta_i(r)+\bar{\beta_i}(r)\left(2-3\frac{R^{2}}{r^{2}} \right)\right] \ , \nonumber \\
  K_{5}(r,R) \; & = &\;  \beta_i(r) \frac{d \log \beta_i}{d \log r}+\bar{\beta_i}(r) \left[\frac{d \log \bar{\beta_i}}{d \log r} \left(2-3\frac{R^{2}}{r^{2}}\right)+6 \, \frac{R^{2}}{r^{2}} \right] \ , 
\end{eqnarray}
with the auxiliary function $\bar{\beta_i}(r)$ given by:
\begin{equation}
  \bar{\beta_i}(r)= \beta_i(r) \left[1+\beta_i(r)-\frac{1}{2} \frac{d\log \beta_i}{d\log r}\right] \ .
  \label{eq:betabar}
\end{equation}


\section{Computing $J$-factors: the easy-peasy pieces}\label{App:JfactorComputation}

\begin{figure}[b!] 
\centering
  \includegraphics[scale=0.3]{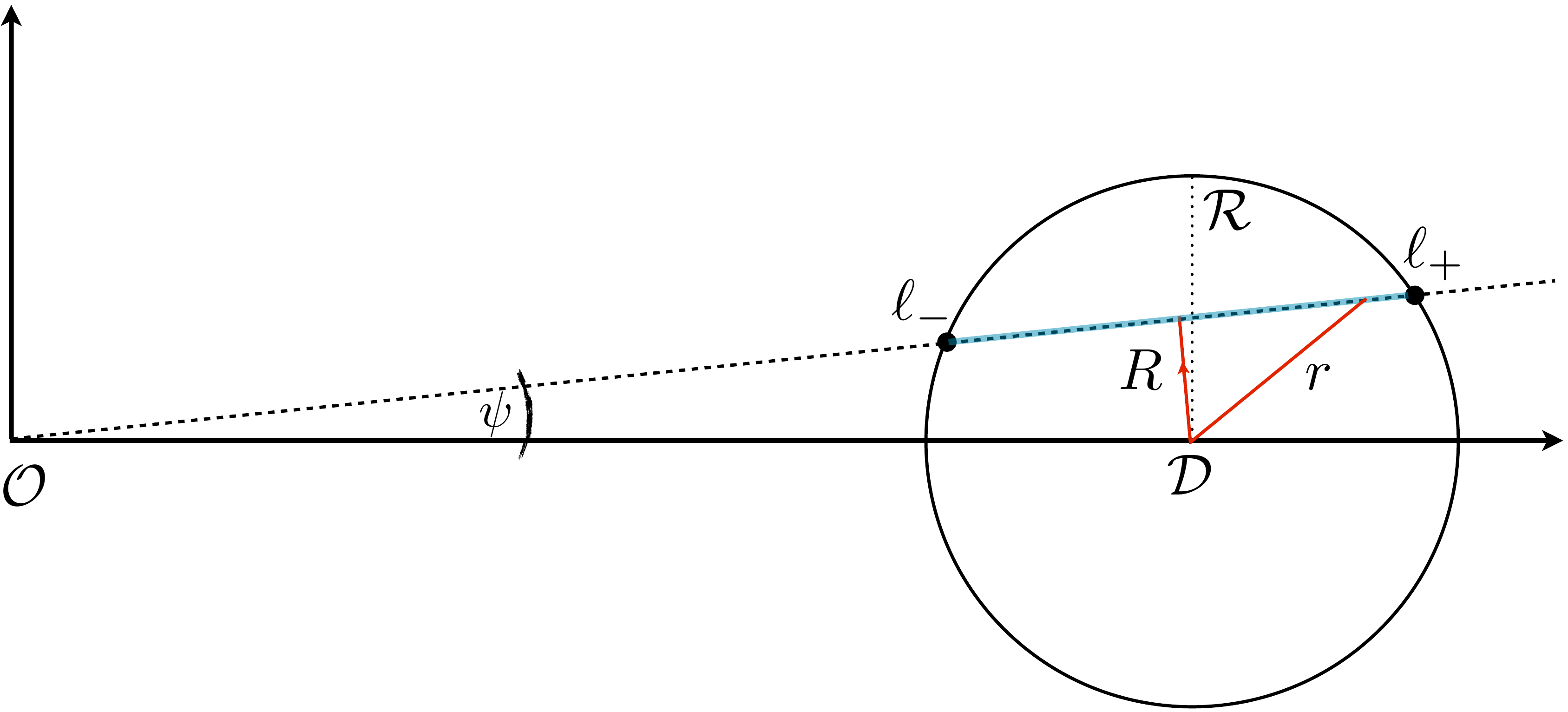}
   \caption{\textit{L.o.s. integration (light blue segment) in $(R, r)$ coordinates (in red) of an observer $\mathcal{O}$  with angular aperture $\psi$, placed at a distance $\mathcal{D}$ from a spherical system of finite size $\mathcal{R}$. } }
   \label{fig:jfactorpic}
\end{figure}

The angular + line-of-sight integral of a spherically symmetric source takes a simple form when the observer looks at the center of the source. In such a case, introducing a change of coordinates that fully exploits the symmetry of the problem, the $J$-factor computation reduces to a single radial integration over the DM density squared times an appropriate radial window function. We briefly resume here the steps to perform this convenient mapping. \\ 
Consider an observer $\mathcal{O}$ pointing towards the center of the astrophysical system under study, located at the distance $\mathcal{D}$, with an angular acceptance $\Delta\Omega$. In the coordinate system centered  on $\mathcal{O}$ -- see Fig.~\ref{fig:jfactorpic} -- Eq.~(\ref{eq:j_psi}) can be written as:
\begin{equation}
  \label{eq:jpsi}
  J= 2 \pi \int_{0}^{\cos \psi_{\textrm{max}}} d \cos \psi \int_{\ell_{-}(\psi,\mathcal{R})}^{\ell_{+}(\psi,\mathcal{R})} 
  d\ell \ \rho^{2} \left[r(\psi,\ell) \right] \ ,
\end{equation}
where $\psi_{\textrm{max}}$ is obtained from $\Delta\Omega = 2 \pi \big(1  - \cos \psi_{\textrm{max}} \big)$ and $\mathcal{R}$ is the radial boundary for the spherical system. In Eq.~(\ref{eq:jpsi}) the explicit expression for $r(\psi,\ell)$ is given by the geometrical relation:
\begin{equation}
  \label{eq:r_ell_psi}
  r^{2}=\ell^{2} + \mathcal{D}^{2} - 2 \, \ell  \, \mathcal{D}  \cos \psi  \ ,
\end{equation}
and  the extremes of integration $\ell_{+}$ and $\ell_{-}$ are the solutions of the equation above:
\begin{equation}
  \label{eq:ell_+_ell_-}
  \ell_{\pm}(\psi,r)= \mathcal{D} \cos \psi  \pm \sqrt{r^{2}-\mathcal{D}^{2}\sin^{2} \psi } \ .
\end{equation}
The values of  $\psi_{\textrm{max}}$, $\mathcal{R}$ and $\mathcal{D}$ correspond to the set of data needed to determine the $J$-factor. Note that one can also trade line-of-sight angles for line-of-sight projected radii, replacing $\psi_{\textrm{max}}$ with $R_{\textrm{max}}$ as given by $\tan \psi_{\textrm{max}}  = R_{\textrm{max}}/\sqrt{\mathcal{D}^{2}-R^{2}_{\textrm{max}}}$. 
In light of the axial symmetry along the line of sight of $\mathcal{O}$, it is sufficient to perform a two-dimensional mapping in order to move to the coordinate system centered on the
 halo density. Following Fig.~\ref{fig:jfactorpic}, the new set of coordinates $(R,r)$ we want to introduce is related to the starting pair $(\ell,\psi)$ by: 
\begin{equation}
  \label{eq:2Dmapping}
   d\big( \cos \psi \big) =  dR  \frac{\partial}{\partial R} \sqrt{1-\frac{R^{2}}{\mathcal{D}^{2}}} 
   \quad \quad {\rm and} \quad \quad
   \pm \, d \ell =  d r \frac{\partial}{\partial r} \big( \sqrt{\mathcal{D}^{2}-R^{2}} \pm \sqrt{r^{2}-R^{2}}\big) \,.
\end{equation}
Thus, we can easily rewrite Eq.~(\ref{eq:jpsi}) in the new coordinate system as:
\begin{equation}
  \label{eq:newjpsi}
  J= \frac{4\pi}{\mathcal{D}^{2}} \int_{0}^{R_{\textrm{max}}}  \frac{d R\,R}{\sqrt{1-\frac{R^{2}}{\mathcal{D}^{2}}}} \int_{R}^{\mathcal{R}} 
   \frac{d r}{\sqrt{1-\frac{R^{2}}{r^{2}}}} \rho^{2}(r) \ ,
\end{equation}
and exchanging the order of integration we get: 
\begin{equation}
  \label{eq:reordnewjpsi}
  J= \frac{4\pi}{\mathcal{D}^{2}} \left[ \int_{0}^{R_{\textrm{max}}} dr \rho^{2}(r)  \int_{0}^{r}  \frac{d R\,R}{\sqrt{\Big(1-\frac{R^{2}}{\mathcal{D}^{2}}\Big)\Big(1-\frac{R^{2}}{r^{2}}\Big)}} + 
  \int_{R_{\textrm{max}}}^{\mathcal{R}} dr \rho^{2}(r)  \int_{0}^{R_{\textrm{max}}}  \frac{d R\,R}{\sqrt{\Big(1-\frac{R^{2}}{\mathcal{D}^{2}}\Big)\Big(1-\frac{R^{2}}{r^{2}}\Big)}} \right] \,,
\end{equation}
where now the $R$ integral can be performed analytically in both terms. Introducing the dimensionless radial function:
\begin{equation}
 \mathcal{W}(r;s,t) \equiv \frac{1}{\mathcal{D}^{2}} \int_{s^{2}}^{t^{2}} \frac{dR^{2}}{2 \, \sqrt{\Big(1-\frac{R^{2}}{\mathcal{D}^{2}}\Big)\Big(1-\frac{R^{2}}{r^{2}}\Big)}} = \frac{r}{\mathcal{D}} \, 
  \log\left( \frac{\sqrt{\mathcal{D}^{2}-t^{2}}-\sqrt{r^{2}-t^{2}}}{\sqrt{\mathcal{D}^{2}-s^{2}}-\sqrt{r^{2}-s^{2}}} \right) \,,
\end{equation} 
we obtain then that the $J$-factor can be evaluated by a single integral formula:
\begin{equation}
   \label{eq:Jradial}
 J = 4 \pi \, \int_{0}^{\mathcal{R}} dr \, \widetilde{J}(r) \, ,
\end{equation}
where the integrand $\widetilde{J}$ is a function of the radial coordinate $r$ and carries the dimension of $\rho^{2}(r)$:
\begin{equation}
 \label{eq:Jtilde}
 \widetilde{J}(r) =  \rho^{2}(r)  \left[ \, \mathcal{W}(r;0,r) \, \theta_{H}(R_{\textrm{max}}-r) +  \mathcal{W}(r;0,R_{\textrm{max}}) \, \theta_{H}(r-R_{\textrm{max}})  \, \right] \, ,
\end{equation}
with $\theta_{H}(r)$ being the Heaviside step function. \\
In case of $r \le R_{\textrm{max}}$ and $R_{\textrm{max}} \ll \mathcal{D}$, the function $\mathcal{W}(r;0,r)$ is very well approximated by $r^{2}/\mathcal{D}^{2}$, while for greater radii $\mathcal{W}(r;0,R_{\textrm{max}})$ rapidly decreases towards the limit $1- \sqrt{1-R_{\textrm{max}}^{2}/\mathcal{D}^{2}} $.\\
Thus, choosing an angular aperture $\psi_{\textrm{max}}$ so that $R_{\textrm{max}} / \mathcal{D}~\ll~1$, since usually $\rho(r)$ rapidly approaches  0 with increasing radius, Eq.~(\ref{eq:Jradial}) can be numerically approximated as:
\begin{equation}
   \label{eq:Joversimplified}
 J \simeq \frac{4 \pi}{\mathcal{D}^{2}} \, \int_{0}^{\mathcal{R}} dr \, r^{2} \, \rho^{2}(r) \, .
\end{equation}
In Section \ref{sec:UMi} we presented all the results  for the computation of the J-factor making use of the full expression in Eq.~(\ref{eq:Jradial})-(\ref{eq:Jtilde}). However, the approximation above performs better than the per mille level already for $\mathcal{D} \sim \mathcal{O}(10^{2})$ kpc, $\psi_{\textrm{max}} \lesssim 1^{\circ}$.
Note that if we would have studied the case of decaying DM particles (see e.g. \cite{Baring:2015sza}) rather than DM pair annihilation, the analogous of Eq.~(\ref{eq:Joversimplified}) would have allowed us to focus directly on the mass profile, Eq.~(\ref{eq:ourinv}), without the need to compute the density profile and to argue the validity of its extrapolation to inner radii.

\end{appendix}

\bibliography{dSph_limits}

\providecommand{\href}[2]{#2}\begingroup\raggedright\begin{thebibliography}{10}

\bibitem{Strigari:2006rd}
L.~E. Strigari, S.~M. Koushiappas, J.~S. Bullock and M.~Kaplinghat,
  \emph{{Precise constraints on the dark matter content of Milky Way dwarf
  galaxies for gamma-ray experiments}},
  \href{http://dx.doi.org/10.1103/PhysRevD.75.083526}{\emph{Phys. Rev.} {\bf
  D75} (2007) 083526}, [\href{http://arxiv.org/abs/astro-ph/0611925}{{\tt
  astro-ph/0611925}}].

\bibitem{Strigari:2007at}
L.~E. Strigari, S.~M. Koushiappas, J.~S. Bullock, M.~Kaplinghat, J.~D. Simon
  et~al., \emph{{The Most Dark Matter Dominated Galaxies: Predicted Gamma-ray
  Signals from the Faintest Milky Way Dwarfs}},
  \href{http://arxiv.org/abs/0709.1510}{{\tt 0709.1510}}.

\bibitem{2002MNRAS.333..697L}
E.~L. {{\L}okas}, \emph{{Dark matter distribution in dwarf spheroidal
  galaxies}},
  \href{http://dx.doi.org/10.1046/j.1365-8711.2002.05457.x}{\emph{\mnras} {\bf
  333} (July, 2002) 697--708},
  [\href{http://arxiv.org/abs/astro-ph/0112023}{{\tt astro-ph/0112023}}].

\bibitem{Battaglia:2013wqa}
G.~Battaglia, A.~Helmi and M.~Breddels, \emph{{Internal kinematics and
  dynamical models of dwarf spheroidal galaxies around the Milky Way}},
  \href{http://dx.doi.org/10.1016/j.newar.2013.05.003}{\emph{New Astron. Rev.}
  {\bf 57} (2013) 52--79}, [\href{http://arxiv.org/abs/1305.5965}{{\tt
  1305.5965}}].

\bibitem{WalkerReviewMWdSphs}
M.~{Walker}, \emph{{Dark Matter in the Galactic Dwarf Spheroidal Satellites}},
  p.~1039.
\newblock 2013.
\newblock 10.1007/978-94-007-5612-0$\_$20.

\bibitem{2012AJ_McConnachie}
A.~W. {McConnachie}, \emph{{The Observed Properties of Dwarf Galaxies in and
  around the Local Group}},
  \href{http://dx.doi.org/10.1088/0004-6256/144/1/4}{\emph{\aj} {\bf 144}
  (July, 2012) 4}, [\href{http://arxiv.org/abs/1204.1562}{{\tt 1204.1562}}].

\bibitem{Ackermann:2015zua}
{\scshape Fermi-LAT} collaboration, M.~Ackermann et~al., \emph{{Searching for
  Dark Matter Annihilation from Milky Way Dwarf Spheroidal Galaxies with Six
  Years of Fermi-LAT Data}},  \href{http://arxiv.org/abs/1503.02641}{{\tt
  1503.02641}}.

\bibitem{Ahnen:2016qkx}
{\scshape Fermi-LAT, MAGIC} collaboration, M.~L. Ahnen et~al., \emph{{Limits to
  dark matter annihilation cross-section from a combined analysis of MAGIC and
  Fermi-LAT observations of dwarf satellite galaxies}},
  \href{http://dx.doi.org/10.1088/1475-7516/2016/02/039}{\emph{JCAP} {\bf 1602}
  (2016) 039}, [\href{http://arxiv.org/abs/1601.06590}{{\tt 1601.06590}}].

\bibitem{Martinez:2009jh}
G.~D. Martinez, J.~S. Bullock, M.~Kaplinghat, L.~E. Strigari and R.~Trotta,
  \emph{{Indirect Dark Matter Detection from Dwarf Satellites: Joint
  Expectations from Astrophysics and Supersymmetry}},
  \href{http://dx.doi.org/10.1088/1475-7516/2009/06/014}{\emph{JCAP} {\bf 0906}
  (2009) 014}, [\href{http://arxiv.org/abs/0902.4715}{{\tt 0902.4715}}].

\bibitem{Cirelli:2015bda}
M.~Cirelli, T.~Hambye, P.~Panci, F.~Sala and M.~Taoso, \emph{{Gamma ray tests
  of Minimal Dark Matter}},
  \href{http://dx.doi.org/10.1088/1475-7516/2015/10/026}{\emph{JCAP} {\bf 1510}
  (2015) 026}, [\href{http://arxiv.org/abs/1507.05519}{{\tt 1507.05519}}].

\bibitem{Liem:2016xpm}
S.~Liem, G.~Bertone, F.~Calore, R.~R. de~Austri, T.~M.~P. Tait, R.~Trotta
  et~al., \emph{{Effective Field Theory of Dark Matter: a Global Analysis}},
  \href{http://arxiv.org/abs/1603.05994}{{\tt 1603.05994}}.

\bibitem{1983ApJ...266L..21L}
D.~N.~C. {Lin} and S.~M. {Faber}, \emph{{Some implications of nonluminous
  matter in dwarf spheroidal galaxies}},
  \href{http://dx.doi.org/10.1086/183971}{\emph{\apjl} {\bf 266} (Mar., 1983)
  L21--L25}.

\bibitem{Diez-Tejedor:2014naa}
A.~Diez-Tejedor, A.~X. Gonzalez-Morales and S.~Profumo, \emph{{Dwarf spheroidal
  galaxies and Bose-Einstein condensate dark matter}},
  \href{http://dx.doi.org/10.1103/PhysRevD.90.043517}{\emph{Phys. Rev.} {\bf
  D90} (2014) 043517}, [\href{http://arxiv.org/abs/1404.1054}{{\tt
  1404.1054}}].

\bibitem{Domcke:2014kla}
V.~Domcke and A.~Urbano, \emph{{Dwarf spheroidal galaxies as degenerate gas of
  free fermions}},
  \href{http://dx.doi.org/10.1088/1475-7516/2015/01/002}{\emph{JCAP} {\bf 1501}
  (2015) 002}, [\href{http://arxiv.org/abs/1409.3167}{{\tt 1409.3167}}].

\bibitem{Zavala:2012us}
J.~Zavala, M.~Vogelsberger and M.~G. Walker, \emph{{Constraining
  Self-Interacting Dark Matter with the Milky Way's dwarf spheroidals}},
  \href{http://dx.doi.org/10.1093/mnrasl/sls053}{\emph{Monthly Notices of the
  Royal Astronomical Society: Letters} {\bf 431} (2013) L20--L24},
  [\href{http://arxiv.org/abs/1211.6426}{{\tt 1211.6426}}].

\bibitem{Kaplinghat:2015aga}
M.~Kaplinghat, S.~Tulin and H.-B. Yu, \emph{{Dark Matter Halos as Particle
  Colliders: Unified Solution to Small-Scale Structure Puzzles from Dwarfs to
  Clusters}},
  \href{http://dx.doi.org/10.1103/PhysRevLett.116.041302}{\emph{Phys. Rev.
  Lett.} {\bf 116} (2016) 041302}, [\href{http://arxiv.org/abs/1508.03339}{{\tt
  1508.03339}}].

\bibitem{Adhikari:2016bei}
R.~Adhikari et~al., \emph{{A White Paper on keV Sterile Neutrino Dark Matter}},
   \href{http://arxiv.org/abs/1602.04816}{{\tt 1602.04816}}.

\bibitem{Cirelli:2010xx}
M.~Cirelli, G.~Corcella, A.~Hektor, G.~Hutsi, M.~Kadastik, P.~Panci et~al.,
  \emph{{PPPC 4 DM ID: A Poor Particle Physicist Cookbook for Dark Matter
  Indirect Detection}}, \href{http://dx.doi.org/10.1088/1475-7516/2012/10/E01,
  10.1088/1475-7516/2011/03/051}{\emph{JCAP} {\bf 1103} (2011) 051},
  [\href{http://arxiv.org/abs/1012.4515}{{\tt 1012.4515}}].

\bibitem{2015ApJ...807...50B}
K.~{Bechtol}, A.~{Drlica-Wagner}, E.~{Balbinot}, A.~{Pieres}, J.~D. {Simon},
  B.~{Yanny} et~al., \emph{{Eight New Milky Way Companions Discovered in
  First-year Dark Energy Survey Data}},
  \href{http://dx.doi.org/10.1088/0004-637X/807/1/50}{\emph{\apj} {\bf 807}
  (July, 2015) 50}, [\href{http://arxiv.org/abs/1503.02584}{{\tt 1503.02584}}].

\bibitem{Ackermann2011}
{\scshape Fermi-LAT collaboration} collaboration, M.~Ackermann et~al.,
  \emph{{Constraining Dark Matter Models from a Combined Analysis of Milky Way
  Satellites with the Fermi Large Area Telescope}},
  \href{http://dx.doi.org/10.1103/PhysRevLett.107.241302}{\emph{Phys.Rev.Lett.}
  {\bf 107} (2011) 241302}, [\href{http://arxiv.org/abs/1108.3546}{{\tt
  1108.3546}}].

\bibitem{Ackermann:2013yva}
{\scshape Fermi-LAT} collaboration, M.~Ackermann et~al., \emph{{Dark matter
  constraints from observations of 25 Milky Way satellite galaxies with the
  Fermi Large Area Telescope}},
  \href{http://dx.doi.org/10.1103/PhysRevD.89.042001}{\emph{Phys. Rev.} {\bf
  D89} (2014) 042001}, [\href{http://arxiv.org/abs/1310.0828}{{\tt
  1310.0828}}].

\bibitem{Martinez:2013els}
G.~D. Martinez, \emph{{A Robust Determination of Milky Way Satellite Properties
  using Hierarchical Mass Modeling}},
  \href{http://arxiv.org/abs/1309.2641}{{\tt 1309.2641}}.

\bibitem{Bonnivard:2014kza}
V.~Bonnivard, C.~Combet, D.~Maurin and M.~G. Walker, \emph{{Spherical Jeans
  analysis for dark matter indirect detection in dwarf spheroidal galaxies -
  Impact of physical parameters and triaxiality}},
  \href{http://dx.doi.org/10.1093/mnras/stu2296}{\emph{Mon. Not. Roy. Astron.
  Soc.} {\bf 446} (2015) 3002--3021},
  [\href{http://arxiv.org/abs/1407.7822}{{\tt 1407.7822}}].

\bibitem{Hayashi:2012si}
K.~Hayashi and M.~Chiba, \emph{{Probing non-spherical dark halos in the
  Galactic dwarf galaxies}},
  \href{http://dx.doi.org/10.1088/0004-637X/755/2/145}{\emph{Astrophys.J.} {\bf
  755} (2012) 145}, [\href{http://arxiv.org/abs/1206.3888}{{\tt 1206.3888}}].

\bibitem{Hayashi:2015yfa}
K.~Hayashi and M.~Chiba, \emph{{Structural properties of non-spherical dark
  halos in Milky Way and Andromeda dwarf spheroidal galaxies}},
  \href{http://dx.doi.org/10.1088/0004-637X/810/1/22}{\emph{Astrophys. J.} {\bf
  810} (2015) 22}, [\href{http://arxiv.org/abs/1507.07620}{{\tt 1507.07620}}].

\bibitem{Walker:2011fs}
M.~G. Walker, C.~Combet, J.~A. Hinton, D.~Maurin and M.~I. Wilkinson,
  \emph{{Dark matter in the classical dwarf spheroidal galaxies: a robust
  constraint on the astrophysical factor for gamma-ray flux calculations}},
  \href{http://dx.doi.org/10.1088/2041-8205/733/2/L46}{\emph{Astrophys. J.}
  {\bf 733} (2011) L46}, [\href{http://arxiv.org/abs/1104.0411}{{\tt
  1104.0411}}].

\bibitem{Charbonnier:2011ft}
A.~Charbonnier et~al., \emph{{Dark matter profiles and annihilation in dwarf
  spheroidal galaxies: prospectives for present and future gamma-ray
  observatories - I. The classical dSphs}},
  \href{http://dx.doi.org/10.1111/j.1365-2966.2011.19387.x}{\emph{Mon. Not.
  Roy. Astron. Soc.} {\bf 418} (2011) 1526--1556},
  [\href{http://arxiv.org/abs/1104.0412}{{\tt 1104.0412}}].

\bibitem{GeringerSameth:2011iw}
A.~Geringer-Sameth and S.~M. Koushiappas, \emph{{Exclusion of canonical WIMPs
  by the joint analysis of Milky Way dwarfs with Fermi}},
  \href{http://dx.doi.org/10.1103/PhysRevLett.107.241303}{\emph{Phys. Rev.
  Lett.} {\bf 107} (2011) 241303}, [\href{http://arxiv.org/abs/1108.2914}{{\tt
  1108.2914}}].

\bibitem{Mazziotta:2012ux}
M.~N. Mazziotta, F.~Loparco, F.~de~Palma and N.~Giglietto, \emph{{A
  model-independent analysis of the Fermi Large Area Telescope gamma-ray data
  from the Milky Way dwarf galaxies and halo to constrain dark matter
  scenarios}},
  \href{http://dx.doi.org/10.1016/j.astropartphys.2012.07.005}{\emph{Astropart.
  Phys.} {\bf 37} (2012) 26--39}, [\href{http://arxiv.org/abs/1203.6731}{{\tt
  1203.6731}}].

\bibitem{Geringer-Sameth:2014yza}
A.~Geringer-Sameth, S.~M. Koushiappas and M.~Walker, \emph{{Dwarf galaxy
  annihilation and decay emission profiles for dark matter experiments}},
  \href{http://dx.doi.org/10.1088/0004-637X/801/2/74}{\emph{Astrophys.J.} {\bf
  801} (2015) 74}, [\href{http://arxiv.org/abs/1408.0002}{{\tt 1408.0002}}].

\bibitem{Geringer-Sameth:2014qqa}
A.~Geringer-Sameth, S.~M. Koushiappas and M.~G. Walker, \emph{{Comprehensive
  search for dark matter annihilation in dwarf galaxies}},
  \href{http://dx.doi.org/10.1103/PhysRevD.91.083535}{\emph{Phys. Rev.} {\bf
  D91} (2015) 083535}, [\href{http://arxiv.org/abs/1410.2242}{{\tt
  1410.2242}}].

\bibitem{Bonnivard:2015xpq}
V.~Bonnivard et~al., \emph{{Dark matter annihilation and decay in dwarf
  spheroidal galaxies: The classical and ultrafaint dSphs}},
  \href{http://dx.doi.org/10.1093/mnras/stv1601}{\emph{Mon. Not. Roy. Astron.
  Soc.} {\bf 453} (2015) 849}, [\href{http://arxiv.org/abs/1504.02048}{{\tt
  1504.02048}}].

\bibitem{Mamon:2010MNRAS}
G.~A. {Mamon} and G.~{Bou{\'e}}, \emph{{Kinematic deprojection and mass
  inversion of spherical systems of known velocity anisotropy}},
  \href{http://dx.doi.org/10.1111/j.1365-2966.2009.15817.x}{\emph{\mnras} {\bf
  401} (Feb., 2010) 2433--2450}, [\href{http://arxiv.org/abs/0906.4971}{{\tt
  0906.4971}}].

\bibitem{Wolf:2009tu}
J.~Wolf, G.~D. Martinez, J.~S. Bullock, M.~Kaplinghat, M.~Geha et~al.,
  \emph{{Accurate Masses for Dispersion-supported Galaxies}},
  {\emph{Mon.Not.Roy.Astron.Soc.} {\bf 406} (2010) 1220},
  [\href{http://arxiv.org/abs/0908.2995}{{\tt 0908.2995}}].

\bibitem{1992ApJDejongheMerritt}
H.~{Dejonghe} and D.~{Merritt}, \emph{{Inferring the mass of spherical stellar
  systems from velocity moments}},
  \href{http://dx.doi.org/10.1086/171368}{\emph{\apj} {\bf 391} (June, 1992)
  531--549}.

\bibitem{bt08}
J.~{Binney} and S.~{Tremaine}, \emph{{Galactic Dynamics: Second Edition}}.
\newblock Princeton University Press, 2008.

\bibitem{Strigari:2013iaa}
L.~E. Strigari, \emph{{Galactic Searches for Dark Matter}},
  \href{http://dx.doi.org/10.1016/j.physrep.2013.05.004}{\emph{Phys. Rept.}
  {\bf 531} (2013) 1--88}, [\href{http://arxiv.org/abs/1211.7090}{{\tt
  1211.7090}}].

\bibitem{Irwin1995}
M.~{Irwin} and D.~{Hatzidimitriou}, \emph{{Structural parameters for the
  Galactic dwarf spheroidals}}, {\emph{\mnras} {\bf 277} (Dec., 1995)
  1354--1378}.

\bibitem{Binney&Mamon82}
J.~{Binney} and G.~A. {Mamon}, \emph{{M/L and velocity anisotropy from
  observations of spherical galaxies, or must M87 have a massive black hole}},
  {\emph{\mnras} {\bf 200} (July, 1982) 361--375}.

\bibitem{plummer}
H.~C. {Plummer}, \emph{{On the problem of distribution in globular star
  clusters}}, {\emph{\mnras} {\bf 71} (Mar., 1911) 460--470}.

\bibitem{king}
I.~{King}, \emph{{The structure of star clusters. I. an empirical density
  law}}, \href{http://dx.doi.org/10.1086/108756}{\emph{\aj} {\bf 67} (Oct.,
  1962) 471}.

\bibitem{sersic}
J.~L. {Sersic}, \emph{{Atlas de galaxias australes}}.
\newblock 1968.

\bibitem{Navarro:1995iw}
J.~F. Navarro, C.~S. Frenk and S.~D.~M. White, \emph{{The Structure of cold
  dark matter halos}}, \href{http://dx.doi.org/10.1086/177173}{\emph{Astrophys.
  J.} {\bf 462} (1996) 563--575},
  [\href{http://arxiv.org/abs/astro-ph/9508025}{{\tt astro-ph/9508025}}].

\bibitem{Navarro:1996gj}
J.~F. Navarro, C.~S. Frenk and S.~D.~M. White, \emph{{A Universal density
  profile from hierarchical clustering}},
  \href{http://dx.doi.org/10.1086/304888}{\emph{Astrophys. J.} {\bf 490} (1997)
  493--508}, [\href{http://arxiv.org/abs/astro-ph/9611107}{{\tt
  astro-ph/9611107}}].

\bibitem{Salucci:2000ps}
P.~Salucci and A.~Burkert, \emph{{Dark matter scaling relations}},
  \href{http://dx.doi.org/10.1086/312747}{\emph{Astrophys. J.} {\bf 537} (2000)
  L9--L12}, [\href{http://arxiv.org/abs/astro-ph/0004397}{{\tt
  astro-ph/0004397}}].

\bibitem{Salucci:2011ee}
P.~Salucci, M.~I. Wilkinson, M.~G. Walker, G.~F. Gilmore, E.~K. Grebel, A.~Koch
  et~al., \emph{{Dwarf spheroidal galaxy kinematics and spiral galaxy scaling
  laws}}, \href{http://dx.doi.org/10.1111/j.1365-2966.2011.20144.x}{\emph{Mon.
  Not. Roy. Astron. Soc.} {\bf 420} (2012) 2034},
  [\href{http://arxiv.org/abs/1111.1165}{{\tt 1111.1165}}].

\bibitem{burkert}
A.~{Burkert}, \emph{{The Structure of Dark Matter Halos in Dwarf Galaxies}},
  \href{http://dx.doi.org/10.1086/309560}{\emph{\apjl} {\bf 447} (July, 1995)
  L25}, [\href{http://arxiv.org/abs/astro-ph/9504041}{{\tt astro-ph/9504041}}].

\bibitem{1979PAZh577O}
L.~P. {Osipkov}, \emph{{Spherical systems of gravitating bodies with an
  ellipsoidal velocity distribution}}, {\emph{Pisma v Astronomicheskii Zhurnal}
  {\bf 5} (Feb., 1979) 77--80}.

\bibitem{1985AJ90.1027M}
D.~{Merritt}, \emph{{Spherical stellar systems with spheroidal velocity
  distributions}}, \href{http://dx.doi.org/10.1086/113810}{\emph{\aj} {\bf 90}
  (June, 1985) 1027--1037}.

\bibitem{Baes:2007tx}
M.~Baes and E.~Van~Hese, \emph{{Dynamical models with a general anisotropy
  profile}}, \href{http://dx.doi.org/10.1051/0004-6361:20077672}{\emph{Astron.
  Astrophys.} {\bf 471} (2007) 419},
  [\href{http://arxiv.org/abs/0705.4109}{{\tt 0705.4109}}].

\bibitem{1990AJ.....99.1548M}
M.~R. {Merrifield} and S.~M. {Kent}, \emph{{Fourth moments and the dynamics of
  spherical systems}}, \href{http://dx.doi.org/10.1086/115438}{\emph{\aj} {\bf
  99} (May, 1990) 1548--1557}.

\bibitem{Lokas:2004sw}
E.~L. Lokas, G.~A. Mamon and F.~Prada, \emph{{Dark matter distribution in the
  Draco dwarf from velocity moments}},
  \href{http://dx.doi.org/10.1111/j.1365-2966.2005.09497.x}{\emph{Mon. Not.
  Roy. Astron. Soc.} {\bf 363} (2005) 918},
  [\href{http://arxiv.org/abs/astro-ph/0411694}{{\tt astro-ph/0411694}}].

\bibitem{2009MNRAS.394L.102L}
E.~L. {{\L}okas}, \emph{{The mass and velocity anisotropy of the Carina,
  Fornax, Sculptor and Sextans dwarf spheroidal galaxies}},
  \href{http://dx.doi.org/10.1111/j.1745-3933.2009.00620.x}{\emph{\mnras} {\bf
  394} (Mar., 2009) L102--L106}, [\href{http://arxiv.org/abs/0901.0715}{{\tt
  0901.0715}}].

\bibitem{Richardson:2012ig}
T.~Richardson and M.~Fairbairn, \emph{{Analytical Solutions to the
  Mass-Anisotropy Degeneracy with Higher Order Jeans Analysis: A General
  Method}}, \href{http://dx.doi.org/10.1093/mnras/stt686}{\emph{Mon. Not. Roy.
  Astron. Soc.} {\bf 432} (2013) 3361--3380},
  [\href{http://arxiv.org/abs/1207.1709}{{\tt 1207.1709}}].

\bibitem{Richardson:2013lja}
T.~Richardson and M.~Fairbairn, \emph{{Cores in Classical Dwarf Spheroidal
  Galaxies? A Dispersion-Kurtosis Jeans Analysis Without Restricted
  Anisotropy}},  \href{http://arxiv.org/abs/1305.0670}{{\tt 1305.0670}}.

\bibitem{Walker:2011zu}
M.~G. Walker and J.~Penarrubia, \emph{{A Method for Measuring (Slopes of) the
  Mass Profiles of Dwarf Spheroidal Galaxies}},
  \href{http://dx.doi.org/10.1088/0004-637X/742/1/20}{\emph{Astrophys. J.} {\bf
  742} (2011) 20}, [\href{http://arxiv.org/abs/1108.2404}{{\tt 1108.2404}}].

\bibitem{Amorisco:2011hb}
N.~C. Amorisco and N.~W. Evans, \emph{{Dark Matter Cores and Cusps: The Case of
  Multiple Stellar Populations in Dwarf Spheroidals}},
  \href{http://dx.doi.org/10.1111/j.1365-2966.2011.19684.x}{\emph{Mon. Not.
  Roy. Astron. Soc.} {\bf 419} (2012) 184--196},
  [\href{http://arxiv.org/abs/1106.1062}{{\tt 1106.1062}}].

\bibitem{Strigari:2014yea}
L.~E. Strigari, C.~S. Frenk and S.~D.~M. White, \emph{{Dynamical models for the
  Sculptor dwarf spheroidal in a Lambda CDM universe}},
  \href{http://arxiv.org/abs/1406.6079}{{\tt 1406.6079}}.

\bibitem{2016MNRAS.457.844F}
A.~{Fattahi}, J.~F. {Navarro}, T.~{Sawala}, C.~S. {Frenk}, K.~A. {Oman}, R.~A.
  {Crain} et~al., \emph{{The APOSTLE project: Local Group kinematic mass
  constraints and simulation candidate selection}},
  \href{http://dx.doi.org/10.1093/mnras/stv2970}{\emph{\mnras} {\bf 457} (Mar.,
  2016) 844--856}, [\href{http://arxiv.org/abs/1507.03643}{{\tt 1507.03643}}].

\bibitem{2016MNRAS.457.1931S}
T.~{Sawala}, C.~S. {Frenk}, A.~{Fattahi}, J.~F. {Navarro}, R.~G. {Bower}, R.~A.
  {Crain} et~al., \emph{{The APOSTLE simulations: solutions to the Local
  Group's cosmic puzzles}},
  \href{http://dx.doi.org/10.1093/mnras/stw145}{\emph{\mnras} {\bf 457} (Apr.,
  2016) 1931--1943}, [\href{http://arxiv.org/abs/1511.01098}{{\tt
  1511.01098}}].

\bibitem{2016arXiv160205957W}
A.~R. {Wetzel}, P.~F. {Hopkins}, J.-h. {Kim}, C.-A. {Faucher-Giguere},
  D.~{Keres} and E.~{Quataert}, \emph{{Reconciling dwarf galaxies with LCDM
  cosmology: Simulating a realistic population of satellites around a Milky
  Way-mass galaxy}}, {\emph{ArXiv e-prints} (Feb., 2016) },
  [\href{http://arxiv.org/abs/1602.05957}{{\tt 1602.05957}}].

\bibitem{Richardson:2013hga}
T.~Richardson, D.~Spolyar and M.~Lehnert, \emph{{Plan $\beta$: Core or Cusp?}},
  \href{http://dx.doi.org/10.1093/mnras/stu383}{\emph{Mon.Not.Roy.Astron.Soc.}
  {\bf 440} (2014) 1680--1689}, [\href{http://arxiv.org/abs/1311.1522}{{\tt
  1311.1522}}].

\bibitem{strigari2007m}
L.~E. {Strigari}, J.~S. {Bullock}, M.~{Kaplinghat}, J.~{Diemand}, M.~{Kuhlen}
  and P.~{Madau}, \emph{{Redefining the Missing Satellites Problem}},
  \href{http://dx.doi.org/10.1086/521914}{\emph{\apj} {\bf 669} (Nov., 2007)
  676--683}, [\href{http://arxiv.org/abs/0704.1817}{{\tt 0704.1817}}].

\bibitem{Strigari:2008ib}
L.~E. Strigari, J.~S. Bullock, M.~Kaplinghat, J.~D. Simon, M.~Geha, B.~Willman
  et~al., \emph{{A common mass scale for satellite galaxies of the Milky Way}},
  \href{http://dx.doi.org/10.1038/nature07222}{\emph{Nature} {\bf 454} (2008)
  1096--1097}, [\href{http://arxiv.org/abs/0808.3772}{{\tt 0808.3772}}].

\bibitem{2008ApJ672904P}
J.~{Pe{\~n}arrubia}, A.~W. {McConnachie} and J.~F. {Navarro}, \emph{{The Cold
  Dark Matter Halos of Local Group Dwarf Spheroidals}},
  \href{http://dx.doi.org/10.1086/521543}{\emph{\apj} {\bf 672} (Jan., 2008)
  904--913}, [\href{http://arxiv.org/abs/astro-ph/0701780}{{\tt
  astro-ph/0701780}}].

\bibitem{2008ApJ6871460P}
J.~{Pe{\~n}arrubia}, A.~W. {McConnachie} and J.~F. {Navarro}, \emph{{Erratum:
  ''The Cold Dark Matter Halos of Local Group Dwarf Spheroidals'' (ApJ, 672,
  904 [2008])}}, \href{http://dx.doi.org/10.1086/591845}{\emph{\apj} {\bf 687}
  (Nov., 2008) 1460}.

\bibitem{Walker2009}
M.~G. {Walker}, M.~{Mateo}, E.~W. {Olszewski}, J.~{Pe{\~n}arrubia}, N.~{Wyn
  Evans} and G.~{Gilmore}, \emph{{A Universal Mass Profile for Dwarf Spheroidal
  Galaxies?}},
  \href{http://dx.doi.org/10.1088/0004-637X/704/2/1274}{\emph{\apj} {\bf 704}
  (Oct., 2009) 1274--1287}, [\href{http://arxiv.org/abs/0906.0341}{{\tt
  0906.0341}}].

\bibitem{Walker2009err}
M.~G. {Walker}, M.~{Mateo}, E.~W. {Olszewski}, J.~{Pe{\~n}arrubia}, N.~{Wyn
  Evans} and G.~{Gilmore}, \emph{{Erratum:``A Universal Mass Profile For Dwarf
  Spheroidal Galaxies?''}},
  \href{http://dx.doi.org/10.1088/0004-637X/710/1/886}{\emph{\apj} {\bf 710}
  (Feb., 2010) 886--890}.

\bibitem{Amorisco:2010ns}
N.~C. Amorisco and N.~W. Evans, \emph{{Phase-space models of the dwarf
  spheroidals}},
  \href{http://dx.doi.org/10.1111/j.1365-2966.2010.17715.x}{\emph{Mon. Not.
  Roy. Astron. Soc.} {\bf 411} (2011) 2118--2136},
  [\href{http://arxiv.org/abs/1009.1813}{{\tt 1009.1813}}].

\bibitem{Campbell:2016vkb}
D.~J.~R. Campbell, C.~S. Frenk, A.~Jenkins, V.~R. Eke, J.~F. Navarro, T.~Sawala
  et~al., \emph{{Knowing the unknowns: uncertainties in simple estimators of
  dynamical masses}},  \href{http://arxiv.org/abs/1603.04443}{{\tt
  1603.04443}}.

\bibitem{2010MNRAS.408.1070C}
L.~{Ciotti} and L.~{Morganti}, \emph{{How general is the global density
  slope-anisotropy inequality?}},
  \href{http://dx.doi.org/10.1111/j.1365-2966.2010.17184.x}{\emph{\mnras} {\bf
  408} (Oct., 2010) 1070--1074}, [\href{http://arxiv.org/abs/1006.2344}{{\tt
  1006.2344}}].

\bibitem{2011ApJ...726...80V}
E.~{Van Hese}, M.~{Baes} and H.~{Dejonghe}, \emph{{On the Universality of the
  Global Density Slope-Anisotropy Inequality}},
  \href{http://dx.doi.org/10.1088/0004-637X/726/2/80}{\emph{\apj} {\bf 726}
  (Jan., 2011) 80}, [\href{http://arxiv.org/abs/1010.4301}{{\tt 1010.4301}}].

\bibitem{An:2005tm}
J.~H. An and N.~W. Evans, \emph{{A cusp slope-central anisotropy theorem}},
  \href{http://dx.doi.org/10.1086/501040}{\emph{Astrophys. J.} {\bf 642} (2006)
  752--758}, [\href{http://arxiv.org/abs/astro-ph/0511686}{{\tt
  astro-ph/0511686}}].

\bibitem{Zhao:1995cp}
H.~Zhao, \emph{{Analytical models for galactic nuclei}},
  \href{http://dx.doi.org/10.1093/mnras/278.2.488}{\emph{Mon. Not. Roy. Astron.
  Soc.} {\bf 278} (1996) 488--496},
  [\href{http://arxiv.org/abs/astro-ph/9509122}{{\tt astro-ph/9509122}}].

\bibitem{Evans:2008ik}
N.~W. Evans, J.~An and M.~G. Walker, \emph{{Cores and Cusps in the Dwarf
  Spheroidals}},
  \href{http://dx.doi.org/10.1111/j.1745-3933.2008.00596.x}{\emph{Mon. Not.
  Roy. Astron. Soc.} {\bf 393} (2009) 50},
  [\href{http://arxiv.org/abs/0811.1488}{{\tt 0811.1488}}].

\bibitem{2014AJ_dwarfs_BH_nearby}
E.~C. {Moran}, K.~{Shahinyan}, H.~R. {Sugarman}, D.~O. {V{\'e}lez} and
  M.~{Eracleous}, \emph{{Black Holes At the Centers of Nearby Dwarf Galaxies}},
  \href{http://dx.doi.org/10.1088/0004-6256/148/6/136}{\emph{\aj} {\bf 148}
  (Dec., 2014) 136}, [\href{http://arxiv.org/abs/1408.4451}{{\tt 1408.4451}}].

\bibitem{2013MmSAI_dwarfs_BH_nearby}
A.~A. {Nucita}, L.~{Manni}, F.~{De Paolis}, G.~{Ingrosso} and D.~{Vetrugno},
  \emph{{The high energy search for IMBHs in close dSph Milky Way
  satellites.}}, {\emph{Memorie della Societa Astronomica Italiana} {\bf 84}
  (2013) 645}.

\bibitem{Gonzalez-Morales:2014eaa}
A.~X. Gonzalez-Morales, S.~Profumo and F.~S. Queiroz, \emph{{Effect of Black
  Holes in Local Dwarf Spheroidal Galaxies on Gamma-Ray Constraints on Dark
  Matter Annihilation}},
  \href{http://dx.doi.org/10.1103/PhysRevD.90.103508}{\emph{Phys. Rev.} {\bf
  D90} (2014) 103508}, [\href{http://arxiv.org/abs/1406.2424}{{\tt
  1406.2424}}].

\bibitem{Wanders:2014xia}
M.~Wanders, G.~Bertone, M.~Volonteri and C.~Weniger, \emph{{No WIMP Mini-Spikes
  in Dwarf Spheroidal Galaxies}},
  \href{http://dx.doi.org/10.1088/1475-7516/2015/04/004}{\emph{JCAP} {\bf 1504}
  (2015) 004}, [\href{http://arxiv.org/abs/1409.5797}{{\tt 1409.5797}}].

\bibitem{Cholis:2012am}
I.~Cholis and P.~Salucci, \emph{{Extracting limits on Dark Matter annihilation
  from gamma-ray observations towards dwarf spheroidal galaxies}},
  \href{http://dx.doi.org/10.1103/PhysRevD.86.023528}{\emph{Phys.Rev.} {\bf
  D86} (2012) 023528}, [\href{http://arxiv.org/abs/1203.2954}{{\tt
  1203.2954}}].

\bibitem{carrera2002}
R.~{Carrera}, A.~{Aparicio}, D.~{Mart{\'{\i}}nez-Delgado} and
  J.~{Alonso-Garc{\'{\i}}a}, \emph{{The Star Formation History and Spatial
  Distribution of Stellar Populations in the Ursa Minor Dwarf Spheroidal
  Galaxy}}, \href{http://dx.doi.org/10.1086/340702}{\emph{\aj} {\bf 123} (June,
  2002) 3199--3209}, [\href{http://arxiv.org/abs/astro-ph/0203300}{{\tt
  astro-ph/0203300}}].

\bibitem{Bellazzini2002}
M.~{Bellazzini}, F.~R. {Ferraro}, L.~{Origlia}, E.~{Pancino}, L.~{Monaco} and
  E.~{Oliva}, \emph{{The Draco and Ursa Minor Dwarf Spheroidal Galaxies: A
  Comparative Study}}, \href{http://dx.doi.org/10.1086/344794}{\emph{\aj} {\bf
  124} (Dec., 2002) 3222--3240},
  [\href{http://arxiv.org/abs/astro-ph/0209391}{{\tt astro-ph/0209391}}].

\bibitem{Olszewski1985}
E.~W. {Olszewski} and M.~{Aaronson}, \emph{{The URSA Minor dwarf galaxy - Still
  an old stellar system}}, \href{http://dx.doi.org/10.1086/113925}{\emph{\aj}
  {\bf 90} (Nov., 1985) 2221--2238}.

\bibitem{Cudworth1986}
K.~M. {Cudworth}, E.~W. {Olszewski} and R.~A. {Schommer}, \emph{{Proper motions
  and bright-star photometry in the Ursa Minor dwarf galaxy}},
  \href{http://dx.doi.org/10.1086/114210}{\emph{\aj} {\bf 92} (Oct., 1986)
  766--776}.

\bibitem{Piatek2005}
S.~{Piatek}, C.~{Pryor}, P.~{Bristow}, E.~W. {Olszewski}, H.~C. {Harris},
  M.~{Mateo} et~al., \emph{{Proper Motions of Dwarf Spheroidal Galaxies from
  Hubble Space Telescope Imaging. III. Measurement for Ursa Minor}},
  \href{http://dx.doi.org/10.1086/430532}{\emph{\aj} {\bf 130} (July, 2005)
  95--115}, [\href{http://arxiv.org/abs/astro-ph/0503620}{{\tt
  astro-ph/0503620}}].

\bibitem{Nemec1988}
J.~M. {Nemec}, A.~{Wehlau} and C.~{Mendes de Oliveira}, \emph{{Variable stars
  in the Ursa Minor dwarf galaxy}},
  \href{http://dx.doi.org/10.1086/114830}{\emph{\aj} {\bf 96} (Aug., 1988)
  528--559}.

\bibitem{Mighell1999}
K.~J. {Mighell} and C.~J. {Burke}, \emph{{WFPC2 Observations of the Ursa Minor
  Dwarf Spheroidal Galaxy}}, \href{http://dx.doi.org/10.1086/300923}{\emph{\aj}
  {\bf 118} (July, 1999) 366--380},
  [\href{http://arxiv.org/abs/astro-ph/9903065}{{\tt astro-ph/9903065}}].

\bibitem{MINUIT}
F.~James and M.~Roos, \emph{{Minuit: A System for Function Minimization and
  Analysis of the Parameter Errors and Correlations}},
  \href{http://dx.doi.org/10.1016/0010-4655(75)90039-9}{\emph{Comput. Phys.
  Commun.} {\bf 10} (1975) 343--367}.

\bibitem{NumRecipes1992}
W.~H. Press, S.~A. Teukolsky, W.~T. Vetterling and B.~P. Flannery,
  \emph{{Numerical Recipes in FORTRAN: The Art of Scientific Computing}}.
\newblock 1992.

\bibitem{Ullio:2001fb}
P.~Ullio, H.~Zhao and M.~Kamionkowski, \emph{{A Dark matter spike at the
  galactic center?}},
  \href{http://dx.doi.org/10.1103/PhysRevD.64.043504}{\emph{Phys. Rev.} {\bf
  D64} (2001) 043504}, [\href{http://arxiv.org/abs/astro-ph/0101481}{{\tt
  astro-ph/0101481}}].

\bibitem{2009ApJL_UMi_BH}
V.~{Lora}, F.~J. {S{\'a}nchez-Salcedo}, A.~C. {Raga} and A.~{Esquivel},
  \emph{{An Upper Limit on the Mass of the Black Hole in Ursa Minor Dwarf
  Galaxy}}, \href{http://dx.doi.org/10.1088/0004-637X/699/2/L113}{\emph{\apjl}
  {\bf 699} (July, 2009) L113--L117},
  [\href{http://arxiv.org/abs/0906.0951}{{\tt 0906.0951}}].

\bibitem{2013_UMi_BH}
A.~A. {Nucita}, F.~{De Paolis}, L.~{Manni} and G.~{Ingrosso}, \emph{{Hint for a
  faint intermediate mass black hole in the Ursa Minor dwarf galaxy}},
  \href{http://dx.doi.org/10.1016/j.newast.2013.03.003}{\emph{New Astronomy}
  {\bf 23} (Oct., 2013) 107--112}.

\bibitem{Caldwell:2008fw}
A.~Caldwell, D.~Kollar and K.~Kroninger, \emph{{BAT: The Bayesian Analysis
  Toolkit}}, \href{http://dx.doi.org/10.1016/j.cpc.2009.06.026}{\emph{Comput.
  Phys. Commun.} {\bf 180} (2009) 2197--2209},
  [\href{http://arxiv.org/abs/0808.2552}{{\tt 0808.2552}}].

\bibitem{Baring:2015sza}
M.~G. Baring, T.~Ghosh, F.~S. Queiroz and K.~Sinha, \emph{{New Limits on the
  Dark Matter Lifetime from Dwarf Spheroidal Galaxies using Fermi-LAT}},
  \href{http://arxiv.org/abs/1510.00389}{{\tt 1510.00389}}.

\end{thebibliography}\endgroup
\bibliographystyle{JHEP}

\end{document}